\documentclass[longauth]{aa}  

\makeatletter
\renewcommand*\aa@pageof{, page \thepage{} of \pageref*{LastPage}}
\makeatother

\renewenvironment{thebibliography}[1]{
  \begin{oldthebibliography}{#1}
    \setlength{\itemsep}{0em}
    \setlength{\parskip}{0em}
}
{
  \end{oldthebibliography}
}

\usepackage{euclid}
\usepackage{graphicx}
\usepackage{subcaption}
\usepackage{txfonts}
\usepackage[pdfencoding=auto,psdextra]{hyperref}
\hypersetup{
    colorlinks=true,
    linkcolor=blue,
    filecolor=magenta,      
    urlcolor=blue,
    citecolor=blue
}
\graphicspath{ {figures/} }
\DeclareGraphicsExtensions{.pdf, .PDF}

\newcommand{\vlos}{v_{||}}
\newcommand{\vlosrand}{\tilde{v}_{||}}
\newcommand{\sigmalos}{\sigma_{\vlos}}
\newcommand{\aperp}{\alpha_\perp}
\newcommand{\apar}{\alpha_\parallel}
\newcommand{\zeff}{z_{\rm eff}}

\newcommand{\voxel}{\texttt{voxel}\xspace}
\newcommand{\zobov}{\texttt{ZOBOV}\xspace}
\newcommand{\Revolver}{\texttt{REVOLVER}\xspace}

\usepackage[nameinlink,noabbrev]{cleveref}
\crefname{equation}{Eq.}{Eqs.}
\crefname{section}{Sect.}{Sects.}
\crefname{figure}{Fig.}{Figs.}
\crefname{table}{Table}{Tables}
\crefname{appendix}{Appendix}{Appendices}
\Crefname{figure}{Figure}{Figures}
\Crefname{equation}{Equation}{Equations}
\Crefname{section}{Section}{Sections}
\Crefname{table}{Table}{Tables}

\begin{document} 

  \title{\Euclid: Cosmology forecasts from the void-galaxy cross-correlation function with reconstruction\thanks{This paper is published on behalf of the Euclid Consortium.}}

    \newcommand{\orcid}[1]{} 
    \author{S.~Radinović$^{1}$\thanks{E-mail: \url{sladana.radinovic@astro.uio.no}}, S.~Nadathur\orcid{0000-0001-9070-3102}$^{2}$, H.-A.~Winther$^{1}$, W.J.~Percival\orcid{0000-0002-0644-5727}$^{3,4,5}$, A.~Woodfinden\orcid{0000-0002-5887-3205}$^{3,4}$, E.~Massara\orcid{0000-0002-0637-8042}$^{3,4}$, E.~Paillas\orcid{0000-0002-4637-2868}$^{3,4}$, S.~Contarini\orcid{0000-0002-9843-723X}$^{6,7,8}$, N.~Hamaus\orcid{0000-0002-0876-2101}$^{9}$, A.~Kovacs\orcid{0000-0002-5825-579X}$^{10,11,12}$
% ...
}

    \institute{$^{1}$ Institute of Theoretical Astrophysics, University of Oslo, P.O. Box 1029 Blindern, 0315 Oslo, Norway\\
    $^{2}$ Institute of Cosmology and Gravitation, University of Portsmouth, Portsmouth PO1 3FX, UK\\
    $^{3}$ Centre for Astrophysics, University of Waterloo, Waterloo, Ontario N2L 3G1, Canada\\
    $^{4}$ Department of Physics and Astronomy, University of Waterloo, Waterloo, Ontario N2L 3G1, Canada\\
    $^{5}$ Perimeter Institute for Theoretical Physics, Waterloo, Ontario N2L 2Y5, Canada\\
    $^{6}$ Dipartimento di Fisica e Astronomia "Augusto Righi" - Alma Mater Studiorum Universit\`{a} di Bologna, via Piero Gobetti 93/2, 40129 Bologna, Italy\\
    $^{7}$ INFN-Sezione di Bologna, Viale Berti Pichat 6/2, 40127 Bologna, Italy\\
    $^{8}$ INAF-Osservatorio di Astrofisica e Scienza dello Spazio di Bologna, Via Piero Gobetti 93/3, 40129 Bologna, Italy\\
    $^{9}$ Universit\"ats-Sternwarte M\"unchen, Fakult\"at f\"ur Physik, Ludwig-Maximilians-Universit\"at M\"unchen, Scheinerstrasse 1, 81679 M\"unchen, Germany\\
    $^{10}$ Instituto de Astrof\'isica de Canarias, Calle V\'ia L\'actea s/n, 38204, San Crist\'obal de La Laguna, Tenerife, Spain\\
    $^{11}$ Departamento de Astrof\'{i}sica, Universidad de La Laguna, 38206, La Laguna, Tenerife, Spain\\
    $^{12}$ MTA-CSFK Lend\"ulet Large-Scale Structure Research Group, Konkoly-Thege Mikl\'os \'ut 15-17, H-1121 Budapest, Hungary\\
    $^{13}$ Center for Computational Astrophysics, Flatiron Institute, 162 5th Avenue, 10010, New York, NY, USA\\
    $^{14}$ The Cooper Union for the Advancement of Science and Art, 41 Cooper Square, New York, NY 10003, USA\\
    $^{15}$ Department of Astrophysical Sciences, Peyton Hall, Princeton University, Princeton, NJ 08544, USA\\
    $^{16}$ INFN-Padova, Via Marzolo 8, 35131 Padova, Italy\\
    $^{17}$ Dipartimento di Fisica e Astronomia "G.Galilei", Universit\'a di Padova, Via Marzolo 8, 35131 Padova, Italy\\
    $^{18}$ Univ Lyon, Univ Claude Bernard Lyon 1, CNRS/IN2P3, IP2I Lyon, UMR 5822, 69622, Villeurbanne, France\\
    $^{19}$ INAF-Osservatorio Astrofisico di Torino, Via Osservatorio 20, 10025 Pino Torinese (TO), Italy\\
    $^{20}$ Dipartimento di Fisica, Universit\`{a} di Genova, Via Dodecaneso 33, 16146, Genova, Italy\\
    $^{21}$ INFN-Sezione di Roma Tre, Via della Vasca Navale 84, 00146, Roma, Italy\\
    $^{22}$ Department of Physics "E. Pancini", University Federico II, Via Cinthia 6, 80126, Napoli, Italy\\
    $^{23}$ Dipartimento di Fisica, Universit\'a degli Studi di Torino, Via P. Giuria 1, 10125 Torino, Italy\\
    $^{24}$ INFN-Sezione di Torino, Via P. Giuria 1, 10125 Torino, Italy\\
    $^{25}$ INAF-IASF Milano, Via Alfonso Corti 12, 20133 Milano, Italy\\
    $^{26}$ INAF-Osservatorio Astronomico di Roma, Via Frascati 33, 00078 Monteporzio Catone, Italy\\
    $^{27}$ INFN-Sezione di Roma, Piazzale Aldo Moro, 2 - c/o Dipartimento di Fisica, Edificio G. Marconi, 00185 Roma, Italy\\
    $^{28}$ Institut de F\'{i}sica d'Altes Energies (IFAE), The Barcelona Institute of Science and Technology, Campus UAB, 08193 Bellaterra (Barcelona), Spain\\
    $^{29}$ Port d'Informaci\'{o} Cient\'{i}fica, Campus UAB, C. Albareda s/n, 08193 Bellaterra (Barcelona), Spain\\
    $^{30}$ INAF-Osservatorio Astronomico di Capodimonte, Via Moiariello 16, 80131 Napoli, Italy\\
    $^{31}$ INFN section of Naples, Via Cinthia 6, 80126, Napoli, Italy\\
    $^{32}$ Dipartimento di Fisica e Astronomia "Augusto Righi" - Alma Mater Studiorum Universit\'a di Bologna, Viale Berti Pichat 6/2, 40127 Bologna, Italy\\
    $^{33}$ Centre National d'Etudes Spatiales, Toulouse, France\\
    $^{34}$ Institut national de physique nucl\'eaire et de physique des particules, 3 rue Michel-Ange, 75794 Paris C\'edex 16, France\\
    $^{35}$ Institute for Astronomy, University of Edinburgh, Royal Observatory, Blackford Hill, Edinburgh EH9 3HJ, UK\\
    $^{36}$ ESAC/ESA, Camino Bajo del Castillo, s/n., Urb. Villafranca del Castillo, 28692 Villanueva de la Ca\~nada, Madrid, Spain\\
    $^{37}$ European Space Agency/ESRIN, Largo Galileo Galilei 1, 00044 Frascati, Roma, Italy\\
    $^{38}$ Institute of Physics, Laboratory of Astrophysics, Ecole Polytechnique F\'{e}d\'{e}rale de Lausanne (EPFL), Observatoire de Sauverny, 1290 Versoix, Switzerland\\
    $^{39}$ Departamento de F\'isica, Faculdade de Ci\^encias, Universidade de Lisboa, Edif\'icio C8, Campo Grande, PT1749-016 Lisboa, Portugal\\
    $^{40}$ Instituto de Astrof\'isica e Ci\^encias do Espa\c{c}o, Faculdade de Ci\^encias, Universidade de Lisboa, Campo Grande, 1749-016 Lisboa, Portugal\\
    $^{41}$ Universit\'e Paris-Saclay, CNRS, Institut d'astrophysique spatiale, 91405, Orsay, France\\
    $^{42}$ Department of Astronomy, University of Geneva, ch. d'Ecogia 16, 1290 Versoix, Switzerland\\
    $^{43}$ Universit\'e Paris-Saclay, Universit\'e Paris Cit\'e, CEA, CNRS, Astrophysique, Instrumentation et Mod\'elisation Paris-Saclay, 91191 Gif-sur-Yvette, France\\
    $^{44}$ Institute of Space Sciences (ICE, CSIC), Campus UAB, Carrer de Can Magrans, s/n, 08193 Barcelona, Spain\\
    $^{45}$ Institut d'Estudis Espacials de Catalunya (IEEC), Carrer Gran Capit\'a 2-4, 08034 Barcelona, Spain\\
    $^{46}$ INAF-Osservatorio Astronomico di Trieste, Via G. B. Tiepolo 11, 34143 Trieste, Italy\\
    $^{47}$ Aix-Marseille Universit\'e, CNRS/IN2P3, CPPM, Marseille, France\\
    $^{48}$ INAF-Osservatorio Astronomico di Padova, Via dell'Osservatorio 5, 35122 Padova, Italy\\
    $^{49}$ Max Planck Institute for Extraterrestrial Physics, Giessenbachstr. 1, 85748 Garching, Germany\\
    $^{50}$ Jet Propulsion Laboratory, California Institute of Technology, 4800 Oak Grove Drive, Pasadena, CA, 91109, USA\\
    $^{51}$ Technical University of Denmark, Elektrovej 327, 2800 Kgs. Lyngby, Denmark\\
    $^{52}$ Cosmic Dawn Center (DAWN), Denmark\\
    $^{53}$ Max-Planck-Institut f\"ur Astronomie, K\"onigstuhl 17, 69117 Heidelberg, Germany\\
    $^{54}$ Mullard Space Science Laboratory, University College London, Holmbury St Mary, Dorking, Surrey RH5 6NT, UK\\
    $^{55}$ Department of Physics, P.O. Box 64, 00014 University of Helsinki, Finland\\
    $^{56}$ Helsinki Institute of Physics, Gustaf H{\"a}llstr{\"o}min katu 2, University of Helsinki, Helsinki, Finland\\
    $^{57}$ NOVA optical infrared instrumentation group at ASTRON, Oude Hoogeveensedijk 4, 7991PD, Dwingeloo, The Netherlands\\
    $^{58}$ Argelander-Institut f\"ur Astronomie, Universit\"at Bonn, Auf dem H\"ugel 71, 53121 Bonn, Germany\\
    $^{59}$ Department of Physics, Institute for Computational Cosmology, Durham University, South Road, DH1 3LE, UK\\
    $^{60}$ Universit\'e Paris Cit\'e, CNRS, Astroparticule et Cosmologie, 75013 Paris, France\\
    $^{61}$ University of Applied Sciences and Arts of Northwestern Switzerland, School of Engineering, 5210 Windisch, Switzerland\\
    $^{62}$ Institut d'Astrophysique de Paris, 98bis Boulevard Arago, 75014, Paris, France\\
    $^{63}$ Institut d'Astrophysique de Paris, UMR 7095, CNRS, and Sorbonne Universit\'e, 98 bis boulevard Arago, 75014 Paris, France\\
    $^{64}$ CEA Saclay, DFR/IRFU, Service d'Astrophysique, Bat. 709, 91191 Gif-sur-Yvette, France\\
    $^{65}$ European Space Agency/ESTEC, Keplerlaan 1, 2201 AZ Noordwijk, The Netherlands\\
    $^{66}$ Leiden Observatory, Leiden University, Niels Bohrweg 2, 2333 CA Leiden, The Netherlands\\
    $^{67}$ Kapteyn Astronomical Institute, University of Groningen, PO Box 800, 9700 AV Groningen, The Netherlands\\
    $^{68}$ Department of Physics and Astronomy, University of Aarhus, Ny Munkegade 120, DK-8000 Aarhus C, Denmark\\
    $^{69}$ AIM, CEA, CNRS, Universit\'{e} Paris-Saclay, Universit\'{e} de Paris, 91191 Gif-sur-Yvette, France\\
    $^{70}$ Space Science Data Center, Italian Space Agency, via del Politecnico snc, 00133 Roma, Italy\\
    $^{71}$ Institute of Space Science, Bucharest, 077125, Romania\\
    $^{72}$ Departamento de F\'isica, FCFM, Universidad de Chile, Blanco Encalada 2008, Santiago, Chile\\
    $^{73}$ Institut de Ciencies de l'Espai (IEEC-CSIC), Campus UAB, Carrer de Can Magrans, s/n Cerdanyola del Vall\'es, 08193 Barcelona, Spain\\
    $^{74}$ Aix-Marseille Universit\'e, CNRS, CNES, LAM, Marseille, France\\
    $^{75}$ Centro de Investigaciones Energ\'eticas, Medioambientales y Tecnol\'ogicas (CIEMAT), Avenida Complutense 40, 28040 Madrid, Spain\\
    $^{76}$ Instituto de Astrof\'isica e Ci\^encias do Espa\c{c}o, Faculdade de Ci\^encias, Universidade de Lisboa, Tapada da Ajuda, 1349-018 Lisboa, Portugal\\
    $^{77}$ Universidad Polit\'ecnica de Cartagena, Departamento de Electr\'onica y Tecnolog\'ia de Computadoras, 30202 Cartagena, Spain\\
    $^{78}$ Institut de Recherche en Astrophysique et Plan\'etologie (IRAP), Universit\'e de Toulouse, CNRS, UPS, CNES, 14 Av. Edouard Belin, 31400 Toulouse, France\\
    $^{79}$ Infrared Processing and Analysis Center, California Institute of Technology, Pasadena, CA 91125, USA\\
    $^{80}$ Junia, EPA department, 59000 Lille, France}
\date{\today}

\authorrunning{S. Radinović et al.}
\titlerunning{Cosmology from the void-galaxy correlation in \Euclid}


\abstract{We have investigated the cosmological constraints that can be expected from measurement of the cross-correlation of galaxies with cosmic voids identified in the \Euclid spectroscopic survey, which will include spectroscopic information for tens of millions of galaxies over $15\,000$ deg$^2$ of the sky in the redshift range $0.9\leq z<1.8$. We have done this using simulated measurements obtained from the Flagship mock catalogue, the official \Euclid mock that closely matches the expected properties of the spectroscopic dataset.
To mitigate anisotropic selection-bias effects, we have used a velocity field reconstruction method to remove large-scale redshift-space distortions from the galaxy field before void-finding. This allowed us to accurately model contributions to the observed anisotropy of the cross-correlation function arising from galaxy velocities around voids as well as from the Alcock--Paczynski effect, and we studied the dependence of constraints on the efficiency of reconstruction.
We find that \Euclid voids will be able to constrain the ratio of the transverse comoving distance $D_{\rm M}$ and Hubble distance $D_{\rm H}$ to a relative precision of about $0.3\%$, and the growth rate $f\sigma_8$ to a precision of between $5\%$ and $8\%$ in each of the four redshift bins covering the full redshift range. In the standard cosmological model, this translates to a statistical uncertainty $\Delta\Omega_\mathrm{m}=\pm0.0028$ on the matter density parameter from voids, which is better than what can be achieved from either \Euclid galaxy clustering and weak lensing individually. We also find that voids alone can measure the dark energy equation of state to a $6\%$ precision.
}


\maketitle


\section{Introduction}

\Euclid \citep{Laureijs:2011} is an upcoming space mission that aims to explore the nature of dark matter and dark energy through observations of galaxy clustering and weak lensing. Clustering statistics will mainly be extracted from the spectroscopic dataset, which will contain accurate spectroscopic redshifts for tens of millions of galaxies, spanning a wide redshift range ($0.9\leq z<1.8$) and covering almost a third of the sky. The primary analyses of such data usually focus on the galaxy auto-correlation statistics, in particular the two-point correlation function (2PCF) and its Fourier space counterpart, the power spectrum. These analyses allow for the measurement of the characteristic baryon acoustic oscillation (BAO) feature \citep[e.g.][]{Alam-BOSS:2017,Alam-eBOSS:2021}, together with the signatures of galaxy velocities which give rise to redshift-space distortions \citep[RSD;][]{Kaiser:1987}. However, non-Gaussianity introduced by non-linear gravitational evolution means that two-point statistics do not fully characterise all of the information available from galaxy surveys. In addition, models based on perturbation theory start to break down at small scales (at $r\lesssim 25\, \hMpc$ in the 2PCF, or $k\gtrsim 0.15\, h{\rm Mpc}^{-1}$ in the power spectrum); therefore, information from these smaller scales is often excluded from analysis to avoid systematic biases in the modelling \citep{BOSS:2016psr, BOSS:2016ntk}.

Several alternative clustering summary statistics have been used to extract additional information from the galaxy distribution, such as polyspectra \citep[e.g.][]{Verde:2002, Scoccimarro:2001, Gil-Marin:2016, Philcox:2022a} and $N$-point correlation functions \citep[e.g.][]{Wang:2004, Nichol:2006, Marin:2013, Guo:2015, Slepian:2017}. Other proposed statistics include marked correlation functions or power spectra \citep[e.g.][]{White:2016, Satpathy:2019, Massara:2022a}, counts-in-cells statistics \citep[e.g.][]{Yang:2011, Uhlemann:2017}, density-dependent clustering \citep[e.g.][]{Tinker:2007, Chiang:2014, Bayer:2021, Bonnaire:2022, Paillas:2022}, 
anisotropic clustering \citep[e.g.][]{Paz:2011fj, Kazin:2012, Correa:2018vge},
wavelet-based methods \citep[e.g.][]{Valogiannis:2022b, Valogiannis:2022a}, non-linear transformations of the galaxy field \citep[e.g.][]{Neyrinck:2009, Carron:2014, Wolk:2015}, and density field reconstruction \citep[e.g.][]{Wang:2022}.

With the advent of large galaxy redshift surveys, cosmic voids -- large low-density regions in the galaxy distribution -- have emerged as interesting probes of cosmology in many contexts, for example using the void size function \citep[e.g.][]{Pisani:2015, Nadathur:2016a, Correa:2018vge, Contarini:2022}, gravitational lensing by voids \citep[e.g.][]{Sanchez:2016, Raghunathan:2020, Bonici:2022}, secondary cosmic microwave background (CMB) anisotropies \citep[e.g.][]{Granett:2008,Nadathur:2016b,Alonso:2018,Kovacs:2019}, or void clustering \citep[e.g.][]{Chan:2014,Kitaura:2016b,Zhao:2021}. Readers can refer to \citet{Pisani_WP} for a comprehensive list.

The anisotropic distribution of galaxies around voids is a particularly interesting observable \citep{Lavaux:2012} that has been shown to be a valuable source of cosmological information, providing parameter constraints that are highly complementary to those obtained from galaxy clustering \citep[e.g.][]{Nadathur:2019c, Hamaus:2022, Woodfinden:2022}. Individual voids have irregular shapes and arbitrary orientations on the sky, but, given a large enough sample of voids, the assumption of statistical isotropy of the Universe (together with some additional assumptions discussed in more detail in \cref{sec:Theory}) should give rise to a spherically symmetric distribution of galaxies around voids on average \citep{Ryden:1995}. Two effects that spoil this symmetry are distortions due to the Alcock--Paczyński (AP) effect \citep{Alcock:1979}, arising from the choice of the fiducial cosmological model used to convert observed galaxy redshifts and angles on the sky into distances, and RSD arising from the peculiar velocities of galaxies around these voids. 

As voids are underdense environments, the RSD contributions to the anisotropy can be relatively successfully modelled using linear theory \citep{Hamaus:2014a,Cai:2016a,Nadathur:2019a,Paillas:2021} without needing to exclude small scales. This provides constraints on the growth rate of structure parametrised by $f\sigma_8$ or $\beta$ \citep[e.g.][]{Paz:2013,Hamaus:2016,Hawken:2017,Nadathur:2019c,Achitouv:2019,Hawken:2020,Aubert20a,Woodfinden:2022}. More importantly, \cite{Hamaus:2015} and \cite{Nadathur:2019c} show that the anisotropies from RSD contributions can be easily distinguished from those arising due to the AP effect. This allows the stacked void profile -- equivalent to the void-galaxy cross-correlation function (CCF) -- to be used as a `standard sphere' for the AP test, as originally advocated by \citet{Lavaux:2012}. This observable can then be used to measure the AP parameter, which is the ratio of the transverse comoving distance $D_{\rm M}(z)$ to the Hubble distance $D_{\rm H}(z)$ as a function of redshift. The constraints obtained on this quantity by applying this method to the BOSS and eBOSS surveys are a factor of $1.7$ to $3.5$ more precise than the constraints from galaxy clustering and the BAO in the same data \citep{Hamaus:2016,Nadathur:2019c,Nadathur:2020b}. This precise AP measurement underpins the power of the void-galaxy CCF as a cosmological probe: \citet{Nadathur:2020a} showed that the combination of voids and BAO measurements alone provides greater than $10\sigma$ evidence for late-time cosmic acceleration, independent of the cosmic microwave background or supernovae.

In this work we aim to forecast the constraints on the AP parameter $D_{\rm M}/D_{\rm H}$ and the growth rate $f\sigma_8$ that can be obtained from measurements of the void-galaxy CCF by \Euclid. This paper is one of a set of four companion papers published on behalf of the Euclid Consortium, which forecast the expected constraints from cosmic voids using different observables. \cite{Contarini:2022} explored the use of the void size function, \cite{Bonici:2022} investigated the cross-correlation of voids with weak lensing, while this paper and \cite{Hamaus:2022} examine different methods of studying the void-galaxy CCF. The primary difference between our work and that of \cite{Hamaus:2022} is that we use a velocity-field reconstruction method to approximately remove large-scale galaxy RSD effects before running the void-finding pipeline \citep{Nadathur:2019b} in order to mitigate against systematic errors due to anisotropic selection effects. On the other hand, instead of applying reconstruction, \cite{Hamaus:2022} modify terms in the theoretical model used for the fits, and so measure a very different pattern of anisotropy in the CCF.

The structure of this paper is as follows: in \cref{sec:Theory} we provide a theoretical description of the void-galaxy cross-correlation function, as well as the AP test. In \cref{sec:Data} we describe the Flagship galaxy mock and the reconstruction method we apply and which enables us to recover the approximate real-space galaxy positions. We also describe the void finder and detail the properties of the void catalogue. In \cref{sec:Method} we describe our pipeline, obtaining the data vectors and translating this into constraints on parameters of interest. \cref{sec:Results} describes the results we obtain, which are then summarised and put into a larger context in \cref{sec:Conclusions}.


\section{Theory}
\label{sec:Theory}


\subsection{The void-galaxy cross-correlation function in redshift space}

The void-galaxy CCF $\xi$ represents the excess probability of finding a galaxy at a given separation from a void centre. We use the notation $\xi^{\rm rr}$ to denote this quantity when the cross-correlation is performed between the real-space positions of galaxies and the positions of voids identified in this real-space galaxy distribution. For convenience, we refer to $\xi^{\rm rr}$ as the real-space void-galaxy CCF. For the same fixed void centres, the cross-correlation with the redshift-space galaxy positions is then denoted $\xi^{\rm rs}$, and will be referred to as the redshift-space CCF. Here $\rm r$ refers to real space, $\rm s$ to redshift space, the first superscript refers to the voids, and the second to the galaxies. We denote the vectors from the void centre to the galaxy in real and redshift space by $\vec r$ and $\vec s$ respectively, and decompose them into their components perpendicular and parallel to the line-of-sight direction, $\vec s = \vec s_\perp + \vec s_{||}$. Thus $\vec s_\perp=\vec r_\perp$ while
\begin{equation}
    \label{eq:spar}
    \vec s_{||} = \vec r_{||} + \frac{\vec \vlos}{aH}\,,
\end{equation}
where $a$ is the scale factor, $H$ the Hubble rate, $\vlos$ is the component of the galaxy peculiar velocity along the line of sight \citep[not the pairwise velocity; see e.g.][]{Massara:2022b}, and the line of sight is directed through the void centre.

Since, by construction, the number of void-galaxy pairs is conserved under the mapping between real and redshift space, $\xi^{\rm rs}$ and $\xi^{\rm rr}$ are related by a convolution with the (position-dependent) velocity distribution:
\begin{equation}
    \label{eq:CCF_mapping_streaming}
	1 + \xi^{\rm rs}(\vec s) = 
			\int [1 + \xi^{\rm rr}(\vec r)] \, 
			P(\vlos,\vec r) \, \diff\vlos \,.
\end{equation}
This is recognisable as the streaming model \citep[e.g.][]{Peebles:1979}, which was used to describe the void-galaxy CCF by \cite{Paz:2013} and also, for example, by \cite{Cai:2016a} and \cite{Paillas:2021}. Although -- for reasons explained below -- we have chosen to hold fixed the void positions identified in the real-space galaxy field, \cref{eq:CCF_mapping_streaming} would be equally valid if both the CCFs were defined with respect to the void positions in the redshift-space galaxy field, that is if $\xi^{\rm rs}$ and $\xi^{\rm rr}$ were replaced by $\xi^{\rm ss}$ and $\xi^{\rm sr}$ respectively. Indeed this is the approach taken in a number of other papers \citep{Paz:2013, Hamaus:2016, Achitouv:2017b}. However, the number of voids in a size-selected sample is not conserved under the RSD mapping \citep{Chuang2017, Nadathur:2019b,Correa:2022}, so \cref{eq:CCF_mapping_streaming} cannot be used to relate CCFs obtained from running void-finding independently in the two spaces, for example $\xi^{\rm ss}$ and $\xi^{\rm rr}$. 

We can view the galaxy peculiar velocities as being composed of a term describing a coherent velocity outflow from the void centre, plus a random velocity component along a random direction. Given the assumption of statistical isotropy in real space, when averaged over sufficiently large numbers of voids, the coherent outflow $\vec v(r)=v_{\rm r}(r)\,\hat{\vec r}$ must be directed radially outwards from the void centre and be spherically symmetric about it. Thus the line-of-sight component of the galaxy velocity can be written as
\begin{equation}
    \label{eq:line-of-sight_velocity}
	\vlos(r, \mu_{\rm r}) = v_{\rm r}(r) \, \mu_{\rm r} + \vlosrand \,,
\end{equation}
where $\mu_{\rm r}\equiv\cos(\theta)$ is the cosine of the angle $\theta$ between $\vec r$ and the line of sight, and $\vlosrand$ is a zero-mean random variable describing the projection of the stochastic velocity component along the line of sight. 

Using \cref{eq:line-of-sight_velocity} to change variables, \cite{Woodfinden:2022} show that \cref{eq:CCF_mapping_streaming} is equivalent to
\begin{equation}
\label{eq:CCF_mapping_dispersion}
	1 + \xi^{\rm rs}(\vec s) = 
			\int [1 + \xi^{\rm rr}(\vec r)] \, J_{\rm \vec r, \vec s} \, 
			P(\vlosrand,\vec r)\, \diff\vlosrand \,,
\end{equation}
where
\begin{equation}
\label{eq:Jacobian}
	J_{\rm \vec r, \vec s} \equiv  \abs{\frac{\diff\vec r}{\diff\vec s}} = 
		\left[1 + \frac{v_{\rm r}}{r\,a\,H} + \frac{r\, v_{\rm r}^\prime - v_{\rm r}}{r\,a\,H} \mu_{\rm r}^2\right]\inv
\end{equation}
is the Jacobian of the transformation from real to redshift space, prime denotes derivatives with respect to $r$, and $P(\vlosrand,{\vec r})$ now describes the distribution of $\vlosrand$ about zero. \Cref{eq:CCF_mapping_dispersion} is the same model described by \citet{Nadathur:2019a}. Several other works \citep[e.g.][]{Cai:2016a,Hawken:2020, Aubert20a} use approximations to this expression, first by assuming a Dirac delta function form for the distribution $P(\vlosrand,{\vec r})$, which reduces the integral equation to 
\begin{equation}
    \label{eq:Kaiser}
    1 + \xi^{\rm rs}(\vec s) = [1 + \xi^{\rm rr}(\vec r)] \, J_{\rm \vec r, \vec s},
\end{equation}
and then approximating the square brackets in \cref{eq:Jacobian} by a series expansion, which one can choose to truncate at some desired order. \citet{Hamaus:2020} introduce additional nuisance parameters to \cref{eq:Kaiser}, and \cite{Hamaus:2022} further modify some terms of the Jacobian expansion.

A key assumption for all models based on \cref{eq:CCF_mapping_dispersion} is that the coherent velocity term $v_{\rm r}(r)$ is spherically symmetric and radially directed, which depends crucially on the assumption that in real space there is no preferred orientation direction for the large number of individually asymmetric voids. This follows from statistical isotropy only provided there is no orientation-dependent selection effect introduced in creating the void sample. \citet{Nadathur:2019b} argued that exactly such an effect is introduced for voids identified in redshift space, since regions with velocity outflows oriented along the line of sight direction appear to have lower densities in redshift space. This means that elongated voids aligned along the line-of-sight direction are more likely to be selected as void candidates if void-finding is performed on the redshift-space galaxy positions. 

This orientation-dependent bias has a large effect on the measured CCFs \citep{Nadathur:2019b, Correa:2022, Paillas:2022}. The CCF with real-space galaxy positions, $\xi^{\rm sr}$, becomes strongly anisotropic, the coherent velocity outflow $v_{\rm r}$ no longer has spherical symmetry, and the multipoles of the CCF with redshift-space galaxy positions, $\xi^{\rm ss}$, have a very different shape. The loss of spherical symmetry in particular introduces significant challenges for the modelling approach described above. To avoid this problem, in this paper we do not consider the case of voids identified in the redshift-space galaxy field. As the real-space galaxy positions are in general not known, we used the reconstruction technique described in \cref{subs:reconstruction} below to approximately recover them before performing void-finding, and used these void positions to approximately measure $\xi^{\rm rs}$ and $\xi^{\rm rr}$.


\subsection{Modelling the velocity distribution}
\label{subs:velocities}

Given the stated assumptions, the convolution described by \cref{eq:CCF_mapping_dispersion} -- or, equivalently, \cref{eq:CCF_mapping_streaming} -- is completely general and exact. In order to make specific predictions, however, we need to specify a model for the velocity field around voids to obtain the radial term $v_{\rm r}(r)$ and the distribution $P(\vlosrand, \vec r)$.

The radial outflow velocity $v_{\rm r}$ itself is commonly modelled \citep[e.g.][]{Cai:2016a, Nadathur:2019c, Woodfinden:2022} using a a linearised form of the continuity equation \citep{Peebles_1980}:  
\begin{equation}
\label{eq:continuity}
	v_{\rm r}(r) = -\frac{1}{3} f\,a\,H\,r\,\Delta \,,
\end{equation}
where
\begin{equation}
\label{eq:density}
	\Delta(r) = \frac{3}{r^3} \int_{0}^{r} \delta(y)\, y^2\, \diff y
\end{equation}
is the fully non-linear enclosed density profile of dark matter, $\delta(r)=\rho/\bar{\rho}-1$ is the matter overdensity, and $f$ is the linear growth rate. \citet{Nadathur:2019c} found that in simulated galaxy catalogues the dark matter profile of voids approximately scales as $\Delta\propto\sigma_8$, where $\sigma_8$ describes the amplitude of the matter density perturbations on $8\,\hMpc$ scales.\footnote{This scaling was demonstrated under conditions where the number density of the galaxies used to trace voids was held constant, and their bias was adjusted to keep $b\sigma_8$ -- and thus the overall galaxy-clustering amplitude -- fixed.} When combined with \cref{eq:continuity}, this implies a scaling $v_{\rm r}\propto f\sigma_8$.

\Cref{eq:continuity} has been found to give a reasonably good description of the matter velocity field around voids by some studies \citep{Hamaus:2014a, Nadathur:2019a}. However, other works \citep{Achitouv:2017b,Paillas:2021} have found that it can be necessary to add correction terms. Moreover, it has been shown by \cite{Massara:2022b} that for some void finders, when void-finding is performed on sparse tracers of the matter density field such as galaxies, an effective velocity bias can arise between the matter velocity which satisfies \cref{eq:continuity}, and the galaxy velocity appearing in \cref{eq:Jacobian}. This can mean that even if \cref{eq:continuity} accurately describes the velocity field for matter, galaxy velocities can differ from this in the inner regions of voids. These considerations will need to be taken into account when improving the velocity modelling beyond that in \cref{eq:continuity}. It is not our aim to attempt this task in this paper, but only to forecast the constraints from the void-galaxy cross-correlation function that could be obtained from such a model in the future. We therefore applied a template-based method that assumes only that the scaling inferred from \cref{eq:continuity} holds. That is, we measured the mean velocity profile $v_{\rm r}^{\rm sim}(r)$ from the Flagship simulation as described in \cref{subs:templates} below, and assume that the velocity in other cosmological models is given by
\begin{equation}
    \label{eq:scaled_vel}
    \frac{v_{\rm r}(r)}{a\, H} = \frac{f\sigma_8}{f^{\rm sim}\sigma_8^{\rm sim}} \paren{\frac{v_{\rm r}(r)}{a\,H}}^{\rm sim}\,.
\end{equation}

For the velocity distribution $P(\vlosrand, \vec r)$, we assume a Gaussian form,
\begin{equation}
\label{eq:velocity_PDF}
	P(\vlosrand,\vec r) = \frac{1}{ \sqrt{2\, \pi\, \sigmalos^2(r)} } 
					  \exp\left( -\frac{\vlosrand^2}{ 2\, \sigmalos^2(r)} \right) \,,
\end{equation}
which has been shown to be a good approximation by \citet{Nadathur:2019a} and \citet{Paillas:2021}. As with $v_{\rm r}(r)$, we assume the width of the dispersion $\sigmalos(r)$ is spherically symmetric about the void centre and take a template-based approach to its modelling. We measured a template dispersion profile $\sigmalos^{\rm sim}(r)$ from the Flagship simulation (\cref{subs:templates}). At large distances from the void centre, this profile asymptotes to a constant, $\sigma_{\rm v}^{\rm sim}\equiv\sigmalos\superscr{sim}(r\rightarrow \infty)$ \citep[see also][]{Hamaus:2015}. To account for changes in other cosmologies, we renormalised this function by an amplitude factor $\sigma_{\rm v}$, taken to be a free parameter to be marginalised over:
\begin{equation}
	\sigmalos(r) =
		\frac{ \sigma_{\rm v} }{ \sigma_{\rm v}\superscr{sim} } \sigmalos\superscr{sim}(r) \,.
\end{equation}
In summary, the template-based approach to modelling the velocity field around voids requires the specification of two template functions, $v_{\rm r}^{\rm sim}(r)$ and $\sigmalos\superscr{sim}(r)$, and depends on two free parameters, $f\sigma_8$ and $\sigma_{\rm v}$.


\subsection{The real-space correlation function}
\label{subs:xir}

Although the velocity model has been specified as in \cref{subs:velocities}, the RSD mapping of the void-galaxy correlation described by \cref{eq:CCF_mapping_dispersion} only relates $\xi^{\rm rs}$ to its counterpart in real space, $\xi^{\rm rr}$. Though some empirical fitting formulae have been used \citep{Paz:2013, Hamaus:2014a, Hawken:2017}, models which attempt to predict $\xi^{\rm rr}$ directly from the cosmological parameters do not exist at present \footnote{Any such model, if developed, would have to account for the selection criteria by which the void sample was chosen, and would thus be specific to a particular void finder, whereas \cref{eq:CCF_mapping_dispersion} is completely general.}. Previous studies have therefore taken different approaches to obtain the real-space CCF. \cite{Hamaus:2020,Hamaus:2022} apply an inversion algorithm \citep{Pisani:2014} to recover it from the redshift-space CCF projected along the line of sight (implictly assuming that the real-space correlation function is spherically symmetric about the void centre). \citet{Nadathur:2019c,Nadathur:2020b} and \citet{Woodfinden:2022} instead measure the average $\xi^{\rm rr}$ from a large number of mocks that match the galaxy clustering seen in the data sample: this is effectively a template-based method similar to that described for $v_{\rm r}(r)$ and $\sigmalos(r)$ above. 

In this paper, we use the reconstruction-based estimators described in \cref{subs:reconstruction} to estimate the real-space CCF $\xi^{\rm rr}(\vec r)$ from the simulation data. While we have argued that in the absence of orientation-dependent selection effects, statistical isotropy should imply spherical symmetry, $\xi^{\rm rr}(\vec r)=\xi^{\rm rr}(r)$, when measured in the true background cosmology, this will in general not be true in the presence of the AP distortions discussed in \cref{subs:AP}, and could also be spoiled by imperfections in the reconstruction-based RSD removal step. Our implementation of \cref{eq:CCF_mapping_dispersion} therefore uses the full $\xi^{\rm rr}(\vec r)$ without imposing an assumption of isotropy.

Our treatment of $\xi^{\rm rr}$ here differs from that in \citet{Nadathur:2019c} and \citet{Woodfinden:2022} because, unlike in those works, we currently only have access to the single representative Flagship simulation here, so we cannot construct a template based on the mean of thousands of mocks (this situation will change by the time of the final \Euclid data analyses). This means that our estimates of $\xi^{\rm rr}$ are necessarily much noisier. As it is measured from the same data as the redshift-space correlation function $\xi^{\rm rs}$ we wish to fit, correlations between the theory model and data vector may also be introduced, and we consider the implications of this in \cref{sec:Appendix}.


\subsection{Alcock--Paczyński effect}
\label{subs:AP}

Besides RSD, another source of anisotropies in the observed void-galaxy cross-correlation function is the AP effect. This arises from the conversion of observed galaxy angular positions and redshifts to distances in order to apply the modelling developed above. This conversion requires the assumption of a fiducial cosmological model which may differ from the true background and thus lead to incorrectly inferred distances, introducing anisotropies.

To parameterise this effect, we introduce two AP scaling parameters describing the ratio of distances parallel and perpendicular to the line of sight to those in the fiducial model,
\begin{equation}
	\label{eq:AP_alphas}
    \apar  = \frac{D_{\rm H}(z) }{D_{\rm H}\superscr{fid}(z) } \,, \quad
    \aperp = \frac{ D_{\rm M}(z) }{ D_{\rm M}\superscr{fid}(z) } \,,
\end{equation}
where $D_\mathrm{M}(z)$ is the comoving angular diameter distance and $D_\mathrm{H}(z) = c/H(z)$ is the Hubble distance at redshift $z$. In this model the correlation function scales as 
\begin{equation}
    \label{eq:AP_xi_scaling}
    \xi^{\rm rs}(s_\perp, s_\parallel) = \xi^{\rm rs,\mathrm{fid}}\left(\aperp s_\perp^\mathrm{fid}, \apar s_\parallel^\mathrm{fid}\right)\,,
\end{equation}
where the superscript $^{\mathrm{fid}}$ denotes quantities calculated in the fiducial cosmology.

The parameters $\apar$ and $\aperp$ can also be recast into the following combinations:
\begin{equation}
	\label{eq:AP_alpha}
    \alpha = \aperp^{2/3}\apar^{1/3} \,,
\end{equation}
which describes an isotropic volume dilation, and 
\begin{equation}
	\label{eq:AP_epsilon}
    \epsilon = \frac{\aperp}{\apar} \,,
\end{equation}
describing anisotropic distortions. The dilation parameter $\alpha$ can be measured using a standard ruler, such as the BAO scale \citep[e.g.][]{Blake:2003, Seo:2003}, by comparing a measured distance scale to one predicted by fundamental physics models. 
In our analysis, however, the calibrated template functions used depend on the void size cut applied to the sample. This cut is numerically identical to the one applied to the data. However, in general if the template cosmology differs from the truth, due to AP dilation the same numerical cut might not select the same population of voids, leading to a shift in the mean void size. Therefore, we conservatively choose to avoid using the observed void size as an absolute ruler by marginalising over values of $\alpha$ in order to reduce the possible effects of the template cosmology. To achieve this we isotropically rescaled all of the template profiles $\xi^{\rm rr}(\vec r)$, $v_{\rm r}(r)$ and $\sigmalos(r)$ with $\alpha$, as described by \citet{Nadathur:2019c}. This rescaling removes all sensitivity of the model to $\alpha$. The model then depends only on the AP distortion parameter $\epsilon$, via its effect on the quadrupole and (to a lesser extent) hexadecapole moments.


\subsection{Multipole decomposition}
\label{subs:multipoles}

The model described above applies generally to the full three-dimensional correlation function, which may be written as $\xi^{\rm rs}(s, \mu_{\rm s})$, where $s=|\vec s|$ and $\mu_{\rm s}\equiv s_{||}/s$. In practice it is preferable to compress the information into a small number of Legendre multipoles, as
\begin{equation}
    \label{eq:multipole_decomposition}
	\xi^{\rm rs}_\ell(s) = \frac{2\ell+1}{2} 
	              \int_{-1}^{1} P_\ell(\mu_{\rm s}) \, \xi^{\rm rs}(s,\mu_{\rm s}) \, \diff\mu_{\rm s} \,,
\end{equation}
where $P_\ell(\mu)$ are the Legendre polynomials of order $\ell$. Symmetry dictates that only the even multipoles are non-zero, and we restrict our attention to the monopole, quadrupole and hexadecapole moments $(\ell=0, 2, 4)$ only, as these contain almost all of the measurable information in the correlation. 

The real-space cross-correlation function $\xi^{\rm rr}$ can also be decomposed into multipoles analogously to \cref{eq:multipole_decomposition}. As mentioned in \cref{subs:xir}, a commonly used approximation, justified by the assumption of spherical symmetry in real space, is that only the monopole term $\xi^{\rm rr}_0$ is non-zero. However, this assumption can be spoiled by the AP effect or imperfect reconstruction of real-space galaxy positions, so for generality we keep track of the high-order multipoles of $\xi^{\rm rr}$ as well.

\begin{figure*}
	\centering
	\includegraphics[width=0.93\hsize]{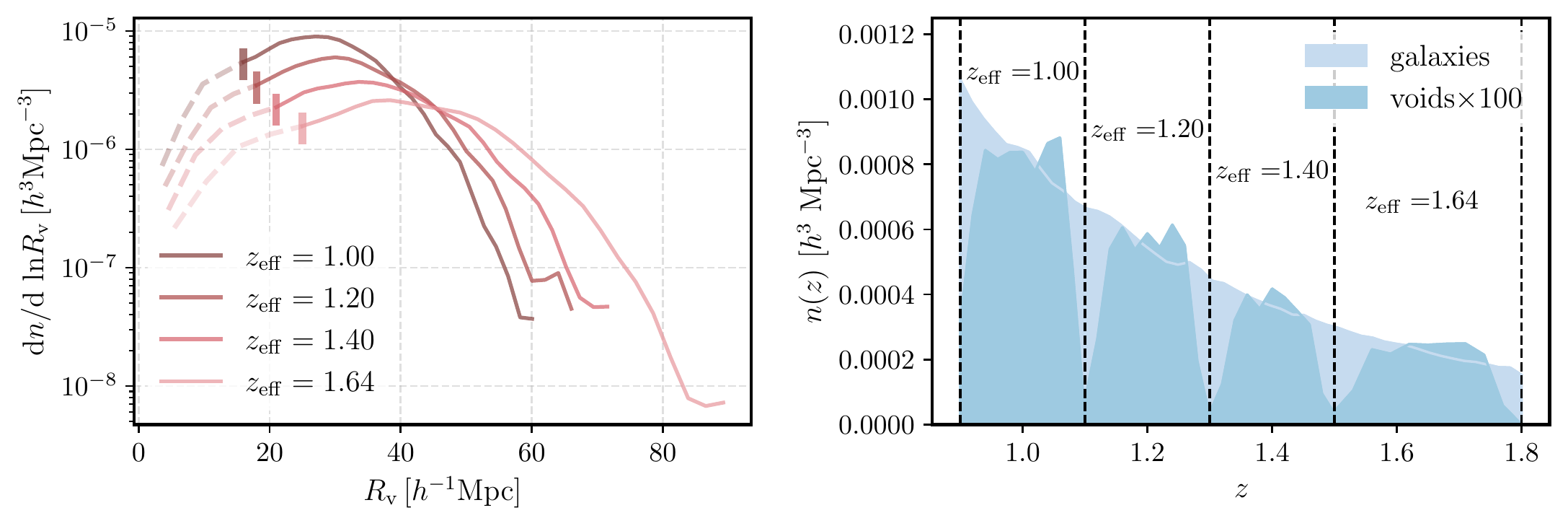}
	\caption{\textit{Left}: Distribution of void sizes in each of the four redshift bins of the Flagship mock catalogue. The vertical bars indicate the void size cuts applied to each catalogue, and dashed lines represent the voids which are discarded. \textit{Right}: Comoving number density $n(z)$ of galaxies and voids as function of redshift in Flagship. For voids, $n(z)$ is multiplied by 100 for visibility. The edges of the redshift bins are shown by dashed vertical lines. The decrease in the number of voids near the edges of each bin arises because void-finding is performed on each bin individually as discussed in the text; some void candidates near the redshift edges of the bin are then missed.}
	\label{fig:galaxies_voids}
\end{figure*}


\section{Data}
\label{sec:Data}


\subsection{Galaxy catalogue}
\label{subs:galaxies}

We used the \Euclid Flagship galaxy mock (Castander et al., in preparation), which is constructed from an {\it N}-body simulation with 2 trillion particles of mass $m_{\rm p}\approx 2\expo 9\, h\inv\si\solarmass$. The simulation was run using \texttt{PKDGRAV3} \citep{Potter:2016ttn} in a box of side $L=3780\, \hMpc$, with a flat $\Lambda$CDM cosmology and  
    matter density $\Omega_{\rm m}=0.319$, 
    baryon density $\Omega_{\rm b}=0.049$, 
    dark energy density $\Omega_\Lambda=0.681$, 
    scalar spectral index $n_{\rm s}=0.96$, 
    Hubble parameter $h=H_0/(100\;\mathrm{km}\,{\rm s}^{-1}\,{\rm Mpc}^{-1})=0.67$, and 
    the RMS value of density fluctuations on $8\,\hMpc$ scales $\sigma_8 = 0.83$. 
These values were chosen to be very close to those obtained by \cite{Planck2020}. 
The halo catalogue was constructed on the fly, with halos identified using \texttt{Rockstar} \citep{Behroozi:2013}, and then populated with a halo occupation distribution (HOD) model which had been calibrated to reproduce observables such as galaxy luminosity and clustering statistics as a function of galaxy luminosity and colour. This was used to create a lightcone mock covering an octant (5157 deg$^2$) of the sky, which was then cut to match the expected observed H$\alpha$ flux $f_{\rm{H}\alpha}>2 \expo{-16} \, \mathrm{erg}\, \si\second\inv \si\cm^{-2}$, and expected redshift range $0.9\le z < 1.8$ for the \Euclid spectroscopic galaxy sample \citep{Laureijs:2011, costille_2018}. To simulate the expected sample completeness of \Euclid, we randomly downsampled the final catalogue, retaining 60\% of all objects.

For each mock galaxy in this catalogue we record both the true redshift (corresponding to the background Hubble expansion) and the observed redshift, which includes the Doppler effect due to the galaxy peculiar velocities. We also consider the effects of redshift errors by adding an additional random component drawn from a Gaussian with RMS $\sigma_{z}=0.001$.\footnote{\Euclid requirements conservatively specify that redshift error should be less than $\sigma_{z}=0.001(1+z)$, but as the Near-Infrared Spectrometer and Photometer (NISP) instrument resolution will be determined by the point spread function and pixel size, the wavelength error and thus redshift error are expected to be constant with redshift. A fuller exploration of the effects of redshift errors, with less restrictive assumptions and including line misidentification, will be presented in future work.} The final resulting galaxy catalogue contains about $6.5\expo{6}$ mostly central galaxies. For the analysis described below we divided this sample into four non-overlapping redshift bins as illustrated in \cref{fig:galaxies_voids}, chosen to match those used in \Euclid forecasts by \cite{Blanchard:2020}. The effective redshift of each bin is generically defined as the weighted average redshifts of void-galaxy pairs in the bin, 
\begin{equation}
    z_{\rm eff} = \frac{ \sum _{ij}\left(
                     \frac{Z_i + z_j}{2}
                  \right)W_i\,w_j }{ \sum _{ij} W_i\,w_j }  \,,
\end{equation}
where $Z_i$ is the redshift of the $i$th void centre, $z_j$ is redshift of the $j$th galaxy, and $W_i$ and $w_j$ are optional weights associated with individual voids and galaxies when computing the CCF (as the Flagship mock does not include observational systematics, we do not use weights, that is they are all set to unity).The effective redshifts of the different bins are $\zeff=1.0$, $1.2$, $1.4$ and $1.64$.


\subsection{Reconstruction}
\label{subs:reconstruction}

\begin{figure*}
    \centering
	\includegraphics[width=0.78\hsize]{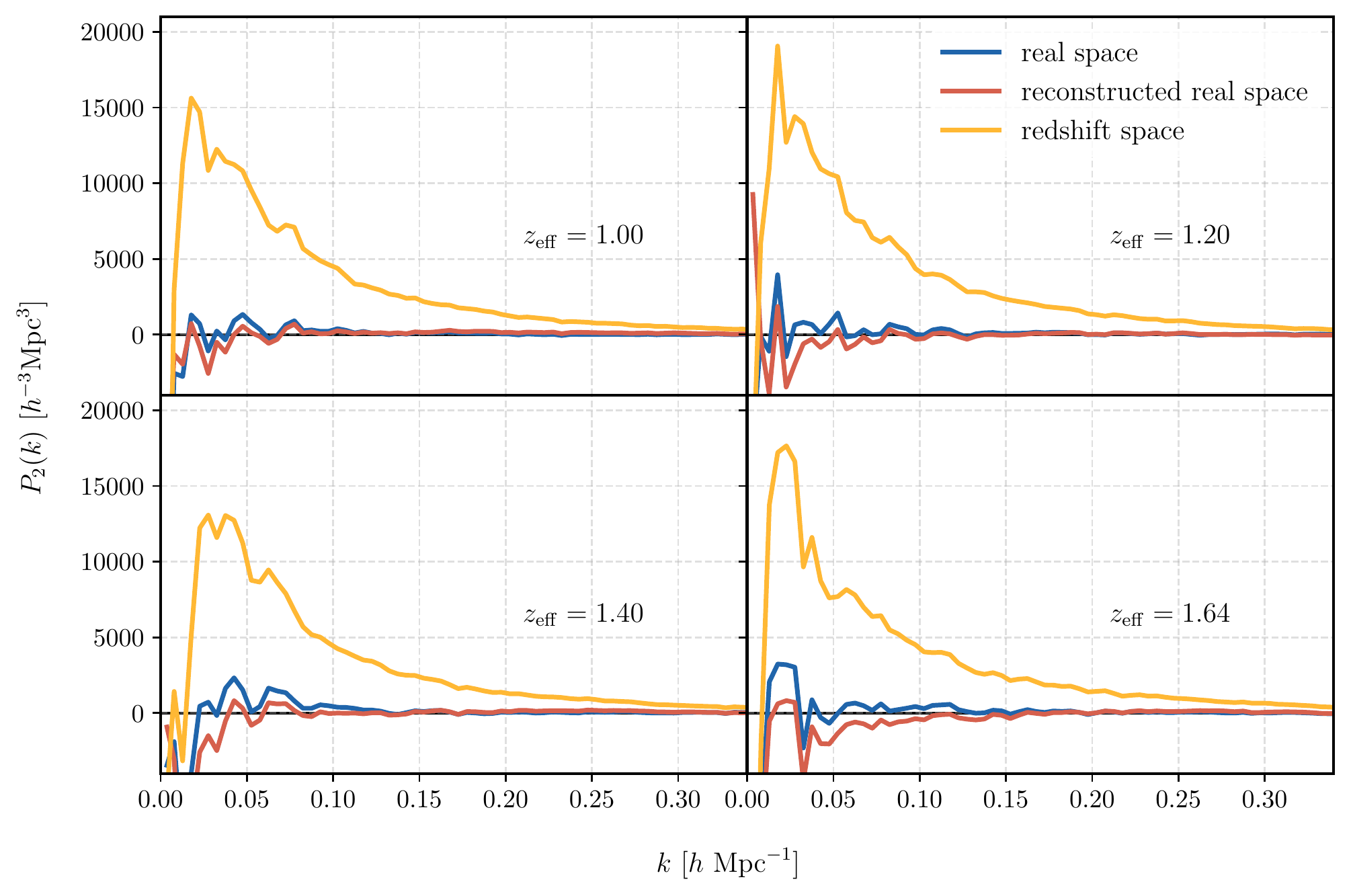}
	\caption{Quadrupole moments of the galaxy-galaxy power spectrum in each redshift bin of the Flagship mock, measured in redshift space (yellow), real space (blue) and after applying reconstruction to remove RSD from the galaxy positions (red). Reconstruction was performed using the fiducial values of $\beta_{\rm fid}$ listed in \cref{tab:catalogue_properties} and using smoothing scales $R_{\rm s}$ of $8$ and $9\,h^{-1}$Mpc for the first two and last two bins, respectively. The similarity of the red and blue lines and their proximity to zero is because reconstruction is successfully removing the large-scale RSD-induced anisotropy in the galaxy positions.}
	\label{fig:pofk_reconstruction}
\end{figure*}

\begin{table*}[]
	\setlength\extrarowheight{5pt}
	\caption{Summary properties of the galaxy and void catalogues used. For each redshift bin we indicate the approximate total number of galaxies $N_{\rm g}$, the number of voids $N_{\rm v}$ after the void size cut $R_{\rm v}>R_{\rm cut}$ was applied, and the fiducial values for the parameters $f\sigma_8$, linear galaxy bias $b$, $\beta=f/b$ and velocity dispersion amplitude $\sigma_{\rm v}$. Void numbers refer to those in the perfect reconstruction case, and vary by a few percent in the other reconstruction cases. }
	\label{tab:catalogue_properties}
	
	\centering
	\begin{tabular*}{0.8\textwidth}{@{\extracolsep{\fill}} ccrccccc}
		\hline\hline
		
		$\zeff$ & 
		$N_{\rm g}$ &
		\multicolumn{1}{c}{$N_{\rm v}$} & 
		$R_{\rm cut} \brackets{\hMpc}$ & 
		$f_{\rm fid}\sigma_{\rm 8,fid}$ &
		$b_{\rm fid}$ &  
		$\beta_{\rm fid}$ & 
		$\sigma_v \brackets{\kms}$ \\
		
		\hline
		1.00    & 
		2.26\expo{6}  & 1.9\expo{4}  & 16  & 0.442 & 1.598  & 0.55  & 321     \\
		1.20    & 
		1.75\expo{6}  & 1.4\expo{4}  & 18  & 0.418 & 1.907  & 0.47  & 318     \\
		1.40    & 
		1.27\expo{6}  & 1.0\expo{4}  & 21  & 0.394 & 2.106  & 0.44  & 310     \\
		1.64    & 
		1.26\expo{6}  & 1.0\expo{4}  & 25  & 0.367 & 2.382  & 0.39  & 295     \\
		\hline
	\end{tabular*}
\end{table*}

As discussed in \cref{sec:Theory}, in order to avoid an orientation-dependent selection bias in the void sample that would invalidate the theory model used, void-finding must be performed on the real-space galaxy distribution without large-scale RSD effects. To achieve this, we used a velocity-field reconstruction method, related to the reconstruction commonly used in BAO analysis \citep{Eisenstein:2007,Padmanabhan:2012}, to approximately remove galaxy RSD before the void-finding step \citep{Nadathur:2019b}. 

Reconstruction was performed using an iterative fast Fourier transform (FFT) algorithm \citep{Burden:2015} which solves the Zeldovich equation in redshift space
\begin{equation}
    \label{eq:Zeldovich}
	\nabla \cdot \vec\Psi + 
	\frac{f}{b} \nabla \cdot [
	\paren{ \vec\Psi \cdot  \vec{\hat{r}} } \vec{\hat{r}} ]
	= - \frac{\delta_{\rm g}}{b} \,,
\end{equation}
for a Lagrangian displacement field $\vec\Psi$, where $f$ is the growth rate, $b$ is the linear galaxy bias, and $\delta_{\rm g}$ is the galaxy overdensity in redshift space. This step is implemented in the \texttt{pyrecon} package\footnote{\url{https://github.com/cosmodesi/pyrecon}}. It is performed on a regular grid, and densities estimated on the grid are first smoothed with a Gaussian kernel of width $R_{\rm s}$ before solving for the displacement. Given the solution to \cref{eq:Zeldovich}, individual galaxy positions are shifted by $-\vec\Psi_{\rm RSD}$ evaluated at their locations, where
\begin{equation}
    \label{eq:shift}
	\vec\Psi_{\rm RSD} = - f \, (\vec\Psi \cdot \vec{\hat{r}}) \, \vec{\hat{r}} \,.
\end{equation}
This shift subtracts the estimated RSD contribution to the observed galaxy redshifts, in order to approximately recover the real-space galaxy field. The algorithm for solving \cref{eq:Zeldovich} is the same as that used for BAO reconstruction \citep[e.g. in the eBOSS analyses][]{Gil-Marin:2020,Bautista:2021} with the difference that the shifts applied to the galaxy positions only attempt to correct for large-scale RSD and not for non-linear evolution. 

If we define a new variable $\vec\psi\equiv b\vec\Psi$, \cref{eq:Zeldovich} can be rewritten as $\nabla \cdot \vec\psi + \beta\nabla \cdot [\paren{ \vec\psi \cdot \vec{\hat{r}}} \vec{\hat{r}} ] = - \delta_{\rm g}$, where $\beta=f/b$ is the RSD parameter. The shifts applied to the galaxy positions in order to remove RSD are then $\vec\Psi_{\rm RSD} = - \beta \, (\vec\psi \cdot \vec{\hat{r}}) \, \vec{\hat{r}}$, and it is clear that the output of our reconstruction procedure depends only on the parameter $\beta$ and not on $f$ and $b$ individually.\footnote{We note that this conclusion applies to our procedure that aims to remove only the RSD effects by shifting by $\vec\Psi_{\rm RSD}$, and not when galaxy positions are shifted by the full displacement field $\vec\Psi$ as is done for BAO analyses.} The fiducial growth rate for the Flagship cosmology can be calculated as $f(z) = \Omega_{\rm m}^{\gamma} (z)$ \citep{Linder:2005in}, where $\Omega_{\rm m}$ is the matter density parameter, and $\gamma=6/11\approx 0.55$. To determine the fiducial linear bias $b$, we measured the galaxy power spectrum in real-space $P_{\rm g}(k)$, calculated the theoretical matter power spectrum $P_{\rm m}(k)$ for Flagship using \texttt{CAMB} \cite{Lewis_2000}, and fit for $b^2=P_{\rm g}(k)/P_{\rm m}(k)$, using only large-scale modes at $k<0.07\, h\,\text{Mpc}^{-1}$.

The fiducial values $b^\mathrm{fid}$ and $\beta^\mathrm{fid}$ obtained in this way are reported in \cref{tab:catalogue_properties} for each redshift bin. However, the growth rate $f$ is a parameter in any fit we perform, and so we do not want to fix $\beta$ to this fiducial value, but instead to marginalise over it. To achieve this in a computationally feasible way, we pre-computed the reconstruction on a grid of $\beta$ values in the range $\left[\beta^{\rm fid}-0.15, \beta^{\rm fid}+0.15\right]$ around the fiducial value. The data vectors obtained from each of these approximations to the real-space galaxy field consist of the monopole, quadrupole and hexadecapole (all ranging from 0 to 120 $\hMpc$), and are interpolated over when varying $\beta$ in the analysis.

As the Zeldovich approximation breaks down on small scales, reconstruction also depends on the width of the Gaussian smoothing kernel, $R_{\rm s}$, which dictates the smallest scales used. Reconstruction with a smoothing radius that is too small will over-correct the RSD, due to contributions from non-linear small-scale fluctuations, while a smoothing radius that is too big will fail to remove all of them. To determine the optimal smoothing radius to use, we computed the quadrupole of the post-reconstruction galaxy power spectrum, $P_2(k)$, to use as a proxy. In real space this quantity is consistent with zero, while RSD introduce significant anisotropies in redshift space (\cref{fig:pofk_reconstruction}). In each redshift bin we tested reconstruction with multiple smoothing radii at the fiducial value $\beta^\mathrm{fid}$, and pick the one that results in a quadrupole which is roughly consistent with zero. This gives $R_{\rm s}=8\,\hMpc$ for the first two redshift bins, and $R_{\rm s}=9\,\hMpc$ for the last two. 

The reconstruction method outlined here cannot perfectly remove all redshift-space effects in the galaxy distribution. We are interested in quantifying the impact of these inaccuracies, as well as those caused by redshift errors, on the void-finding and data analysis pipeline below. We therefore repeated our analysis pipeline in three different scenarios, labelled as: 
\begin{enumerate} 
    \item \textbf{perfect reconstruction}, using the true real-space galaxy positions from the simulation; 
    \item \textbf{realistic reconstruction}, applying reconstruction to the observed redshifts to obtain approximate real-space galaxy positions; 
    \item and \textbf{realistic reconstruction with} $\sigma_{z}$, applying reconstruction to observed redshifts including redshift errors. 
\end{enumerate}


\subsection{Void finding}
\label{subs:voids}

Voids were found in the real space or post-reconstruction galaxy distribution using the \voxel algorithm, implemented in \Revolver\footnote{\url{https://github.com/seshnadathur/Revolver}} \citep{Revolver}. \voxel is a watershed-based void-finding algorithm that operates on a particle-mesh interpolation of the galaxy density field. In the particle-mesh assignment step, tracers (in our case galaxies) are first placed onto a grid to estimate the local density field, a step which is much faster than commonly used tessellation algorithms, and scales linearly with the number of tracers, $\mathcal{O}(N)$ (as opposed to $\mathcal{O}(N^{3/2})$ for tesselation). The mesh size is determined by requiring each cubic grid cell, or voxel, to have a side length $a_\mathrm{vox}=0.5\left(\frac{4\pi\bar n}{3}\right)^{-1/3}$, where $\bar{n}$ is the approximate value for the mean number density of galaxies. As $\bar{n}$ varies by a factor of about $5$ over the Flagship redshift range, in order to maintain an appropriate resolution in the highest density bin while maintaining a reasonable number of voxels for computational efficiency, we decided to perform the void-finding separately in each of the four redshift bins, using a different $a_\mathrm{vox}$ for each. The \texttt{voxel} algorithm (like all void-finder algorithms) unavoidably misses some voids near the redshift edges of the galaxy sample, so this analysis choice leads to the drops in void numbers around redshifts $z=1.1$, 1.3 and 1.5 seen in \cref{fig:galaxies_voids}. An alternative approach could involve the use of overlapping redshift bins, which would require a more delicate treatment of correlation between redshifts. We have found that this does not improve our final constraints enough to justify the complexity of the analysis.

In order to correct for the survey window and redshift-dependent or angular selection effects, this density estimate is normalised by the density of a random unclustered catalogue of points which has the same window and selection effects imposed, projected on the same grid. We see from this that, in addition to being significantly faster, this method of estimating the galaxy density field can then more easily account for complex survey masks and systematic effects than popular tessellation-based density estimation methods used in other void-finding codes, such as those based on \zobov \citep{Neyrinck_2008}. For our purposes with the Flagship catalogue, we created a random catalogue with 50 times as many points as the galaxy catalogue, covering the Flagship octant and following the same redshift distribution $n(z)$.

The estimated galaxy density field is then smoothed with a Gaussian filter of smoothing length $r_{\rm s}=\bar{n}^{-1/3}$, in order to reduce shot-noise fluctuations on smaller scales. Voids are then identified as the sites of local minima in the smoothed density field, and their extents are determined using a watershed method, similar to that of \zobov, to create a final catalogue of non-overlapping voids. Each void is composed of a contiguous set of voxels, which define its volume. Individual voids have irregular shapes and orientations, but we define an effective void radius as the radius of a sphere with the same volume,
\begin{equation}
    R_{\rm v} = \paren{\; \frac{3}{4\,\pi} N_\mathrm{vox}V_\mathrm{vox}\;}^\frac{1}{3} \,,
\end{equation}
where $N_\mathrm{vox}$ is the number of voxels in the void and $V_\mathrm{vox}=a_\mathrm{vox}^3$ is the voxel volume. The centre of each void is identified as the location of the lowest-density voxel within it, being the position of the density minimum. This results in voids with very few galaxies in the centre.

For each reconstruction scenario considered, the full void catalogues obtained in this way contain about $87\,000$ voids (with small variations in number) across all redshift bins. However, we expect that smaller voids will not be adequately described by our approach to modelling the coherent velocity outflow $v_{\rm r}(r)$. This is due to a combination of factors. The dynamics of small voids are more likely to be dominated by the presence of neighbouring large density fluctuations so that the isolated void model is inappropriate \citep[see also][]{Schuster:2023}; they are more likely to be segments of larger voids that appear artificially small because they extend outside of the survey boundaries; and they are also more likely to be entirely spurious detections arising from shot-noise fluctuations \citep{Cousinou:2019}. Therefore, we applied a void size cut to exclude them from the catalogue used for clustering measurements\footnote{This cut is similar to the one applied by \cite{Hamaus:2022}, who use a factor $1.86$ instead of $1.5$.}:
\begin{equation}
    R_{\rm v}>R_{\rm cut} = 1.5 \; \bar{n}^{-\frac{1}{3}} \,.
\end{equation}
Keeping only voids that survive this cut reduces the total numbers by about 40\%. The radius cuts and the final sizes of the void catalogues are presented in \cref{tab:catalogue_properties}.


\section{Method}
\label{sec:Method}


\subsection{Measuring the void-galaxy cross-correlation function}
\label{subs:measurement}

We measured correlation functions using the Python wrapper \texttt{pycorr}\footnote{\url{https://github.com/cosmodesi/pycorr}} for \texttt{Corrfunc} \citep{Corrfunc1, Corrfunc2}. \texttt{Corrfunc} is a pair-counting engine that counts the number of pairs of voids and galaxies within a given distance and angular separation bin. This was then compared to the equivalent pair counts computed using random unclustered catalogues to obtain an estimate of the CCF using the Landy-Szalay estimator \citep{Landy:1993}:
\begin{equation}
    \label{eq:LS}
    \xi(r, \mu) = \frac{ {\rm DD}_{12}(r, \mu) - 
                         {\rm DR}_{12}(r, \mu) - 
                         {\rm DR}_{21}(r, \mu) + 
                         {\rm RR}_{12}(r, \mu)}
                       { {\rm RR}_{12}(r, \mu) } \,.
\end{equation}
Here ${\rm D}$ and ${\rm R}$ refer to the data and unclustered random catalogues respectively, subscripts $1$ and $2$ are used to differentiate between voids and galaxies, and each term $XY$ refers to the number of pairs of species $X$ and $Y$ in the $(r, \mu)$ bin, normalised relative to the total possible number of such pairs. Thus ${\rm DD}_{12}$ refers to normalised number of void-galaxy pairs, ${\rm DR}_{12}$ to the normalised number of pairs of voids and galaxy randoms, and so on. We created the unclustered random catalogues as described in \cref{subs:voids}, such that they have $50$ times as many points as the respective data samples, are uniformly sampled from the Flagship octant footprint and follow the redshift distributions of voids and galaxies respectively (\cref{fig:galaxies_voids}). We measured pair counts and correlation functions $30$ radial bins over $0<r<120\,\hMpc$ and in $200$ angular bins over $-1<\mu<1$, keeping distances in units of $\hMpc$ rather than rescaling them in units of the void size, as is done in some void publications.

It is important to note that shot-noise fluctuations in the galaxy random catalogue used in the mesh density estimation step during void finding (\cref{subs:voids}) will be slightly correlated with the positions of identified voids. We therefore used an independent realisation of the galaxy random catalogue (generated using the same method but with a different random seed) for measuring correlations via \cref{eq:LS}. If instead the same random catalogue were used for both, small artefacts could be propagated to the correlation $\xi$ in the regions very close to the void centres.

We used the estimator in \cref{eq:LS} with the pair separations determined by the positions of voids and galaxies in different combinations of real and redshift space to measure the different correlation functions relevant to our analysis. For instance, in the idealised case of perfect reconstruction, voids are identified in the true real-space galaxy field, and their cross-correlation with the real and redshift-space galaxy positions give $\xi^{\rm rr}$ and $\xi^{\rm rs}$, respectively. On the other hand, when we use realistic reconstruction we use the superscript $\rm p$ to distinguish positions in this post-reconstruction approximation of the real-space galaxy field from the true real-space positions. Voids are found in the post-reconstruction field and, as before, their positions are held fixed to allow measurements of $\xi^{\rm ps}$ and $\xi^{\rm pp}$ with the galaxies in redshift space and approximate real space respectively.

\begin{figure}
    \centering
	\includegraphics[width=\hsize]{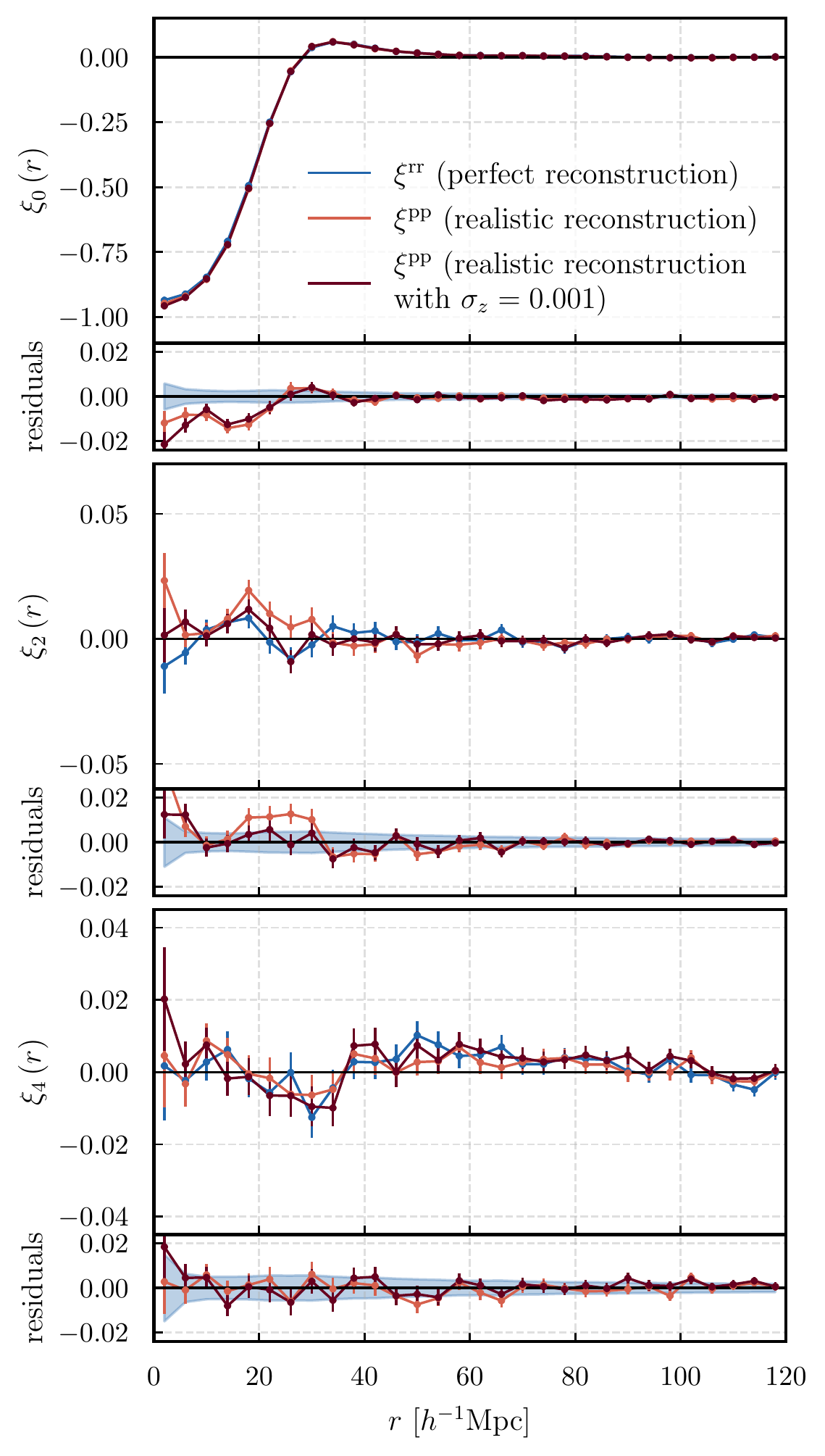}
	\caption{Void-galaxy CCF in real space, measured in the Flagship simulation $\zeff=1.0$ redshift bin. From top to bottom, the panels show the monopole, quadrupole and hexadecapole moments of the CCF. The multipoles of $\xi^{\rm rr}$, corresponding to the perfect reconstruction scenario and measured using the true real-space positions are shown in blue, while light and dark red respectively show the multipoles of $\xi^{\rm pp}$ obtained using reconstruction without and with redshift errors. The inset panels show the residuals with respect to the perfect reconstruction case of $\xi^{\rm rr}_\ell$ for each multipole. In all cases $\xi_2$ and $\xi_4$ are close to zero, indicating that reconstruction successfully removes orientation-dependent selection bias.}
	\label{fig:real_vs_recon}
\end{figure}

\Cref{fig:real_vs_recon} shows the recovered monopole, quadrupole and hexadecapole moments for the true real-space void-galaxy CCF $\xi^{\rm rr}$ in the first Flagship redshift bin, and its comparison with the equivalent multipoles of $\xi^{\rm pp}$ in the realistic reconstruction scenarios with and without including redshift errors. Reconstruction recovers the real-space CCF multipoles well, and anisotropies in all cases are negligible, as evidenced by the quadrupole and hexadecapole moments being close to zero. This demonstrates that our void sample selection successfully avoids the problem of orientation-dependent selection bias discussed in \cref{sec:Theory}. 


\subsection{Constructing template functions} 
\label{subs:templates}

\begin{figure}
	\centering
	\includegraphics[width=0.9\hsize]{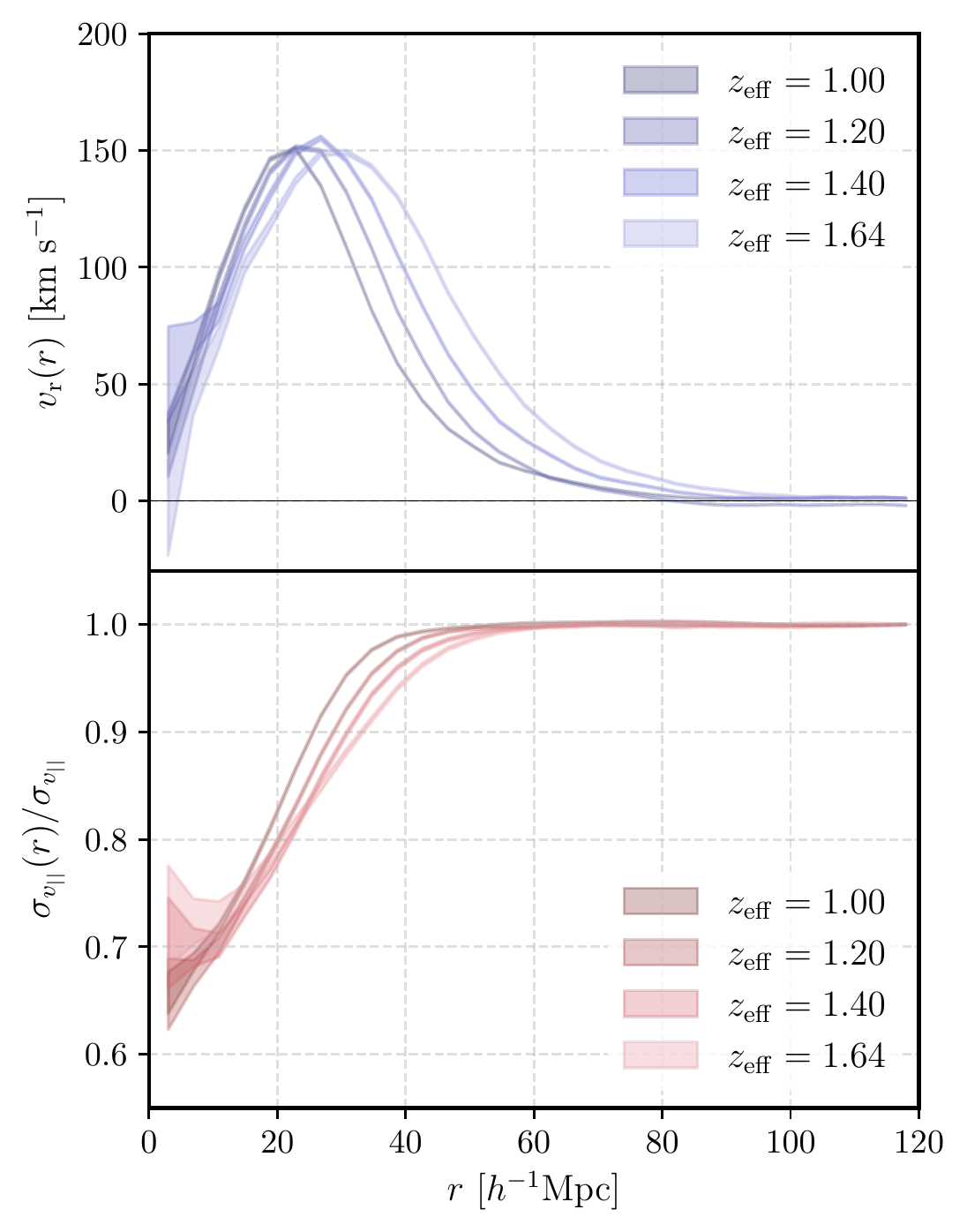}
	\caption{Measured radial outflow velocity profile $v_{\rm r}(r)$ (top), and normalised line-of-sight velocity dispersion $\sigmalos(r)$ (bottom) of galaxies around voids found in the four redshift bins of Flagship. Shading represents the $68.3\%$ ($1\sigma$) measurement uncertainty coming from Poisson noise. The dispersion amplitude $\sigma_{\rm v}$ for the four bins can be found in \cref{tab:catalogue_properties}.}
	\label{fig:templates}
\end{figure}

As discussed in \cref{sec:Theory}, we adopt a template-fitting approach to the analysis, and so we require templates for the three functions $v_{\rm r}(r)$, $\sigmalos(r)$ and $\xi^{\rm rr}(\vec r)$. These were measured from the Flagship simulation data as follows.

To measure the radial velocity profile $v_{\rm r}(r)$, we measured the mean radial component of galaxy peculiar velocities (given in the simulation box rest frame) in spherical shells around the void centre, where the average was taken over all void-galaxy pairs in that separation bin. We used 30 linearly spaced radial bins, up to $r=120\,\hMpc$. The mean radial velocity profiles at different redshifts are shown in the top panel of \cref{fig:templates}. The velocity is positive over almost the entire radial range, corresponding to an outflow from the void centre as expected. The profile shows a peak in the vicinity of the edge of the void and at larger radii it goes to zero, showing that there is no net bulk motion far outside of the void, as expected.

We also measured the width of the line-of-sight velocity PDF, $\sigmalos(r)$, from the distribution of the galaxy peculiar velocities in the same spherical shells. The full vector form of \cref{eq:line-of-sight_velocity} can be written as 
\begin{equation}
    \label{eq:vector_vel}
    \vec v(\vec r) = v_{\rm r}(r)\,\vec{\hat{r}} + \tilde{v}\,\vec{\hat{n}}\,,
\end{equation}
where the stochastic velocity component has random magnitude $\tilde{v}$ and is directed along a random direction $\vec{\hat{n}}$. The quantity we wish to estimate is $\sigmalos^2(r)=\langle\tilde{v}^2\left(\vec{\hat{n}}\cdot\vec{\hat{r}}\right)^2\rangle$ and from \cref{eq:vector_vel} we obtain \citep{Fiorini:2022}
\begin{equation}
    \label{eq:sigmav1}
    \sigmalos(r) = \frac{\sqrt{\langle|\mathbf{v}|^2\rangle - v_{\rm r}^2}}{\sqrt{3}}\,,
\end{equation}
where $\langle|\mathbf{v}|^2\rangle$ and $v_{\rm r}(r)$ are quantities independent of $\mu_{\rm r}$, and can be measured in radial bins. 

The bottom panel of \cref{fig:templates} shows the normalised template functions for $\sigmalos(r)$ measured for each of the four redshift bins. The measured amplitudes at large distances, $\sigmalos$, are listed in \cref{tab:catalogue_properties}, where we see that the dispersion is higher at lower redshift, consistent with the growth of large-scale structure. From the figure we also see that the dispersion decreases in the void interior, where the potential is steep and the bulk outflow motion dominates. 

The final template function required is that for $\xi^{\rm rr}(\vec r)$. In the two realistic reconstruction scenarios we test, we assume that the true real-space information is not available, so that $\xi^{\rm rr}$ cannot be directly measured. Instead we used the quantity $\xi^{\rm pp}(\vec r)$, which could be measured using the approximate real-space catalogues, as an estimator for $\xi^{\rm rr}(\vec r)$. This is the same approach as used by \cite{Nadathur:2019c, Nadathur:2020b} and \cite{Woodfinden:2022} except that here we do not have a large number of mock catalogues over which we can average this estimator. This greatly increases the noise in the estimate, which is especially relevant when attempting to determine the dependence of $\xi^{\rm pp}$ on the reconstruction parameter $\beta$. For the current work we therefore decided to fix this estimator to $\xi^{\rm pp}(\beta^\mathrm{fid})$ evaluated at the fiducial $\beta$ value given in \cref{tab:catalogue_properties}. This approach will be revisited for a future full \Euclid analysis, by which time we expect a full complement of mocks to be available.

As mentioned in \cref{subs:xir}, our general code implementation seeks to preserve information on possible anisotropies in $\xi^{\rm pp}(\vec r)$ which may be introduced by the AP effect or imperfections in reconstruction, without imposing the assumption of spherical symmetry, $\xi^{\rm pp}(\vec r)=\xi^{\rm pp}(r)$. In practice we achieve this by measuring the monopole $\xi^{\rm pp}_0(r)$, quadrupole $\xi^{\rm pp}_2(r)$ and hexadecapole $\xi^{\rm pp}_4(r)$ moments and using them to reconstruct $\xi^{\rm pp}(\vec r)$. In the idealised perfect reconstruction scenario, things are much simpler and we just use the true real-space information from Flagship and directly measure the multipoles $\xi^{\rm rr}_0(r)$, $\xi^{\rm rr}_2(r)$ and $\xi^{\rm rr}_4(r)$ from the data. 

\begin{figure}
	\centering
	\includegraphics[width=\hsize]{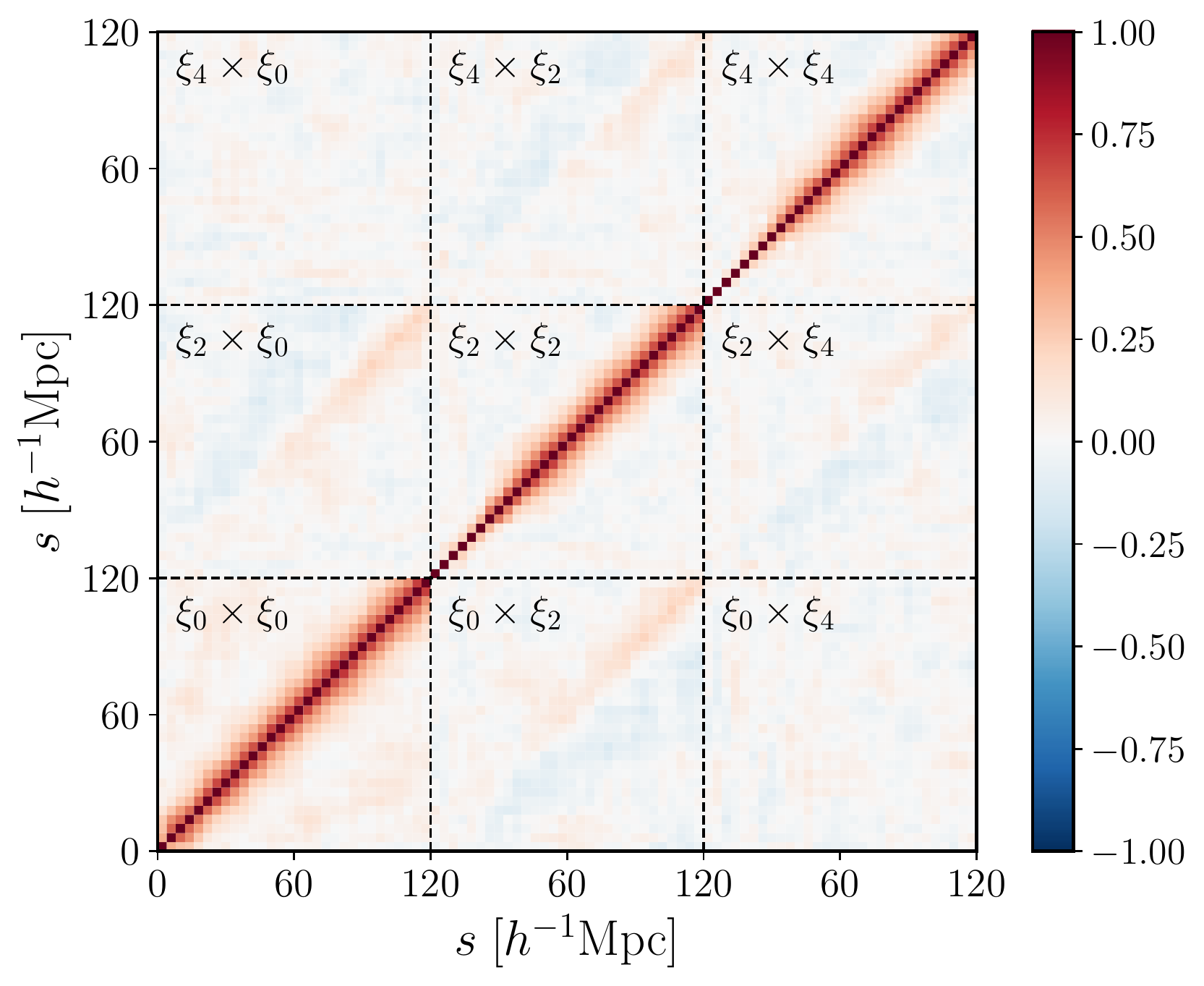}
	\caption{Normalised covariance matrix $\tens{C}^{\rm data}$ for the data vector from the Flagship $\zeff = 1.0$ redshift bin, estimated using jackknife resampling. Dashed lines differentiate between monopole, quadrupole and hexadecapole blocks of the data vector. }
	\label{fig:datacov}
\end{figure}

\begin{figure*}
	\includegraphics[width=\textwidth]{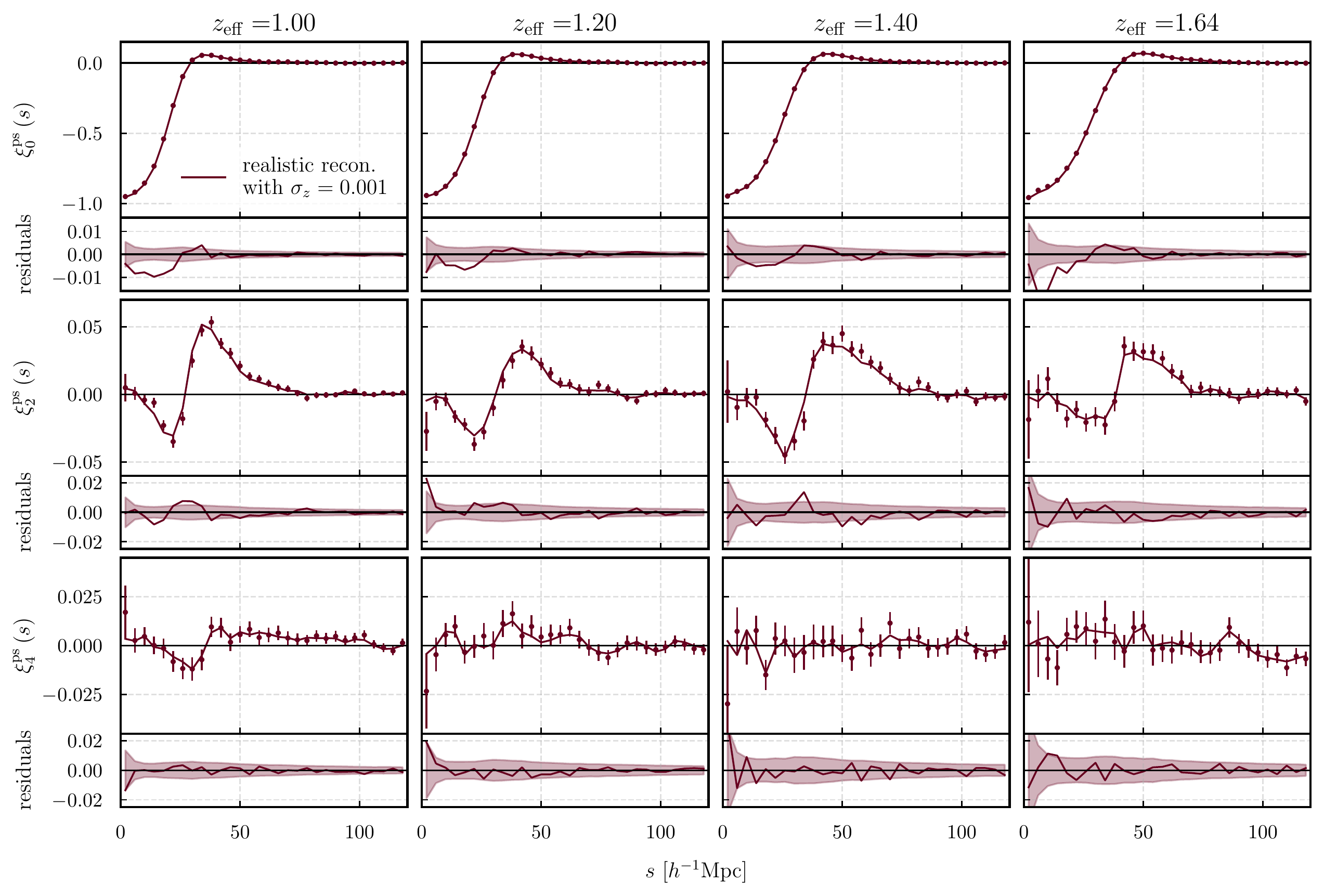}
	\caption{Measured multipole moments of the redshift-space void-galaxy CCF, $\xi^{\rm ps}_\ell$, for realistic reconstruction applied to the catalogue including redshift errors, are shown as the data points. Error bars shown are determined from the diagonal entries of the covariance matrix $\tens{C}^{\rm data}$. The solid lines show the corresponding best-fit models, and insets show the residuals of the model with respect to the data. The rows correspond to (from top to bottom) the monopole, quadrupole and hexadecapole moments ($\ell=0,2, 4$). Columns correspond to the different redshift bins, labelled with the effective redshift $z_{\rm eff}$.}
	\label{fig:realistic_recon_multipoles}
\end{figure*}


\subsection{Model fitting and parameter inference} 
\label{subs:inference}

We compare the model to the measured redshift-space CCF in terms of the compression to multipoles. In each redshift bin, we define the data vector $\vec\xi^{\rm s}\equiv\left(\xi^{\rm rs}_0, \xi^{\rm rs}_2, \xi^{\rm rs}_4\right)$ formed by concatenating the monopole, quadrupole and hexadecapole moments of $\xi^{\rm rs}(s, \mu_{\rm s})$, measured in 30 radial bins. In the two realistic reconstruction scenarios where the true real-space information is not used, we use the CCF $\xi^{\rm ps}(s, \mu_{\rm s})$ measured using the void positions identified in the post-reconstruction, approximate real-space galaxy distribution to construct $\vec\xi^{\rm s}$. In this case the data vector inherits a dependence on the reconstruction parameter $\beta$ from the void positions.

To determine the uncertainty in this measurement of $\vec\xi^{\rm s}$, we used a jackknife resampling method to estimate the covariance matrix. For each of the four redshift bins we divided the Flagship data sample into $N_{\rm JK}=900$ non-overlapping sub-regions of equal volume. We iterated over all sub-boxes, in each case estimating the cross-correlation function $\vec{\xi}^{\rm s(k)}$ by removing all the void-galaxy pairs for which \textit{both} void and galaxy lie in the $k$th sub-box, and reweighting the ones where \textit{either} void \textit{or} galaxy lie within the sub-box using the matched weighting scheme prescribed by \cite{Mohammad:2022} and implemented in \texttt{pycorr}. We then used these realisations to estimate the covariance between the $i$th and $j$th bins of the data vector
\begin{equation}
	\label{eq:covariance}
	\tens{C}_{ij}^{\rm data} =
	\frac{N_{\rm JK} - 1}{N_{\rm JK}} 
	\sum_{k=1}^{N_{\rm JK}} 
	\paren{ \vec{\xi}_{i}^{{\rm s}(k)} - \bar{\vec{\xi}^{\rm s}_{i}} } 
	\paren{ \vec{\xi}_{j}^{{\rm s}(k)} - \bar{\vec{\xi}^{\rm s}_{j}} } \,,
\end{equation}
where $\bar{\vec{\xi}^{\rm s}}_{i}=\frac{1}{N_{\rm JK}}\sum_{k=1}^{N_{\rm JK}}\vec{\xi}_{i}^{{\rm s}(k)}$ is the mean of the $N_{\rm JK}$ jackknife realisations. The correlation structure of the resultant covariance matrix is shown in \cref{fig:datacov}. Due to the limitations of having only the single Flagship mock, there will be an additional correlation between the model prediction and the measured data vector that has been neglected in this estimate of the covariance but is discussed in \cref{sec:Appendix}.

At any point in parameter space, we computed the model prediction for $\vec\xi^{\rm s}$ as described in \cref{sec:Theory} using the public code \texttt{victor}\footnote{\url{https://github.com/seshnadathur/victor}}. We assume that the likelihood has a Gaussian form, 
\begin{equation}
    \label{eq:likelihood}
    \log \mathcal{L} = -\frac{1}{2} \,
                \paren{\vec{\xi}^{\rm s}_{\rm \,theory} - \vec{\xi}^{\rm s}_{\rm \,data}}
                \tens{C}\inv 
                \paren{\vec{\xi}^{\rm s}_{\rm \,theory} - \vec{\xi}^{\rm s}_{\rm \,data}}
                \superscr{T} \,. 
\end{equation}
In the realistic reconstruction scenarios we consider, this likelihood evaluation depends on four parameters: explicitly on $f\sigma_8$, $\sigmalos$ and $\epsilon$, and implicitly on the reconstruction parameter $\beta$ via its effect on the recovered void positions in the post-reconstruction field and thus on the measured correlations. In the idealised case of perfect reconstruction there is no $\beta$ dependence. We sampled the full parameter space using the interface between \texttt{victor} and the Monte Carlo Markov Chain (MCMC) sampling code \texttt{Cobaya}\footnote{\url{https://github.com/CobayaSampler/cobaya}} \citep{Torrado:2020dgo}. Repeating reconstruction, void-finding and CCF measurement at each $\beta$ value along the chain would be too slow for the MCMC so instead we followed the method of previous works \citep{Nadathur:2019c, Nadathur:2020b} and pre-computed the correlations on a grid of $\beta$ values around the fiducial $\beta^\mathrm{fid}$ and interpolated the data vector at the time of evaluation of the likelihood.\footnote{\cite{Woodfinden:2022} instead perform the interpolation at the level of the likelihood itself; both options are implemented in \texttt{victor} and give very similar results.} To avoid extrapolating outside of this grid, the prior on $\beta$ was limited to the same range as the interpolation; we used uninformative flat priors for all of the other parameters. Additionally, convergence of the MCMC chains is checked using the Gelman-Rubin $R-1$ statistic, using the convergence criterion $R-1<0.01$, and 20\% of the initial steps were discarded as burn-in.


\section{Results}
\label{sec:Results}

\begin{table*}[]
	\setlength\extrarowheight{3pt}
	\caption{Marginalised posterior constraints on the AP distortion parameter $\epsilon$ and growth rate $f\sigma_8$ obtained from fits to the Flagship mock data in four redshift bins. We present results in three scenarios: idealised perfect reconstruction (using true real-space galaxy positions), realistic reconstruction and realistic reconstruction performed on a catalogue with added redshift errors. Fits are performed using covariance $\tens{C}^{\rm data}$ estimated for Flagship. The true value of $\epsilon$ is 1 as the data are analysed in the Flagship fiducial cosmology. The true values of $f\sigma_8$ in each redshift bin are as listed in \cref{tab:catalogue_properties}.}
	\label{tab:results}
	\centering
	\begin{tabular*}{0.8\textwidth}{@{\extracolsep{\fill}}lccc}
		
		\hline\hline
		&
		$\zeff$ &		$f\sigma_8$ & 	$\epsilon$ \\
		\hline
		
		Flat prior range & &	
		$\brackets{0.05,1.5}$ 	& $\brackets{0.8,1.2}$ \\
		\hline
		
		Perfect reconstruction &
		$1.00$ &  
		$0.46 \pm 0.04$ & $1.002 \pm 0.005$ \\ &
		$1.20$ & 
		$0.41 \pm 0.05$ & $1.002 \pm 0.006$ \\ &
		$1.40$ & 
		$0.40 \pm 0.05$ & $1.003 \pm 0.006$ \\ &
		$1.64$ & 
		$0.41 \pm 0.04$ & $1.000 \pm 0.006$ \\ 
		\hline
		
		Realistic reconstruction & 
		$1.00$ & 
		$0.49 \pm 0.04$ & $0.995 \pm 0.005$ \\ & 
		$1.20$ & 
		$0.44 \pm 0.05$ & $1.000 \pm 0.005$ \\ & 
		$1.40$ & 
		$0.40 \pm 0.05$ & $1.005 \pm 0.006$ \\ & 
		$1.64$ & 
		$0.43 \pm 0.05$ & $0.999 \pm 0.006$ \\ 
		\hline
		
		Realistic reconstruction & 
		$1.00$ & 
		$0.51 \pm 0.04$ & $0.997 \pm 0.005$ \\ 
		with $\sigma_{z}=0.001$ & 
		$1.20$ & 
		$0.44 \pm 0.05$ & $1.004 \pm 0.005$ \\ & 
		$1.40$ & 
		$0.40 \pm 0.06$ & $1.005 \pm 0.006$ \\ & 
		$1.64$ & 
		$0.44 \pm 0.05$ & $1.000 \pm 0.006$ \\ 
		\hline
		
	\end{tabular*}
\end{table*}

\begin{figure*}
	\centering
	
	\begin{subfigure}[b]{0.48\hsize}
		\raggedleft 
		\includegraphics[width=0.75\hsize]{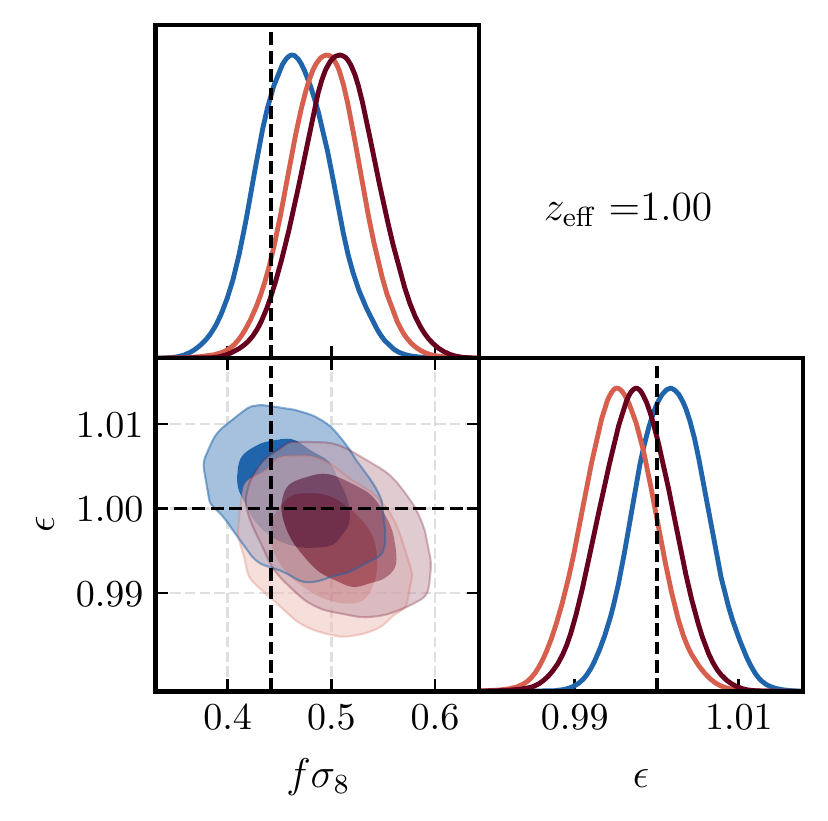}
		\includegraphics[width=0.75\hsize]{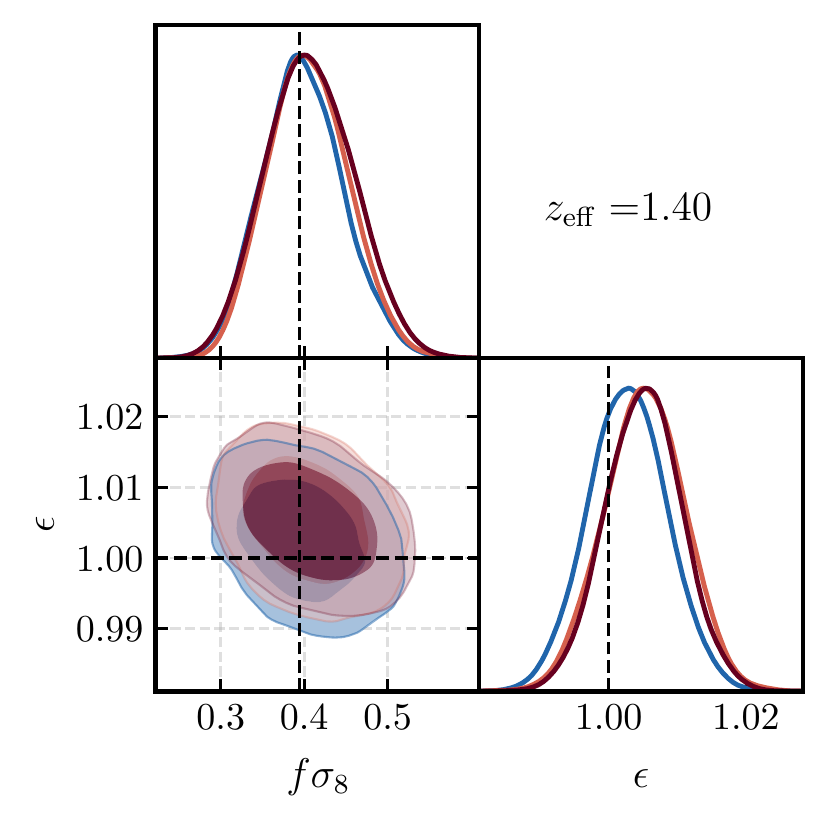}
	\end{subfigure}
	\hfill
	\begin{subfigure}[b]{0.48\hsize}
		\raggedright 
		\includegraphics[width=0.75\hsize]{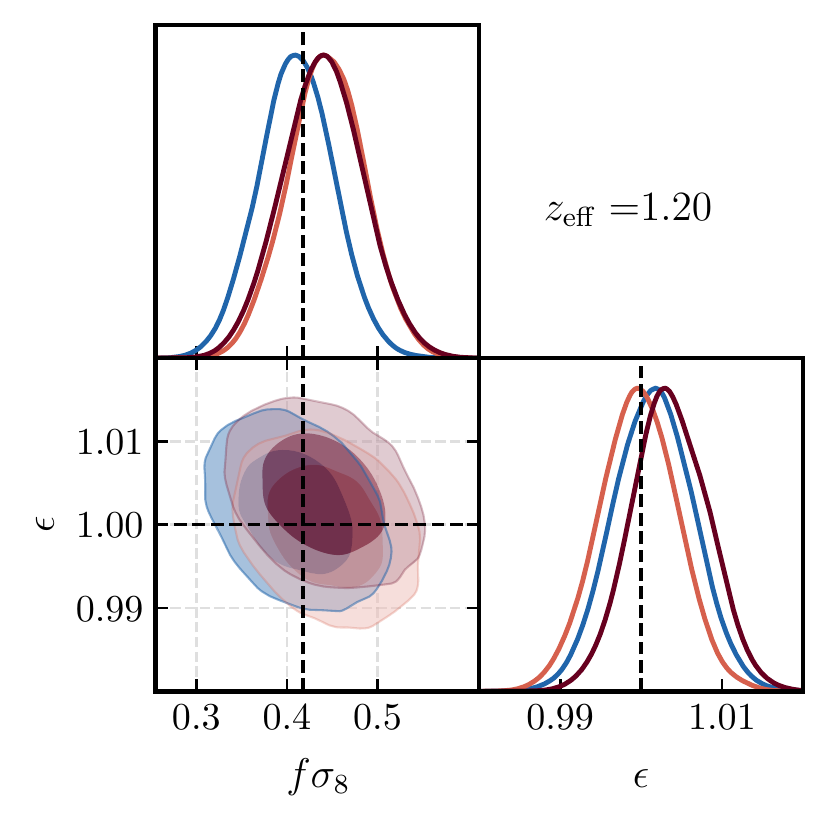}
		\includegraphics[width=0.75\hsize]{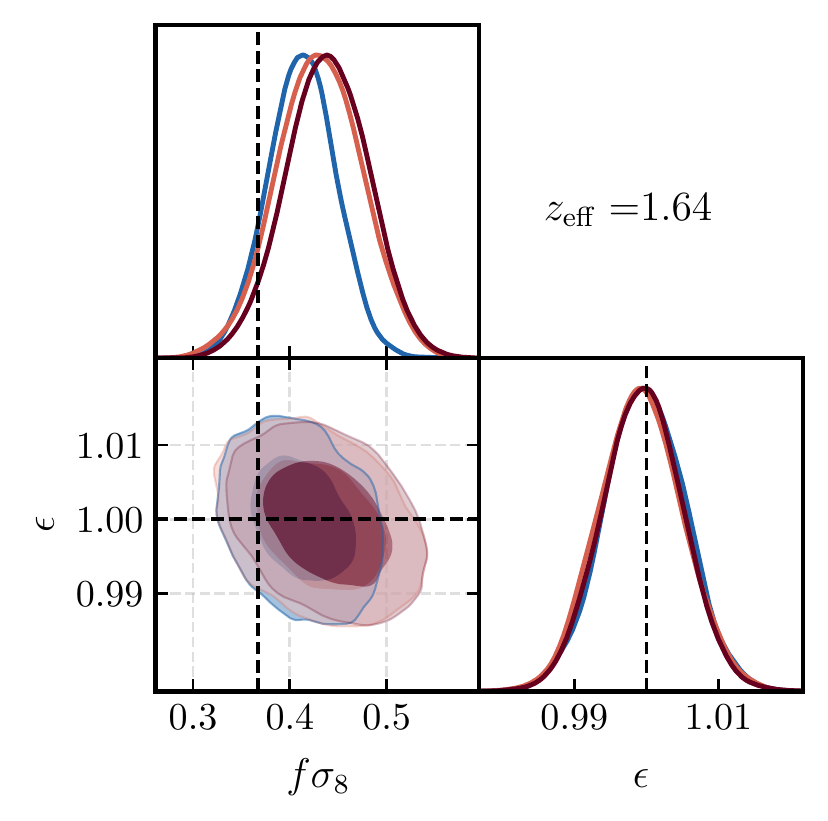}
	\end{subfigure}
	\includegraphics[width=0.85\hsize]{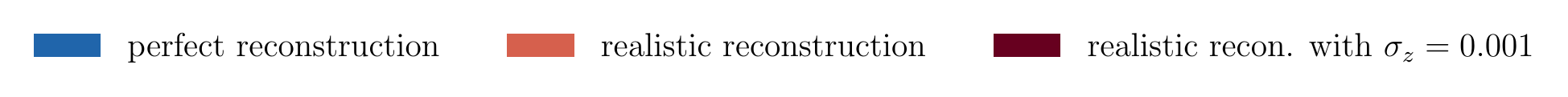}
	
	\caption{Marginalised 1D and 2D posterior constraints for parameters $f\sigma_8$ and $\epsilon$ from fits to the Flagship void-galaxy CCF measurements in four redshift bins. Contours show the $68.3\%$ ($1\sigma$) and $95.5\%$ ($2\sigma$) confidence intervals obtained in the case of perfect reconstruction (blue), realistic reconstruction (light red), and realistic reconstruction with added redshift error (dark red). Dashed crosshairs indicate the true values for the Flagship cosmology. }
	\label{fig:triangle_plot}
\end{figure*}

\begin{figure}
	\centering
	\includegraphics[width=\hsize]{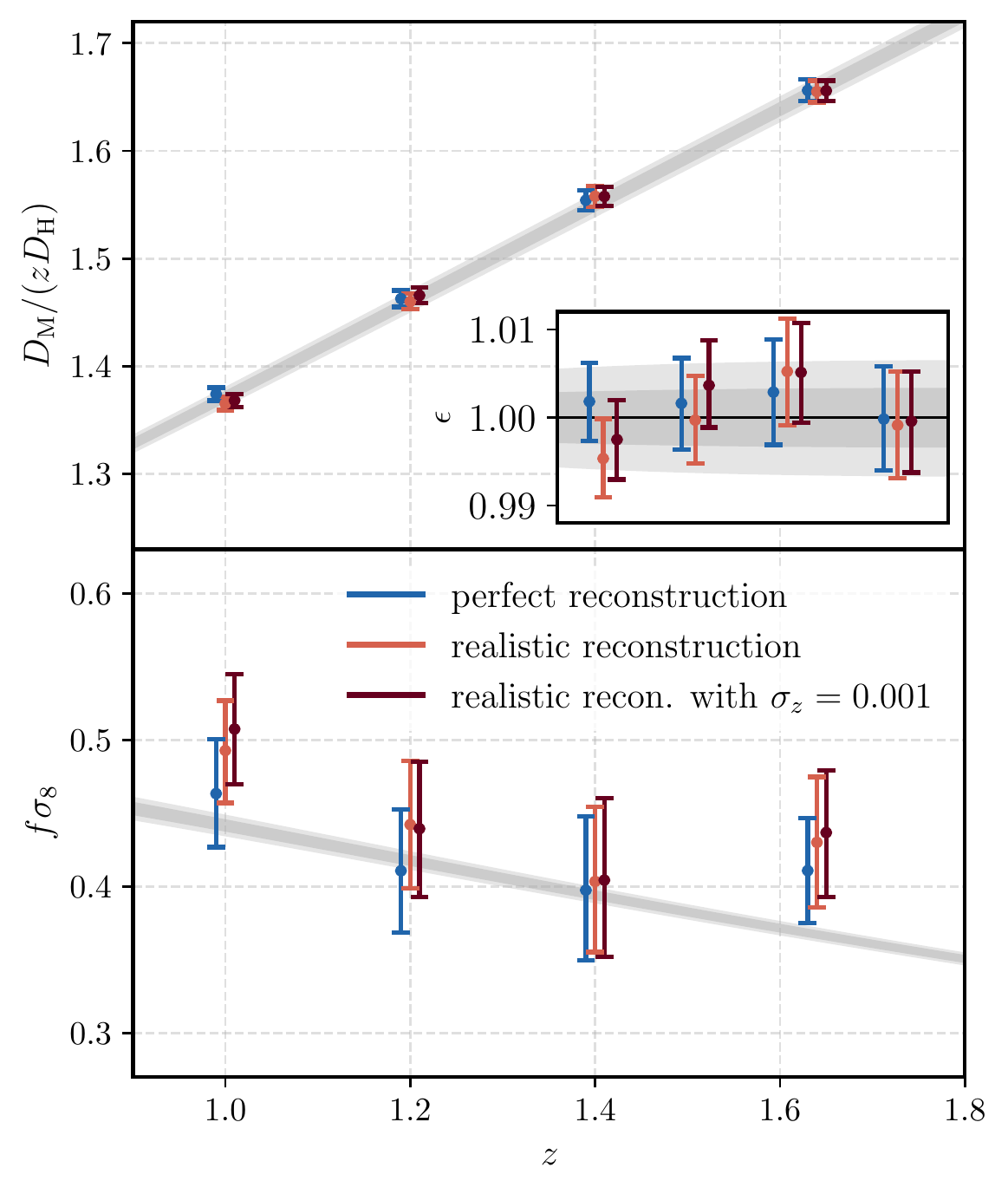}
	
	\caption{Redshift dependence of the measured cosmological parameters obtained from fits to Flagship simulation data. The top panel shows constraints on the ratio of comoving transverse and Hubble distances divided by redshift $z$, $D_{\rm M}/(zD_{\rm H})$, which is derived from $\epsilon$ (inset). The bottom panel shows measurements of the growth rate $f\sigma_8$. Coloured data points indicate the different reconstruction scenarios considered, plotted at slightly shifted redshift locations for clarity. The grey shaded bands represent the $68.3\%$ ($1\sigma$) and $95.5\%$ ($2\sigma$) confidence intervals from fits to \cite{Planck2020}, extrapolated down to these redshifts assuming $\Lambda$CDM, and centred on the Flagship cosmology. }
	
	\label{fig:redshift_evolution}
\end{figure}


\subsection{Fits to Flagship data}
\label{subs:reconstruction_results}

In \cref{fig:realistic_recon_multipoles} we show the measured monopole, quadrupole and hexadecapole moments of the redshift-space CCF $\xi^{\rm ps}$ and the predictions for the best-fit model, together with the residuals, in each of the redshift bins for the realistic reconstruction case and for the catalogue with redshift errors. It can be seen that the model describes the observed multipoles well, although the hexadecapole moment does not contain much information except in the lowest redshift bin where the density of galaxies and voids is highest. This is the most realistic of the scenarios considered, but a similar quality of fit was obtained for the idealised perfect reconstruction scenario and when neglecting redshift errors.

The marginalised one-dimensional posterior constraints on the growth rate $f\sigma_8$ and the AP distortion parameter $\epsilon$ individually obtained from analysis of the Flagship mock are summarised in \cref{tab:results}. For the idealised case of perfect reconstruction, we recover unbiased constraints on both $f\sigma_8$ and $\epsilon$ in all redshift bins. We find a measurement precision on $\epsilon$ of about $0.5$--$0.6\%$ in each individual redshift bin, and the precision on $f\sigma_8$ varies between ${\sim}\,8.5\%$ for the lowest redshift bin and ${\sim}\,12\%$ in the two highest redshift bins. Using realistic reconstruction and adding redshift errors to the mocks does not significantly degrade these constraints, and the central values recovered are consistent to within the stated uncertainty in each case. The reconstruction method used is therefore not causing a significant loss of information.

The marginalised 1D and 2D posteriors on $f\sigma_8$ and $\epsilon$ are shown in \cref{fig:triangle_plot}, where the dashed crosshairs indicate the fiducial values for the Flagship cosmology, which are consistently recovered. Comparing the perfect and realistic reconstruction contours, the largest shift (of ${\sim}\,1\sigma$) is seen in the lowest redshift bin. The introduction of redshift errors does not cause significant additional shifts in any redshift bin.

\Cref{fig:redshift_evolution} shows the marginalised 1D measurements of the growth rate and ratio of the transverse comoving distance $D_{\rm M}$ to the Hubble distance $D_{\rm H}$, derived from the $\epsilon$ constraints, from the Flagship data as a function of redshift. These measurements are compared to the $68.3\%$ ($1\sigma$) and $95.5\%$ ($2\sigma$) confidence intervals on these quantities derived from extrapolating the CMB constraints from \cite{Planck2020} down to these redshifts while assuming a $\Lambda$CDM cosmological model. Results for $\epsilon$ itself are shown in the inset plot. 


\subsection{Full \Euclid volume}
\label{subs:Euclid_full_results}

\begin{figure}
	\centering
	\includegraphics[width=\hsize]{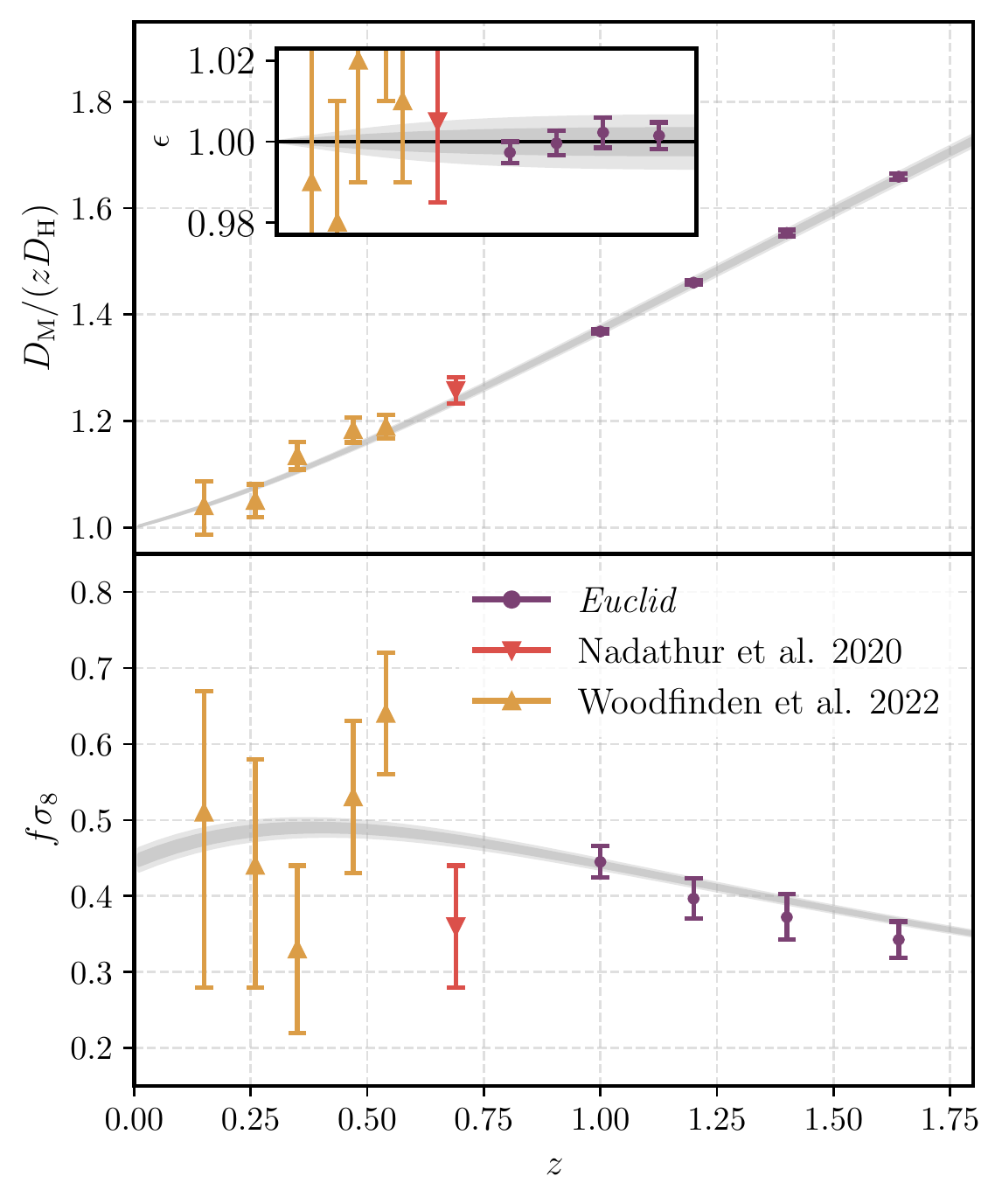}
	\caption{Redshift dependence of the forecast measurements that could be obtained from the $15\,000\;{\rm deg}^2$ \Euclid survey (purple data points). These are generated from a random realisation of the \Euclid CCF (\cref{subs:Euclid_full_results}) and so the central values can be shifted relative to the Flagship results in \cref{fig:redshift_evolution}. Low redshift measurements from current SDSS data obtained using the same analysis method by \cite{Nadathur:2020b} and \cite{Woodfinden:2022} are included for comparison.}
	\label{fig:redshift_evolution_Euclid}
\end{figure}

\begin{figure}
    \centering
    \includegraphics[width=\hsize]{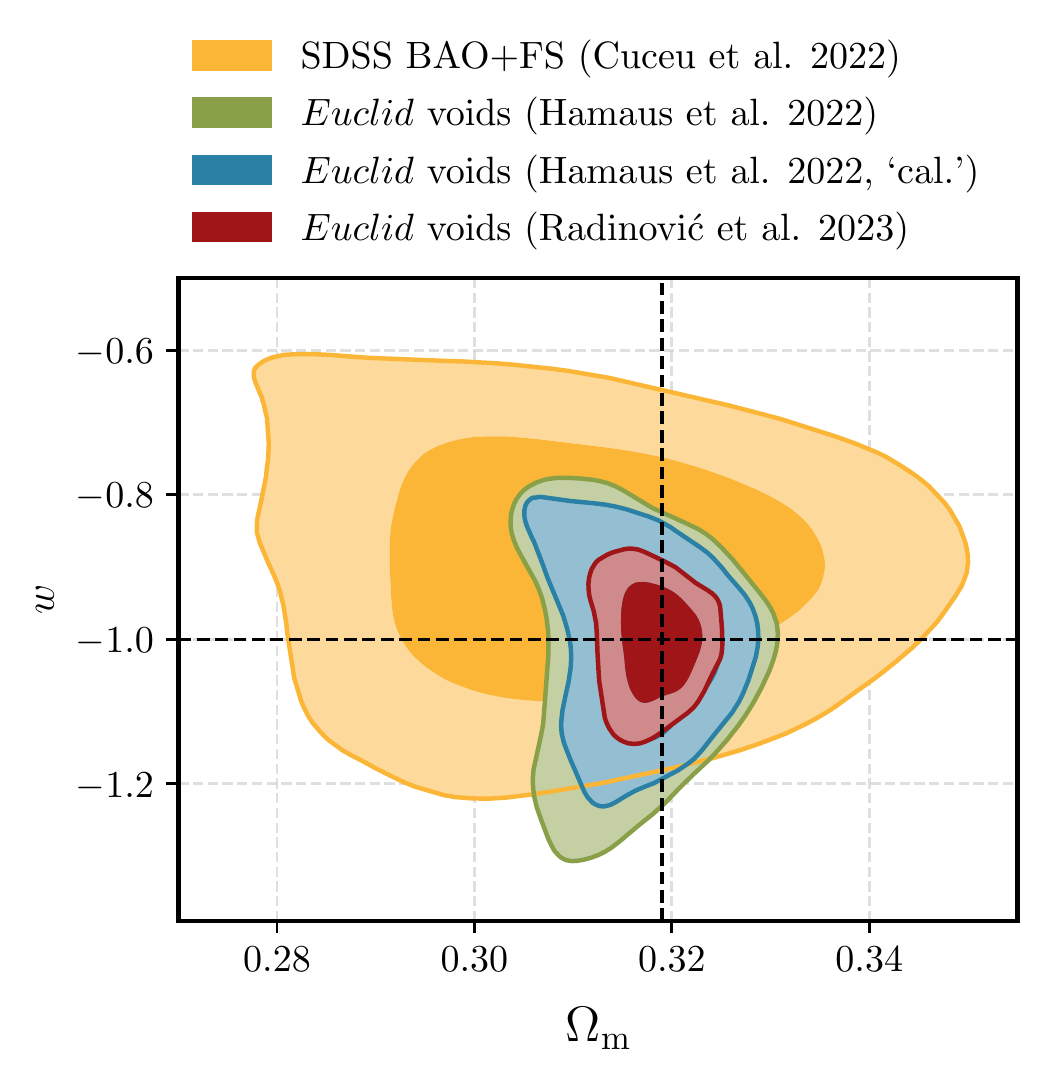}
    \caption{Forecast constraints on the matter density parameter $\Omega_{\rm m}$ and the dark energy equation of state $w$ in a $w$CDM cosmological model from a measurement of the void-galaxy CCF with \Euclid as outlined in this work (red contours), corresponding to $w=-1.00^{+0.06}_{-0.05}$. We also show previous forecasts for the \Euclid void-galaxy CCF -- obtained using a different model and measurement method -- by \cite{Hamaus:2022}, for the `independent' (green contours) and `calibrated' (blue, labelled `cal.' here) cases in that paper, corresponding to whether nuisance parameters are marginalised over or fixed. 
    For context, we show current measurements from current BAO and `full-shape' measurements in the SDSS galaxy, quasar and Lyman-$\alpha$ datasets from the MGS, BOSS and eBOSS surveys (yellow) presented by \cite{Cuceu:2022}.
    }
    \label{fig:triangle_plot_DE}
\end{figure}

\begin{table}[]
	\setlength\extrarowheight{3pt}
	\caption{Forecast marginalised posterior constraints on $\epsilon$ and $f\sigma_8$ from the full \Euclid survey, obtained from fitting a synthetic data vector using an appropriately scaled data covariance matrix. Uncertainties in the two parameters are correlated, with correlation coefficient $\rho$.}
	\label{tab:forecast}
	\centering
	\begin{tabular}{cccc}
		\hline
		$\zeff$ & $f\sigma_8$   & $\epsilon$   & $\rho$   \\ \hline\hline
		1.00    & $0.445^{+0.022}_{-0.021}$ 
                                & $0.9973 \pm 0.0027$ 
                                               & $-0.28$ \\
		1.20    & $0.397^{+0.028}_{-0.027}$ 
                                & $0.9996^{+0.0031}_{-0.0030}$ 
                                               & $-0.23$ \\
		1.40    & $0.372^{+0.031}_{-0.030}$ 
                                & $1.002 \pm 0.004$ 
                                               & $-0.25$ \\
		1.64    & $0.343 \pm 0.024$ 
                                & $1.0014^{+0.0034}_{-0.0033}$ 
                                               & $-0.16$ \\ \hline
	\end{tabular}
\end{table}

The planned final footprint of \Euclid is 15 000 deg$^2$, almost three times larger than that of the Flagship mock used in the analysis above. The constraints we expect from using the full \Euclid volume will therefore be correspondingly better. A simple scaling of the covariance matrix suggests that it should shrink proportional to the increase in the volume, that is by a factor of $2.91$. This corresponds to measurement uncertainties on the binned multipole moments that should be over $40\%$ smaller than those shown in \cref{fig:realistic_recon_multipoles}. We first confirmed this scaling with survey volume by computing covariance matrices for eight volume subsections of the Flagship simulation, ranging from 0.125 to 0.875. We confirmed that the terms of $\tens C$ scale roughly linearly with the survey volume. We therefore obtained the expected covariance matrix $\tens{C}_{ij}^{\rm Euclid}$ for the full \Euclid survey by scaling the one obtained in \cref{subs:inference} by a factor of $1/2.91$. 

To test how this reduction in uncertainties in the measurement of the void-galaxy CCF propagates through to the recovered cosmological parameter constraints, we repeated the parameter inference step with this rescaled covariance and using a synthetic data vector generated assuming this level of measurement noise. To do this, we considered the idealised perfect reconstruction scenario. Using the real-space CCF multipoles $\xi^{\rm rr}_\ell(r)$ measured from Flagship, we computed the fiducial model prediction, $\vec\xi^{\rm s,\mathrm{th}}$, for the redshift-space data vector. We then created a synthetic data vector
\begin{equation}
    \label{eq:synthetic_DV}
    \vec\xi^{\rm s,\mathrm{mock}} = \tens{L}\vec{Z} + \vec\xi^{\rm s,\mathrm{th}}\,,
\end{equation}
where $\tens{L}\tens{L}^\mathrm{T}=\tens{C}^\mathrm{Euclid}$, $\tens{L}$ is obtained by a Cholesky decomposition of the scaled covariance matrix $\tens{C}^\mathrm{Euclid}$, and $\vec{Z}$ is a vector of independent standard normal random variables. We then repeated the parameter inference of \cref{subs:inference} fitting to $\vec\xi^{\rm s,\mathrm{mock}}$ using the rescaled covariance matrix.

The results of this fitting procedure provide the forecast for \Euclid constraints from the void-galaxy CCF measurement, and form the main result of this paper. The forecasts are summarised in \cref{tab:forecast} and shown as a function of redshift in \cref{fig:redshift_evolution_Euclid}, compared to existing constraints from the application of the same void-galaxy method on current data. As expected, we recover the fiducial cosmological parameters up to the $68.3\%$ ($1\sigma$) statistical uncertainty in each redshift bin. Comparing \cref{tab:results} and \cref{tab:forecast} shows that while the forecast uncertainty in $\epsilon$ scales approximately as the $1/\sqrt{2.91}$ factor expected from the simple volume scaling, the forecast constraints on $f\sigma_8$ are slightly more pessimistic. Our final results suggest that with \Euclid voids we will be able to measure the distance ratio $D_{\rm M}/D_{\rm H}$ to a precision of $0.3$--$0.4\%$ in four redshift bins over the range $0.9\leq z<1.8$. This is comparable even to the extrapolated constraints from \Planck which assume $\Lambda$CDM. Voids will also provide a precision of about $5$--$8\%$ in measurement of the growth rate over these redshifts, a large increase in the statistical power of this method compared to current surveys, and a useful consistency check for results from galaxy clustering.

\cref{fig:redshift_evolution_Euclid} highlights the key point that AP constraints from voids will be competitive with those derived from model-dependent extrapolations to low redshift from the CMB (and will exceed those that can be obtained from galaxy clustering alone). AP measurements therefore present the best potential for information gain from using the void-galaxy CCF. On the other hand, while the constraints on $f\sigma_8$ are weaker than those that can be obtained from galaxy clustering alone and do not add much cosmological information, they can serve as useful consistency checks.

\subsection{Cosmology forecasts}
\label{subs:cosmo_forecast}

We now investigate the impact on cosmological model forecasts of the expected measurement precision that can be achieved with \Euclid voids, summarised in \cref{tab:forecast}. \citet{Blanchard:2020} have presented forecasts for the constraints on model parameters that can be achieved with the primary \Euclid probes, namely spectroscopic galaxy clustering, weak lensing, photometric galaxy clustering, and their cross-correlations. In this work, we compare voids to these benchmarks as a separate standalone probe, and do not investigate the combination of voids with other observables (see, for instance, \citealt{Nadathur:2019c,Nadathur:2020b,Bonici:2022} for examples of such combinations). 

In this scenario, voids probe cosmology through the AP measurement of $D_{\rm M}(z)/D_{\rm H}(z)$ and the stringent constraints this places on the background expansion history. In contrast, the growth rate constraints are weaker than those that can be obtained from galaxy clustering. These measurements can be used as tests of consistency of the model in different regimes, but in the absence of any discrepancy they do not add much cosmological information, even when voids are combined with other probes \citep{Nadathur:2020b}. When treating voids as a standalone probe as we do here, they do not provide sufficient constraints to be of interest. We therefore neglect the $f\sigma_8$ results and focus entirely on the geometrical AP constraints. To do so, we centred the $D_{\rm M}(z)/D_{\rm H}(z)$ values at their expected values for the Flagship cosmology and used the measurement uncertainties from \cref{tab:forecast}.

Within the flat $\Lambda$CDM model scenario, measurement of $D_{\rm M}/D_{\rm H}$ at a given redshift directly translates to a constraint on the matter density parameter $\Omega_{\rm m}$ (or, equivalently, on $\Omega_\Lambda=1-\Omega_{\rm m}$). The precision that we can achieve on $D_{\rm M}(z)/D_{\rm H}(z)$ with voids corresponds to $\Omega_{\rm m}=0.3183\pm 0.0028$ (at the $68.3\%$ confidence level), a relative statistical uncertainty of $\pm0.009$. The corresponding forecast relative uncertainty from \Euclid spectroscopic galaxy clustering in the optimistic (pessimistic) forecast scenario is 0.013 (0.021), from weak lensing is 0.012 (0.018), and from the combination of the two is 0.006 (0.009) \citep{Blanchard:2020}. Thus for this model and parameter, \Euclid voids alone seem to outperform both galaxy clustering and weak lensing individually, and look to be competitive with their combination. It is worth noting, however, that this forecast is based only on the statistical constraints, obtained by rescaling a Flagship covariance up to the full Euclid volume.

We also consider a minimal one-parameter extension to the $\Lambda$CDM model known as $w$CDM, in which the cosmological constant $\Lambda$ is replaced by a dark energy component with a constant equation of state that is not restricted to $w=-1$. \cref{fig:triangle_plot_DE} shows the forecast constraints obtained on $\Omega_{\rm m}$ and $w$ in the dark red contours. We recover $w=-1.00^{+0.06}_{-0.05}$ and $\Omega_{\rm m}=0.3183\pm0.0028$, corresponding to a relative precision of 6\% on $w$. For comparison, the yellow contours in \cref{fig:triangle_plot_DE} show the current constraints on these parameters from SDSS at multiple redshifts between $z=0.15$ and $z=2.33$, comprising the BAO results from different galaxy and quasar samples reported by \cite{Alam-eBOSS:2021} together with recent improved high-redshift AP measurements from the `full shape' of the Lyman-$\alpha$ forest clustering \citep{Cuceu:2022}, which give $w=-0.90\pm0.12$. Due to a well-known geometrical degeneracy which is broken by low-redshift AP measurements, the constraints on this model from the CMB (including CMB lensing) are relatively weak, $w=-1.57^{+0.50}_{-0.40}$ \citep{Planck2020}. For completeness, \cref{fig:triangle_plot_DE} also shows two sets of forecasts for the \Euclid void-galaxy CCF obtained by \cite{Hamaus:2022} through a different method for modelling and measurement to that used in this work.

These results highlight the excellent promise of the void-galaxy CCF as a standalone cosmological probe. When removing the assumption of flatness and allowing more general dark energy models with additional degrees of freedom, the void measurement of $D_{\rm M}/D_{\rm H}$ is no longer sufficient to provide useful model constraints on its own. However, the higher precision of the AP measurement still provides large gains in information and improves the figure of merit for dark energy in these models when voids are used in combination with other probes, as shown by \cite{Nadathur:2020a}. Thus this technique retains great potential to enhance \Euclid science.


\section{Conclusions}
\label{sec:Conclusions}

We have presented a forecast of the cosmological constraints that can be obtained from measurement of the anisotropic void-galaxy cross-correlation function (CCF) in the spectroscopic sample of the upcoming \Euclid galaxy survey. The pattern of anisotropies in the redshift-space CCF can be used to simultaneously fit for AP distortions caused by differences between the fiducial analysis model and the true background cosmology, and for RSD caused by the peculiar motions of galaxies around voids. Similar cosmological analyses of the void-galaxy CCF have previously been applied to data from the Sloan Digital Sky Survey \citep{Nadathur:2019c,Nadathur:2020b,Hamaus:2020,Woodfinden:2022}, and have been shown to provide significantly better AP measurements than obtained from traditional galaxy-clustering analyses, including BAO. Precise AP measurements, particularly at high redshift, provide very powerful probes of cosmological models of dark energy \citep{Nadathur:2020a, Cuceu:2022}.

Our forecast is based on an analysis of the lightcone Flagship galaxy mock catalogue, which was calibrated to match the expected properties of the \Euclid spectroscopic galaxy sample, and covers 5157 deg$^2$ of the sky between the redshifts $0.9\leq z<1.8$. We analysed this mock data using the same method previously used by \cite{Nadathur:2019c} and \cite{Woodfinden:2022}. A key component of this method is the use of velocity field reconstruction to approximately remove large-scale RSD from the galaxy distribution before identifying voids. This step is necessary to remove a selection bias in the construction of the void sample that depends on the orientation of the voids to the line-of-sight direction, thus ensuring the validity of the theoretical modelling. To assess the efficiency of the particular algorithm used to perform this reconstruction, we compared the results obtained using the true real-space galaxy field -- corresponding to the idealised perfect reconstruction scenario, that will not be achievable in practice with \Euclid data -- to those obtained with realistic reconstruction, with and without including spectroscopic redshift errors in the mock data. 

We performed the analysis in four non-overlapping redshift bins, compressing the anisotropic CCF information to its first three even-order multipoles -- monopole, quadrupole, and hexadecapole -- which contain all of the practically measurable information. We found that using a realistic reconstruction algorithm did not significantly increase the statistical uncertainty of the recovered marginalised constraints on the cosmological parameters $\epsilon$ and $f\sigma_8$ compared to the idealised scenario, and also did not introduce significant systematic offsets at the level of statistical precision available from the Flagship mock. Realistic spectroscopic redshift errors were found to have a similarly minor effect on the results. However, we caution that the analysis here was limited by the availability of only a single simulation realisation, so we can not guarantee that systematic offsets will not be present when dealing with \Euclid data. A more detailed study of systematic errors should be performed when a larger number of \Euclid mocks become available.

We then extrapolated from the results obtained from the Flagship mock to the full \Euclid survey, covering almost three times larger survey volume, to obtain our forecast for the achievable statistical precision. We found that void-galaxy CCF measurements with \Euclid should be able to achieve a statistical precision of down to $0.3\%$ on the measurement of the distance ratio $D_{\rm M}(z)/D_{\rm H}(z)$ over the redshift range $0.9\leq z<1.8$, and $5$--$8\%$ on the growth rate parameter $f\sigma_8$. Within the context of flat $\Lambda$CDM cosmological models, the statistical uncertainty we achieve on $D_{\rm M}(z)/D_{\rm H}(z)$ translates to a relative statistical uncertainty of $0.009$ on the matter density parameter $\Omega_{{\rm m}}$ which is $\sim30\%$ better than forecasts of what is achievable from both the \Euclid spectroscopic galaxy clustering and weak lensing probes individually, and is comparable to their combination \citep{Blanchard:2020}. In the $w$CDM model scenario, we found that voids alone can measure the dark energy equation of state to a relative precision of 6\%, $w=-1.00^{+0.06}_{-0.05}$.

It should be noted here that the modelling method adopted for this forecast is the template-fitting approach used in several previous works. In this approach template functions describing the real-space CCF, the void velocity profile (or equivalently, matter density profile) and the velocity dispersion profile are first constructed by calibrating against simulation results and then allowed to vary with cosmological parameters, as described in \cref{sec:Theory}. The recent work of \cite{Massara:2022b} suggests that improvements to this template-fitting approach may be required in the future, but our aim is to forecast the constraints possible with currently available techniques so we have not considered these here. We have also not accounted for possible systematic effects which may enter this analysis if the simulations from which the templates are constructed differ strongly from the true cosmological model. A quantitative study of these effects \citep[such as conducted in the context of SDSS by][]{Woodfinden:2022} would require mock catalogues constructed at different cosmologies, but at present we only have access to a single realisation of the \Euclid Flagship mock at a fixed cosmology. A large number of mock catalogues would also allow a better estimate of the covariance matrix than the jackknife resampling technique used here.

In this work, the template we used for the real-space CCF was necessarily estimated from the single realisation of the Flagship data itself. This increases the noise in this estimate, in a manner that is correlated with measurement noise in the data vector. When the final \Euclid data are analysed, we anticipate that large numbers of mock realisations will be available, so that while a similar procedure could be used in the final analysis, it will not be mandatory. If indeed both real-space and redshift-space CCFs are measured from the same data, it would be necessary to account for the correlation between them for a fully self-consistent analysis. We describe how to do this in \cref{sec:Appendix}, although we choose not to include this in the main analysis presented above. However, as shown in \cref{sec:Appendix}, the effect of accounting for this correlation is to {\it{decrease}} the effective statistical uncertainties on the recovered cosmological parameters. The forecasts using the larger statistical uncertainties presented in our headline analysis are therefore expected to be conservative.   

Our treatment of possible observational systematics is quite simple and ignores some effects that we expect will be present in the \Euclid data. In particular, although we have incorporated the effects of a uniformly low completeness of the spectroscopic sample and a Gaussian distribution of redshift errors, we have not considered the purity of the sample and errors due to spectral line misidentification. These effects will be explored in detail in a future work. This paper is one of several companion papers forecasting the constraining power from different void observables in \Euclid \citep{Hamaus:2022,Contarini:2022, Bonici:2022} which all use the same common baseline of assumptions, to allow a consistent comparison across methods. 

\cite{Hamaus:2022} have also investigated forecasts for constraints obtained from the void-galaxy CCF with Flagship, using a different approach. Unlike in our work, that paper did not use reconstruction to try to remove the orientation-dependent sample selection bias in the void catalogues, and thus obtained sharply different measurements of the CCF than those we find here. They then introduced additional nuisance parameters and modified the growth-dependent terms of the theoretical model derived from the conservation equation in \cref{sec:Theory} in order to describe the Flagship data. Other minor differences in approach include the assumption that the matter density profile around voids is simply related to the galaxy profile by the large-scale linear galaxy bias $b$ (instead of our template method for predicting the mean galaxy velocity), and neglecting velocity dispersion around the mean. 

Despite these differences, the forecasts of \cite{Hamaus:2022} are broadly compatible with ours. In their baseline model (which they refer to as `independent'), which marginalises over nuisance parameters as we do here, they predict a $0.5\%$ measurement of the AP parameter, somewhat weaker than our $0.3\%$, and a $4\%$ measurement of the growth rate, slightly better than our value of $5\%$--$8\%$. They further report that if the nuisance parameters are instead fixed to their best-fit values (referred to as the `calibrated' case), these results could improve to $0.4\%$ and $1\%$ precision on the AP parameter and growth rate, respectively. The additional improvement in the AP measurement that is possible from our method can have a large effect on the cosmological constraints obtained, as shown in \cref{fig:triangle_plot_DE}, where uncertainties in $w$ and $\Omega_{\rm m}$ are significantly reduced. Taken together, the results of these two papers showcase both the promise and the robustness of the void-galaxy cross-correlation method applied to \Euclid data.


\begin{acknowledgements}
SR acknowledges support from the Research Council of Norway through project 325113. 
SN acknowledges support from an STFC Ernest Rutherford Fellowship, grant reference ST/T005009/2.
AW acknowledges the support of the Natural Sciences and Engineering Research Council of Canada (NSERC) [funding reference number 547744]. Cette recherche a été financée par le Conseil de recherches en sciences naturelles et en génie du Canada (CRSNG) [numéro de référence 547744]. 
Research at Perimeter Institute is supported in part by the Government of Canada through the Department of Innovation, Science and Economic Development Canada and by the Province of Ontario through the Ministry of Colleges and Universities.
This work has made use of CosmoHub \citep{CosmoHub_Carretero,CosmoHub_Tallada}. CosmoHub has been developed by the Port d’Informació Científica (PIC), maintained through a collaboration of the Institut de Física d’Altes Energies (IFAE) and the Centro de Investigaciones Energéticas, Medioambientales y Tecnológicas (CIEMAT) and the Institute of Space Sciences (CSIC and IEEC), and was partially funded by the ``Plan Estatal de Investigación Científica y Técnica y de Innovación" programme of the Spanish government. 
\AckEC.
For the purpose of open access, the authors have applied a CC BY public copyright licence to any Author Accepted Manuscript version arising.
\end{acknowledgements}


\bibliographystyle{aa}
\bibliography{bibliography}

\begin{thebibliography}{103}
\expandafter\ifx\csname natexlab\endcsname\relax\def\natexlab#1{#1}\fi

\bibitem[{{Achitouv}(2017)}]{Achitouv:2017b}
{Achitouv}, I. 2017, \prd, 96, 083506

\bibitem[{{Achitouv}(2019)}]{Achitouv:2019}
{Achitouv}, I. 2019, \prd, 100, 123513

\bibitem[{{Alam} {et~al.}(2017){Alam}, {Ata}, {Bailey}, {Beutler}, {Bizyaev},
  {Blazek}, {Bolton}, {Brownstein}, {Burden}, {Chuang}, {Comparat}, {Cuesta},
  {Dawson}, {Eisenstein}, {Escoffier}, {Gil-Mar{\'\i}n}, {Grieb}, {Hand}, {Ho},
  {Kinemuchi}, {Kirkby}, {Kitaura}, {Malanushenko}, {Malanushenko}, {Maraston},
  {McBride}, {Nichol}, {Olmstead}, {Oravetz}, {Padmanabhan},
  {Palanque-Delabrouille}, {Pan}, {Pellejero-Ibanez}, {Percival}, {Petitjean},
  {Prada}, {Price-Whelan}, {Reid}, {Rodr{\'\i}guez-Torres}, {Roe}, {Ross},
  {Ross}, {Rossi}, {Rubi{\~n}o-Mart{\'\i}n}, {Saito}, {Salazar-Albornoz},
  {Samushia}, {S{\'a}nchez}, {Satpathy}, {Schlegel}, {Schneider},
  {Sc{\'o}ccola}, {Seo}, {Sheldon}, {Simmons}, {Slosar}, {Strauss}, {Swanson},
  {Thomas}, {Tinker}, {Tojeiro}, {Maga{\~n}a}, {Vazquez}, {Verde}, {Wake},
  {Wang}, {Weinberg}, {White}, {Wood-Vasey}, {Y{\`e}che}, {Zehavi}, {Zhai}, \&
  {Zhao}}]{Alam-BOSS:2017}
{Alam}, S., {Ata}, M., {Bailey}, S., {et~al.} 2017, \mnras, 470, 2617

\bibitem[{{Alam} {et~al.}(2021){Alam}, {Aubert}, {Avila}, {Balland},
  {Bautista}, {Bershady}, {Bizyaev}, {Blanton}, {Bolton}, {Bovy}, {Brinkmann},
  {Brownstein}, {Burtin}, {Chabanier}, {Chapman}, {Choi}, {Chuang}, {Comparat},
  {Cousinou}, {Cuceu}, {Dawson}, {de la Torre}, {de Mattia}, {Agathe}, {des
  Bourboux}, {Escoffier}, {Etourneau}, {Farr}, {Font-Ribera}, {Frinchaboy},
  {Fromenteau}, {Gil-Mar{\'\i}n}, {Le Goff}, {Gonzalez-Morales},
  {Gonzalez-Perez}, {Grabowski}, {Guy}, {Hawken}, {Hou}, {Kong}, {Parker},
  {Klaene}, {Kneib}, {Lin}, {Long}, {Lyke}, {de la Macorra}, {Martini},
  {Masters}, {Mohammad}, {Moon}, {Mueller}, {Mu{\~n}oz-Guti{\'e}rrez}, {Myers},
  {Nadathur}, {Neveux}, {Newman}, {Noterdaeme}, {Oravetz}, {Oravetz},
  {Palanque-Delabrouille}, {Pan}, {Paviot}, {Percival}, {P{\'e}rez-R{\`a}fols},
  {Petitjean}, {Pieri}, {Prakash}, {Raichoor}, {Ravoux}, {Rezaie}, {Rich},
  {Ross}, {Rossi}, {Ruggeri}, {Ruhlmann-Kleider}, {S{\'a}nchez}, {S{\'a}nchez},
  {S{\'a}nchez-Gallego}, {Sayres}, {Schneider}, {Seo}, {Shafieloo}, {Slosar},
  {Smith}, {Stermer}, {Tamone}, {Tinker}, {Tojeiro}, {Vargas-Maga{\~n}a},
  {Variu}, {Wang}, {Weaver}, {Weijmans}, {Y{\`e}che}, {Zarrouk}, {Zhao},
  {Zhao}, \& {Zheng}}]{Alam-eBOSS:2021}
{Alam}, S., {Aubert}, M., {Avila}, S., {et~al.} 2021, \prd, 103, 083533

\bibitem[{{Alcock} \& {Paczynski}(1979)}]{Alcock:1979}
{Alcock}, C. \& {Paczynski}, B. 1979, \nat, 281, 358

\bibitem[{{Alonso} {et~al.}(2018){Alonso}, {Hill}, {Hlo{\v z}ek}, \&
  {Spergel}}]{Alonso:2018}
{Alonso}, D., {Hill}, J.~C., {Hlo{\v z}ek}, R., \& {Spergel}, D.~N. 2018, \prd,
  97, 063514

\bibitem[{{Aubert} {et~al.}(2022){Aubert}, {Cousinou}, {Escoffier}, {Hawken},
  {Nadathur}, {Alam}, {Bautista}, {Burtin}, {Chuang}, {de la Macorra}, {de
  Mattia}, {Gil-Mar{\'\i}n}, {Hou}, {Jullo}, {Kneib}, {Neveux}, {Rossi},
  {Schneider}, {Smith}, {Tamone}, {Vargas Maga{\~n}a}, \& {Zhao}}]{Aubert20a}
{Aubert}, M., {Cousinou}, M.-C., {Escoffier}, S., {et~al.} 2022, \mnras, 513,
  186

\bibitem[{{Bautista} {et~al.}(2021){Bautista}, {Paviot}, {Vargas Maga{\~n}a},
  {de la Torre}, {Fromenteau}, {Gil-Mar{\'\i}n}, {Ross}, {Burtin}, {Dawson},
  {Hou}, {Kneib}, {de Mattia}, {Percival}, {Rossi}, {Tojeiro}, {Zhao}, {Zhao},
  {Alam}, {Brownstein}, {Chapman}, {Choi}, {Chuang}, {Escoffier}, {de la
  Macorra}, {du Mas des Bourboux}, {Mohammad}, {Moon}, {M{\"u}ller},
  {Nadathur}, {Newman}, {Schneider}, {Seo}, \& {Wang}}]{Bautista:2021}
{Bautista}, J.~E., {Paviot}, R., {Vargas Maga{\~n}a}, M., {et~al.} 2021,
  \mnras, 500, 736

\bibitem[{Bayer {et~al.}(2021)Bayer, Villaescusa-Navarro, Massara, Liu,
  Spergel, Verde, Wandelt, Viel, \& Ho}]{Bayer:2021}
Bayer, A.~E., Villaescusa-Navarro, F., Massara, E., {et~al.} 2021, \apj, 919,
  24

\bibitem[{{Behroozi} {et~al.}(2013){Behroozi}, {Wechsler}, \&
  {Wu}}]{Behroozi:2013}
{Behroozi}, P.~S., {Wechsler}, R.~H., \& {Wu}, H.-Y. 2013, \apj, 762, 109

\bibitem[{Beutler {et~al.}(2017)}]{BOSS:2016psr}
Beutler, F. {et~al.} 2017, \mnras, 466, 2242

\bibitem[{{Blake} \& {Glazebrook}(2003)}]{Blake:2003}
{Blake}, C. \& {Glazebrook}, K. 2003, \apj, 594, 665

\bibitem[{{Bonici} {et~al.}(2023){Bonici}, {Carbone}, {Vielzeuf}, {Paganin},
  {Cardone}, {Hamaus}, {Pisani}, {Hawken}, {Kovacs}, {Nadathur}, {Contarini},
  {Verza}, {Tutusaus}, {Marulli}, {Moscardini}, {Aubert}, {Giocoli},
  {Pourtsidou}, {Camera}, {Escoffier}, {Caminata}, {Martinelli}, {Pallavicini},
  {Pettorino}, {Sakr}, {Sapone}, {Testera}, {Tosi}, {Yankelevich}, {Amara},
  {Auricchio}, {Baldi}, {Bonino}, {Branchini}, {Brescia}, {Brinchmann},
  {Capobianco}, {Carretero}, {Castellano}, {Cavuoti}, {Cledassou}, {Congedo},
  {Conversi}, {Copin}, {Corcione}, {Courbin}, {Cropper}, {Da Silva},
  {Degaudenzi}, {Douspis}, {Dubath}, {Duncan}, {Dupac}, {Dusini}, {Ealet},
  {Farrens}, {Ferriol}, {Fosalba}, {Frailis}, {Franceschi}, {Fumana},
  {Gomez-Alvarez}, {Garilli}, {Gillis}, {Grazian}, {Grupp}, {Guzzo}, {Haugan},
  {Holmes}, {Hormuth}, {Hornstrup}, {Jahnke}, {Kummel}, {Kermiche},
  {Kiessling}, {Kilbinger}, {Kunz}, {Kurki-Suonio}, {Laureijs}, {Ligori},
  {Lilje}, {Lloro}, {Maiorano}, {Mansutti}, {Marggraf}, {Markovic}, {Massey},
  {Medinaceli}, {Melchior}, {Meneghetti}, {Meylan}, {Moresco}, {Munari},
  {Niemi}, {Padilla}, {Paltani}, {Pasian}, {Pedersen}, {Percival}, {Pires},
  {Polenta}, {Poncet}, {Popa}, {Raison}, {Rebolo}, {Renzi}, {Rhodes},
  {Rossetti}, {Saglia}, {Sartoris}, {Scodeggio}, {Secroun}, {Seidel},
  {Sirignano}, {Sirri}, {Stanco}, {Starck}, {Surace}, {Tallada-Crespi},
  {Tavagnacco}, {Taylor}, {Tereno}, {Toledo-Moreo}, {Torradeflot}, {Valentijn},
  {Valenziano}, {Wang}, {Weller}, {Zamorani}, {Zoubian}, \&
  {Andreon}}]{Bonici:2022}
{Bonici}, M., {Carbone}, C., {Vielzeuf}, P., {et~al.} 2023, \aap, 670, A47

\bibitem[{{Bonnaire} {et~al.}(2022){Bonnaire}, {Aghanim}, {Kuruvilla}, \&
  {Decelle}}]{Bonnaire:2022}
{Bonnaire}, T., {Aghanim}, N., {Kuruvilla}, J., \& {Decelle}, A. 2022, \aap,
  661, A146

\bibitem[{{Burden} {et~al.}(2015){Burden}, {Percival}, \&
  {Howlett}}]{Burden:2015}
{Burden}, A., {Percival}, W.~J., \& {Howlett}, C. 2015, \mnras, 453, 456

\bibitem[{{Cai} {et~al.}(2016){Cai}, {Taylor}, {Peacock}, \&
  {Padilla}}]{Cai:2016a}
{Cai}, Y.-C., {Taylor}, A., {Peacock}, J.~A., \& {Padilla}, N. 2016, \mnras,
  462, 2465

\bibitem[{{Carretero} {et~al.}(2017){Carretero}, {Tallada}, {Casals}, {Caubet},
  {Castander}, {Blot}, {Alarc{\'o}n}, {Serrano}, {Fosalba}, {Acosta-Silva},
  {Tonello}, {Torradeflot}, {Eriksen}, {Neissner}, \&
  {Delfino}}]{CosmoHub_Carretero}
{Carretero}, J., {Tallada}, P., {Casals}, J., {et~al.} 2017, in Proceedings of
  the European Physical Society Conference on High Energy Physics. 5-12 July,
  488

\bibitem[{{Carron} \& {Szapudi}(2014)}]{Carron:2014}
{Carron}, J. \& {Szapudi}, I. 2014, \mnras, 439, L11

\bibitem[{{Chan} {et~al.}(2014){Chan}, {Hamaus}, \& {Desjacques}}]{Chan:2014}
{Chan}, K.~C., {Hamaus}, N., \& {Desjacques}, V. 2014, \prd, 90, 103521

\bibitem[{{Chiang} {et~al.}(2014){Chiang}, {Wagner}, {Schmidt}, \&
  {Komatsu}}]{Chiang:2014}
{Chiang}, C.-T., {Wagner}, C., {Schmidt}, F., \& {Komatsu}, E. 2014, \jcap,
  2014, 048

\bibitem[{Chuang {et~al.}(2017)Chuang, Kitaura, Liang, Font-Ribera, Zhao,
  McDonald, \& Tao}]{Chuang2017}
Chuang, C.-H., Kitaura, F.-S., Liang, Y., {et~al.} 2017, \prd, 95

\bibitem[{{Contarini} {et~al.}(2022){Contarini}, {Verza}, {Pisani}, {Hamaus},
  {Sahl{\'e}n}, {Carbone}, {Dusini}, {Marulli}, {Moscardini}, {Renzi},
  {Sirignano}, {Stanco}, {Aubert}, {Bonici}, {Castignani}, {Courtois},
  {Escoffier}, {Guinet}, {Kovacs}, {Lavaux}, {Massara}, {Nadathur}, {Pollina},
  {Ronconi}, {Ruppin}, {Sakr}, {Veropalumbo}, {Wandelt}, {Amara}, {Auricchio},
  {Baldi}, {Bonino}, {Branchini}, {Brescia}, {Brinchmann}, {Camera},
  {Capobianco}, {Carretero}, {Castellano}, {Cavuoti}, {Cledassou}, {Congedo},
  {Conselice}, {Conversi}, {Copin}, {Corcione}, {Courbin}, {Cropper}, {Da
  Silva}, {Degaudenzi}, {Dubath}, {Duncan}, {Dupac}, {Ealet}, {Farrens},
  {Ferriol}, {Fosalba}, {Frailis}, {Franceschi}, {Garilli}, {Gillard},
  {Gillis}, {Giocoli}, {Grazian}, {Grupp}, {Guzzo}, {Haugan}, {Holmes},
  {Hormuth}, {Jahnke}, {K{\"u}mmel}, {Kermiche}, {Kiessling}, {Kilbinger},
  {Kunz}, {Kurki-Suonio}, {Laureijs}, {Ligori}, {Lilje}, {Lloro}, {Maiorano},
  {Mansutti}, {Marggraf}, {Markovic}, {Massey}, {Melchior}, {Meneghetti},
  {Meylan}, {Moresco}, {Munari}, {Niemi}, {Padilla}, {Paltani}, {Pasian},
  {Pedersen}, {Percival}, {Pettorino}, {Pires}, {Polenta}, {Poncet}, {Popa},
  {Pozzetti}, {Raison}, {Rhodes}, {Rossetti}, {Saglia}, {Sartoris},
  {Schneider}, {Secroun}, {Seidel}, {Sirri}, {Surace}, {Tallada-Cresp{\'\i}},
  {Taylor}, {Tereno}, {Toledo-Moreo}, {Torradeflot}, {Valentijn}, {Valenziano},
  {Wang}, {Weller}, {Zamorani}, {Zoubian}, {Andreon}, {Maino}, \&
  {Mei}}]{Contarini:2022}
{Contarini}, S., {Verza}, G., {Pisani}, A., {et~al.} 2022, \aap, 667, A162

\bibitem[{Correa {et~al.}(2019)Correa, Paz, Padilla, Ruiz, Angulo, \&
  S\'anchez}]{Correa:2018vge}
Correa, C.~M., Paz, D.~J., Padilla, N.~D., {et~al.} 2019, \mnras, 485, 5761

\bibitem[{{Correa} {et~al.}(2022){Correa}, {Paz}, {Padilla}, {S{\'a}nchez},
  {Ruiz}, \& {Angulo}}]{Correa:2022}
{Correa}, C.~M., {Paz}, D.~J., {Padilla}, N.~D., {et~al.} 2022, \mnras, 509,
  1871

\bibitem[{Costille {et~al.}(2018)Costille, Caillat, Rossin, Pascal, Sanchez,
  Barette, Laurent, Foulon, \& Pari{\`e}s}]{costille_2018}
Costille, A., Caillat, A., Rossin, C., {et~al.} 2018, in {SPIE Astronomical
  Telescopes + Instrumentation 2018}, Vol. 10698, Austin, United States,
  106982B

\bibitem[{{Cousinou} {et~al.}(2019){Cousinou}, {Pisani}, {Tilquin}, {Hamaus},
  {Hawken}, \& {Escoffier}}]{Cousinou:2019}
{Cousinou}, M.~C., {Pisani}, A., {Tilquin}, A., {et~al.} 2019, Astronomy and
  Computing, 27, 53

\bibitem[{{Cuceu} {et~al.}(2023){Cuceu}, {Font-Ribera}, {Nadathur}, {Joachimi},
  \& {Martini}}]{Cuceu:2022}
{Cuceu}, A., {Font-Ribera}, A., {Nadathur}, S., {Joachimi}, B., \& {Martini},
  P. 2023, \prl, 130, 191003

\bibitem[{Eisenstein {et~al.}(2007)Eisenstein, Seo, Sirko, \&
  Spergel}]{Eisenstein:2007}
Eisenstein, D.~J., Seo, H.-j., Sirko, E., \& Spergel, D. 2007, \apj, 664, 675

\bibitem[{{Euclid Collaboration: Blanchard} {et~al.}(2020){Euclid
  Collaboration: Blanchard}, {Camera}, {Carbone}, {Cardone}, {Casas}, {Clesse},
  {Ili{\'c}}, {Kilbinger}, {Kitching}, {Kunz}, {Lacasa}, {Linder}, {Majerotto},
  {Markovi{\v{c}}}, {Martinelli}, {Pettorino}, {Pourtsidou}, {Sakr},
  {S{\'a}nchez}, {Sapone}, {Tutusaus}, {Yahia-Cherif}, {Yankelevich},
  {Andreon}, {Aussel}, {Balaguera-Antol{\'\i}nez}, {Baldi}, {Bardelli},
  {Bender}, {Biviano}, {Bonino}, {Boucaud}, {Bozzo}, {Branchini}, {Brau-Nogue},
  {Brescia}, {Brinchmann}, {Burigana}, {Cabanac}, {Capobianco}, {Cappi},
  {Carretero}, {Carvalho}, {Casas}, {Castander}, {Castellano}, {Cavuoti},
  {Cimatti}, {Cledassou}, {Colodro-Conde}, {Congedo}, {Conselice}, {Conversi},
  {Copin}, {Corcione}, {Coupon}, {Courtois}, {Cropper}, {Da Silva}, {de la
  Torre}, {Di Ferdinando}, {Dubath}, {Ducret}, {Duncan}, {Dupac}, {Dusini},
  {Fabbian}, {Fabricius}, {Farrens}, {Fosalba}, {Fotopoulou}, {Fourmanoit},
  {Frailis}, {Franceschi}, {Franzetti}, {Fumana}, {Galeotta}, {Gillard},
  {Gillis}, {Giocoli}, {G{\'o}mez-Alvarez}, {Graci{\'a}-Carpio}, {Grupp},
  {Guzzo}, {Hoekstra}, {Hormuth}, {Israel}, {Jahnke}, {Keihanen}, {Kermiche},
  {Kirkpatrick}, {Kohley}, {Kubik}, {Kurki-Suonio}, {Ligori}, {Lilje}, {Lloro},
  {Maino}, {Maiorano}, {Marggraf}, {Martinet}, {Marulli}, {Massey},
  {Medinaceli}, {Mei}, {Mellier}, {Metcalf}, {Metge}, {Meylan}, {Moresco},
  {Moscardini}, {Munari}, {Nichol}, {Niemi}, {Nucita}, {Padilla}, {Paltani},
  {Pasian}, {Percival}, {Pires}, {Polenta}, {Poncet}, {Pozzetti}, {Racca},
  {Raison}, {Renzi}, {Rhodes}, {Romelli}, {Roncarelli}, {Rossetti}, {Saglia},
  {Schneider}, {Scottez}, {Secroun}, {Sirri}, {Stanco}, {Starck}, {Sureau},
  {Tallada-Cresp{\'\i}}, {Tavagnacco}, {Taylor}, {Tenti}, {Tereno},
  {Toledo-Moreo}, {Torradeflot}, {Valenziano}, {Vassallo}, {Verdoes Kleijn},
  {Viel}, {Wang}, {Zacchei}, {Zoubian}, \& {Zucca}}]{Blanchard:2020}
{Euclid Collaboration: Blanchard}, A., {Camera}, S., {Carbone}, C., {et~al.}
  2020, \aap, 642, A191

\bibitem[{Fiorini {et~al.}(2022)Fiorini, Koyama, \& Izard}]{Fiorini:2022}
Fiorini, B., Koyama, K., \& Izard, A. 2022, \jcap, 12, 028

\bibitem[{{Gil-Mar{\'\i}n} {et~al.}(2020){Gil-Mar{\'\i}n}, {Bautista},
  {Paviot}, {Vargas-Maga{\~n}a}, {de la Torre}, {Fromenteau}, {Alam},
  {{\'A}vila}, {Burtin}, {Chuang}, {Dawson}, {Hou}, {de Mattia}, {Mohammad},
  {M{\"u}ller}, {Nadathur}, {Neveux}, {Percival}, {Raichoor}, {Rezaie}, {Ross},
  {Rossi}, {Ruhlmann-Kleider}, {Smith}, {Tamone}, {Tinker}, {Tojeiro}, {Wang},
  {Zhao}, {Zhao}, {Brinkmann}, {Brownstein}, {Choi}, {Escoffier}, {de la
  Macorra}, {Moon}, {Newman}, {Schneider}, {Seo}, \& {Vivek}}]{Gil-Marin:2020}
{Gil-Mar{\'\i}n}, H., {Bautista}, J.~E., {Paviot}, R., {et~al.} 2020, \mnras,
  498, 2492

\bibitem[{Gil-Mar{\'{\i}}n {et~al.}(2016)Gil-Mar{\'{\i}}n, Percival, Verde,
  Brownstein, Chuang, Kitaura, Rodr{\'{\i}}guez-Torres, \&
  Olmstead}]{Gil-Marin:2016}
Gil-Mar{\'{\i}}n, H., Percival, W.~J., Verde, L., {et~al.} 2016, \mnras, 465,
  1757

\bibitem[{Granett {et~al.}(2008)Granett, Neyrinck, \& Szapudi}]{Granett:2008}
Granett, B.~R., Neyrinck, M.~C., \& Szapudi, I. 2008, \apj, 683, L99

\bibitem[{{Guo} {et~al.}(2015){Guo}, {Zheng}, {Jing}, {Zehavi}, {Li},
  {Weinberg}, {Skibba}, {Nichol}, {Rossi}, {Sabiu}, {Schneider}, \&
  {McBride}}]{Guo:2015}
{Guo}, H., {Zheng}, Z., {Jing}, Y.~P., {et~al.} 2015, \mnras, 449, L95

\bibitem[{{Hamaus} {et~al.}(2022){Hamaus}, {Aubert}, {Pisani}, {Contarini},
  {Verza}, {Cousinou}, {Escoffier}, {Hawken}, {Lavaux}, {Pollina}, {Wandelt},
  {Weller}, {Bonici}, {Carbone}, {Guzzo}, {Kovacs}, {Marulli}, {Massara},
  {Moscardini}, {Ntelis}, {Percival}, {Radinovi{\'c}}, {Sahl{\'e}n}, {Sakr},
  {S{\'a}nchez}, {Winther}, {Auricchio}, {Awan}, {Bender}, {Bodendorf},
  {Bonino}, {Branchini}, {Brescia}, {Brinchmann}, {Capobianco}, {Carretero},
  {Castander}, {Castellano}, {Cavuoti}, {Cimatti}, {Cledassou}, {Congedo},
  {Conversi}, {Copin}, {Corcione}, {Cropper}, {Da Silva}, {Degaudenzi},
  {Douspis}, {Dubath}, {Duncan}, {Dupac}, {Dusini}, {Ealet}, {Ferriol},
  {Fosalba}, {Frailis}, {Franceschi}, {Franzetti}, {Fumana}, {Garilli},
  {Gillis}, {Giocoli}, {Grazian}, {Grupp}, {Haugan}, {Holmes}, {Hormuth},
  {Jahnke}, {Kermiche}, {Kiessling}, {Kilbinger}, {Kitching}, {K{\"u}mmel},
  {Kunz}, {Kurki-Suonio}, {Ligori}, {Lilje}, {Lloro}, {Maiorano}, {Marggraf},
  {Markovic}, {Massey}, {Maurogordato}, {Melchior}, {Meneghetti}, {Meylan},
  {Moresco}, {Munari}, {Niemi}, {Padilla}, {Paltani}, {Pasian}, {Pedersen},
  {Pettorino}, {Pires}, {Poncet}, {Popa}, {Pozzetti}, {Rebolo}, {Rhodes},
  {Rix}, {Roncarelli}, {Rossetti}, {Saglia}, {Schneider}, {Secroun}, {Seidel},
  {Serrano}, {Sirignano}, {Sirri}, {Starck}, {Tallada-Cresp{\'\i}},
  {Tavagnacco}, {Taylor}, {Tereno}, {Toledo-Moreo}, {Torradeflot}, {Valentijn},
  {Valenziano}, {Wang}, {Welikala}, {Zamorani}, {Zoubian}, {Andreon}, {Baldi},
  {Camera}, {Mei}, {Neissner}, \& {Romelli}}]{Hamaus:2022}
{Hamaus}, N., {Aubert}, M., {Pisani}, A., {et~al.} 2022, \aap, 658, A20

\bibitem[{{Hamaus} {et~al.}(2020){Hamaus}, Pisani, Choi, Lavaux, Wandelt, \&
  Weller}]{Hamaus:2020}
{Hamaus}, N., Pisani, A., Choi, J.-A., {et~al.} 2020, \jcap, 2020, 023

\bibitem[{{Hamaus} {et~al.}(2016){Hamaus}, {Pisani}, {Sutter}, {Lavaux},
  {Escoffier}, {Wandelt}, \& {Weller}}]{Hamaus:2016}
{Hamaus}, N., {Pisani}, A., {Sutter}, P.~M., {et~al.} 2016, \prl, 117, 091302

\bibitem[{{Hamaus} {et~al.}(2015){Hamaus}, {Sutter}, {Lavaux}, \&
  {Wandelt}}]{Hamaus:2015}
{Hamaus}, N., {Sutter}, P.~M., {Lavaux}, G., \& {Wandelt}, B.~D. 2015, \jcap,
  2015, 036

\bibitem[{{Hamaus} {et~al.}(2014){Hamaus}, {Sutter}, \&
  {Wandelt}}]{Hamaus:2014a}
{Hamaus}, N., {Sutter}, P.~M., \& {Wandelt}, B.~D. 2014, \prl, 112, 251302

\bibitem[{{Hawken} {et~al.}(2020){Hawken}, {Aubert}, {Pisani}, {Cousinou},
  {Escoffier}, {Nadathur}, {Rossi}, \& {Schneider}}]{Hawken:2020}
{Hawken}, A.~J., {Aubert}, M., {Pisani}, A., {et~al.} 2020, \jcap, 2020, 012

\bibitem[{{Hawken} {et~al.}(2017){Hawken}, {Granett}, {Iovino}, {Guzzo},
  {Peacock}, {de la Torre}, {Garilli}, {Bolzonella}, {Scodeggio}, {Abbas},
  {Adami}, {Bottini}, {Cappi}, {Cucciati}, {Davidzon}, {Fritz}, {Franzetti},
  {Krywult}, {Le Brun}, {Le F{\`e}vre}, {Maccagni}, {Ma{\l}ek}, {Marulli},
  {Polletta}, {Pollo}, {Tasca}, {Tojeiro}, {Vergani}, {Zanichelli}, {Arnouts},
  {Bel}, {Branchini}, {De Lucia}, {Ilbert}, {Moscardini}, \&
  {Percival}}]{Hawken:2017}
{Hawken}, A.~J., {Granett}, B.~R., {Iovino}, A., {et~al.} 2017, \aap, 607, A54

\bibitem[{{Kaiser}(1987)}]{Kaiser:1987}
{Kaiser}, N. 1987, \mnras, 227, 1

\bibitem[{Kazin {et~al.}(2012)Kazin, Sánchez, \& Blanton}]{Kazin:2012}
Kazin, E.~A., Sánchez, A.~G., \& Blanton, M.~R. 2012, \mnras, 419, 3223

\bibitem[{{Kitaura} {et~al.}(2016){Kitaura}, {Chuang}, {Liang}, {Zhao}, {Tao},
  {Rodr{\'{\i}}guez-Torres}, {Eisenstein}, {Gil-Mar{\'{\i}}n}, {Kneib},
  {McBride}, {Percival}, {Ross}, {S{\'a}nchez}, {Tinker}, {Tojeiro},
  {Vargas-Magana}, \& {Zhao}}]{Kitaura:2016b}
{Kitaura}, F.-S., {Chuang}, C.-H., {Liang}, Y., {et~al.} 2016, \prl, 116,
  171301

\bibitem[{{Kov{\'a}cs} {et~al.}(2019){Kov{\'a}cs}, {S{\'a}nchez},
  {Garc{\'\i}a-Bellido}, {Elvin-Poole}, {Hamaus}, {Miranda}, {Nadathur},
  {Abbott}, {Abdalla}, {Annis}, {Avila}, {Bertin}, {Brooks}, {Burke}, {Carnero
  Rosell}, {Carrasco Kind}, {Carretero}, {Cawthon}, {Crocce}, {Cunha}, {da
  Costa}, {Davis}, {De Vicente}, {DePoy}, {Desai}, {Diehl}, {Doel},
  {Fernandez}, {Flaugher}, {Fosalba}, {Frieman}, {Gazta{\~n}aga}, {Gerdes},
  {Gruendl}, {Gutierrez}, {Hartley}, {Hollowood}, {Honscheid}, {Hoyle},
  {James}, {Krause}, {Kuehn}, {Kuropatkin}, {Lahav}, {Lima}, {Maia}, {March},
  {Marshall}, {Melchior}, {Menanteau}, {Miller}, {Miquel}, {Mohr}, {Plazas},
  {Romer}, {Rykoff}, {Sanchez}, {Scarpine}, {Schindler}, {Schubnell},
  {Sevilla-Noarbe}, {Smith}, {Smith}, {Soares-Santos}, {Sobreira}, {Suchyta},
  {Swanson}, {Tarle}, {Thomas}, {Vikram}, {Weller}, \& {DES
  Collaboration}}]{Kovacs:2019}
{Kov{\'a}cs}, A., {S{\'a}nchez}, C., {Garc{\'\i}a-Bellido}, J., {et~al.} 2019,
  \mnras, 484, 5267

\bibitem[{{Landy} \& {Szalay}(1993)}]{Landy:1993}
{Landy}, S.~D. \& {Szalay}, A.~S. 1993, \apj, 412, 64

\bibitem[{{Laureijs} {et~al.}(2011){Laureijs}, {Amiaux}, {Arduini},
  {Augu{\`e}res}, {Brinchmann}, {Cole}, {Cropper}, {Dabin}, {Duvet}, {Ealet},
  {Garilli}, {Gondoin}, {Guzzo}, {Hoar}, {Hoekstra}, {Holmes}, {Kitching},
  {Maciaszek}, {Mellier}, {Pasian}, {Percival}, {Rhodes}, {Saavedra Criado},
  {Sauvage}, {Scaramella}, {Valenziano}, {Warren}, {Bender}, {Castander},
  {Cimatti}, {Le F{\`e}vre}, {Kurki-Suonio}, {Levi}, {Lilje}, {Meylan},
  {Nichol}, {Pedersen}, {Popa}, {Rebolo Lopez}, {Rix}, {Rottgering},
  {Zeilinger}, {Grupp}, {Hudelot}, {Massey}, {Meneghetti}, {Miller}, {Paltani},
  {Paulin-Henriksson}, {Pires}, {Saxton}, {Schrabback}, {Seidel}, {Walsh},
  {Aghanim}, {Amendola}, {Bartlett}, {Baccigalupi}, {Beaulieu}, {Benabed},
  {Cuby}, {Elbaz}, {Fosalba}, {Gavazzi}, {Helmi}, {Hook}, {Irwin}, {Kneib},
  {Kunz}, {Mannucci}, {Moscardini}, {Tao}, {Teyssier}, {Weller}, {Zamorani},
  {Zapatero Osorio}, {Boulade}, {Foumond}, {Di Giorgio}, {Guttridge}, {James},
  {Kemp}, {Martignac}, {Spencer}, {Walton}, {Bl{\"u}mchen}, {Bonoli},
  {Bortoletto}, {Cerna}, {Corcione}, {Fabron}, {Jahnke}, {Ligori}, {Madrid},
  {Martin}, {Morgante}, {Pamplona}, {Prieto}, {Riva}, {Toledo}, {Trifoglio},
  {Zerbi}, {Abdalla}, {Douspis}, {Grenet}, {Borgani}, {Bouwens}, {Courbin},
  {Delouis}, {Dubath}, {Fontana}, {Frailis}, {Grazian}, {Koppenh{\"o}fer},
  {Mansutti}, {Melchior}, {Mignoli}, {Mohr}, {Neissner}, {Noddle}, {Poncet},
  {Scodeggio}, {Serrano}, {Shane}, {Starck}, {Surace}, {Taylor},
  {Verdoes-Kleijn}, {Vuerli}, {Williams}, {Zacchei}, {Altieri}, {Escudero
  Sanz}, {Kohley}, {Oosterbroek}, {Astier}, {Bacon}, {Bardelli}, {Baugh},
  {Bellagamba}, {Benoist}, {Bianchi}, {Biviano}, {Branchini}, {Carbone},
  {Cardone}, {Clements}, {Colombi}, {Conselice}, {Cresci}, {Deacon}, {Dunlop},
  {Fedeli}, {Fontanot}, {Franzetti}, {Giocoli}, {Garcia-Bellido}, {Gow},
  {Heavens}, {Hewett}, {Heymans}, {Holland}, {Huang}, {Ilbert}, {Joachimi},
  {Jennins}, {Kerins}, {Kiessling}, {Kirk}, {Kotak}, {Krause}, {Lahav}, {van
  Leeuwen}, {Lesgourgues}, {Lombardi}, {Magliocchetti}, {Maguire}, {Majerotto},
  {Maoli}, {Marulli}, {Maurogordato}, {McCracken}, {McLure}, {Melchiorri},
  {Merson}, {Moresco}, {Nonino}, {Norberg}, {Peacock}, {Pello}, {Penny},
  {Pettorino}, {Di Porto}, {Pozzetti}, {Quercellini}, {Radovich}, {Rassat},
  {Roche}, {Ronayette}, {Rossetti}, {Sartoris}, {Schneider}, {Semboloni},
  {Serjeant}, {Simpson}, {Skordis}, {Smadja}, {Smartt}, {Spano}, {Spiro},
  {Sullivan}, {Tilquin}, {Trotta}, {Verde}, {Wang}, {Williger}, {Zhao},
  {Zoubian}, \& {Zucca}}]{Laureijs:2011}
{Laureijs}, R., {Amiaux}, J., {Arduini}, S., {et~al.} 2011, arXiv:1110.3193

\bibitem[{Lavaux \& Wandelt(2012)}]{Lavaux:2012}
Lavaux, G. \& Wandelt, B.~D. 2012, \apj, 754, 109

\bibitem[{{Lewis} {et~al.}(2000){Lewis}, {Challinor}, \&
  {Lasenby}}]{Lewis_2000}
{Lewis}, A., {Challinor}, A., \& {Lasenby}, A. 2000, \apj, 538, 473

\bibitem[{Linder(2005)}]{Linder:2005in}
Linder, E.~V. 2005, \prd, 72, 043529

\bibitem[{{Mar{\'\i}n} {et~al.}(2013){Mar{\'\i}n}, {Blake}, {Poole}, {McBride},
  {Brough}, {Colless}, {Contreras}, {Couch}, {Croton}, {Croom}, {Davis},
  {Drinkwater}, {Forster}, {Gilbank}, {Gladders}, {Glazebrook}, {Jelliffe},
  {Jurek}, {Li}, {Madore}, {Martin}, {Pimbblet}, {Pracy}, {Sharp}, {Wisnioski},
  {Woods}, {Wyder}, \& {Yee}}]{Marin:2013}
{Mar{\'\i}n}, F.~A., {Blake}, C., {Poole}, G.~B., {et~al.} 2013, \mnras, 432,
  2654

\bibitem[{{Massara} {et~al.}(2022){Massara}, {Percival}, {Dalal}, {Nadathur},
  {Radinovi{\'c}}, {Winther}, \& {Woodfinden}}]{Massara:2022b}
{Massara}, E., {Percival}, W.~J., {Dalal}, N., {et~al.} 2022, \mnras, 517, 4458

\bibitem[{Massara {et~al.}(2023)Massara, Villaescusa-Navarro, Hahn, Abidi,
  Eickenberg, Ho, Lemos, Moradinezhad~Dizgah, \& Blancard}]{Massara:2022a}
Massara, E., Villaescusa-Navarro, F., Hahn, C., {et~al.} 2023, Astrophys. J.,
  951, 70

\bibitem[{{Mohammad} \& {Percival}(2022)}]{Mohammad:2022}
{Mohammad}, F.~G. \& {Percival}, W.~J. 2022, \mnras, 514, 1289

\bibitem[{{Nadathur}(2016)}]{Nadathur:2016a}
{Nadathur}, S. 2016, \mnras, 461, 358

\bibitem[{{Nadathur} {et~al.}(2019{\natexlab{a}}){Nadathur}, {Carter}, \&
  {Percival}}]{Nadathur:2019b}
{Nadathur}, S., {Carter}, P., \& {Percival}, W.~J. 2019{\natexlab{a}}, \mnras,
  482, 2459

\bibitem[{{Nadathur} {et~al.}(2019{\natexlab{b}}){Nadathur}, {Carter},
  {Percival}, {Winther}, \& {Bautista}}]{Nadathur:2019c}
{Nadathur}, S., {Carter}, P.~M., {Percival}, W.~J., {Winther}, H.~A., \&
  {Bautista}, J.~E. 2019{\natexlab{b}}, \prd, 100, 023504

\bibitem[{{Nadathur} {et~al.}(2019{\natexlab{c}}){Nadathur}, {Carter},
  {Percival}, {Winther}, \& {Bautista}}]{Revolver}
{Nadathur}, S., {Carter}, P.~M., {Percival}, W.~J., {Winther}, H.~A., \&
  {Bautista}, J.~E. 2019{\natexlab{c}}, ascl:1907.023

\bibitem[{{Nadathur} \& {Crittenden}(2016)}]{Nadathur:2016b}
{Nadathur}, S. \& {Crittenden}, R. 2016, \apjl, 830, L19

\bibitem[{{Nadathur} \& {Percival}(2019)}]{Nadathur:2019a}
{Nadathur}, S. \& {Percival}, W.~J. 2019, \mnras, 483, 3472

\bibitem[{{Nadathur} {et~al.}(2020{\natexlab{a}}){Nadathur}, {Percival},
  {Beutler}, \& {Winther}}]{Nadathur:2020a}
{Nadathur}, S., {Percival}, W.~J., {Beutler}, F., \& {Winther}, H.~A.
  2020{\natexlab{a}}, \prl, 124, 221301

\bibitem[{{Nadathur} {et~al.}(2020{\natexlab{b}}){Nadathur}, {Woodfinden},
  {Percival}, {Aubert}, {Bautista}, {Dawson}, {Escoffier}, {Fromenteau},
  {Gil-Mar{\'\i}n}, {Rich}, {Ross}, {Rossi}, {Maga{\~n}a}, {Brownstein}, \&
  {Schneider}}]{Nadathur:2020b}
{Nadathur}, S., {Woodfinden}, A., {Percival}, W.~J., {et~al.}
  2020{\natexlab{b}}, \mnras, 499, 4140

\bibitem[{Neyrinck(2008)}]{Neyrinck_2008}
Neyrinck, M.~C. 2008, \mnras, 386, 2101

\bibitem[{{Neyrinck} {et~al.}(2009){Neyrinck}, {Szapudi}, \&
  {Szalay}}]{Neyrinck:2009}
{Neyrinck}, M.~C., {Szapudi}, I., \& {Szalay}, A.~S. 2009, \apjl, 698, L90

\bibitem[{{Nichol} {et~al.}(2006){Nichol}, {Sheth}, {Suto}, {Gray}, {Kayo},
  {Wechsler}, {Marin}, {Kulkarni}, {Blanton}, {Connolly}, {Gardner}, {Jain},
  {Miller}, {Moore}, {Pope}, {Pun}, {Schneider}, {Schneider}, {Szalay},
  {Szapudi}, {Zehavi}, {Bahcall}, {Csabai}, \& {Brinkmann}}]{Nichol:2006}
{Nichol}, R.~C., {Sheth}, R.~K., {Suto}, Y., {et~al.} 2006, \mnras, 368, 1507

\bibitem[{Padmanabhan {et~al.}(2012)Padmanabhan, Xu, Eisenstein, Scalzo,
  Cuesta, Mehta, \& Kazin}]{Padmanabhan:2012}
Padmanabhan, N., Xu, X., Eisenstein, D.~J., {et~al.} 2012, \mnras, 427, 2132

\bibitem[{{Paillas} {et~al.}(2021){Paillas}, {Cai}, {Padilla}, \&
  {S{\'a}nchez}}]{Paillas:2021}
{Paillas}, E., {Cai}, Y.-C., {Padilla}, N., \& {S{\'a}nchez}, A.~G. 2021,
  \mnras, 505, 5731

\bibitem[{Paillas {et~al.}(2023)Paillas, Cuesta-Lazaro, Zarrouk, Cai, Percival,
  Nadathur, Pinon, de~Mattia, \& Beutler}]{Paillas:2022}
Paillas, E., Cuesta-Lazaro, C., Zarrouk, P., {et~al.} 2023, \mnras, 522, 606

\bibitem[{{Paz} {et~al.}(2013){Paz}, {Lares}, {Ceccarelli}, {Padilla}, \&
  {Lambas}}]{Paz:2013}
{Paz}, D., {Lares}, M., {Ceccarelli}, L., {Padilla}, N., \& {Lambas}, D.~G.
  2013, \mnras, 436, 3480

\bibitem[{Paz {et~al.}(2011)Paz, Sgro, Merchan, \& Padilla}]{Paz:2011fj}
Paz, D.~J., Sgro, M.~A., Merchan, M., \& Padilla, N. 2011, \mnras, 414, 2029

\bibitem[{{Peebles}(1979)}]{Peebles:1979}
{Peebles}, P.~J.~E. 1979, \aj, 84, 730

\bibitem[{{Peebles}(1980)}]{Peebles_1980}
{Peebles}, P.~J.~E. 1980, {The large-scale structure of the universe}
  (Princeton University Press)

\bibitem[{{Philcox} \& {Ivanov}(2022)}]{Philcox:2022a}
{Philcox}, O. H.~E. \& {Ivanov}, M.~M. 2022, \prd, 105, 043517

\bibitem[{{Pisani} {et~al.}(2014){Pisani}, {Lavaux}, {Sutter}, \&
  {Wandelt}}]{Pisani:2014}
{Pisani}, A., {Lavaux}, G., {Sutter}, P.~M., \& {Wandelt}, B.~D. 2014, \mnras,
  443, 3238

\bibitem[{{Pisani} {et~al.}(2019){Pisani}, {Massara}, {Spergel}, {Alonso},
  {Baker}, {Cai}, {Cautun}, {Davies}, {Demchenko}, {Dor{\'e}}, {Goulding},
  {Habouzit}, {Hamaus}, {Hawken}, {Hirata}, {Ho}, {Jain}, {Kreisch}, {Marulli},
  {Padilla}, {Pollina}, {Sahl{\'e}n}, {Sheth}, {Somerville}, {Szapudi}, {van de
  Weygaert}, {Villaescusa-Navarro}, {Wandelt}, \& {Wang}}]{Pisani_WP}
{Pisani}, A., {Massara}, E., {Spergel}, D.~N., {et~al.} 2019, \baas, 51, 40

\bibitem[{{Pisani} {et~al.}(2015){Pisani}, {Sutter}, {Hamaus}, {Alizadeh},
  {Biswas}, {Wandelt}, \& {Hirata}}]{Pisani:2015}
{Pisani}, A., {Sutter}, P.~M., {Hamaus}, N., {et~al.} 2015, \prd, 92, 083531

\bibitem[{{Planck Collaboration} {et~al.}(2020){Planck Collaboration},
  {Aghanim}, {Akrami}, {Ashdown}, {Aumont}, {Baccigalupi}, {Ballardini},
  {Banday}, {Barreiro}, {Bartolo}, {Basak}, {Battye}, {Benabed}, {Bernard},
  {Bersanelli}, {Bielewicz}, {Bock}, {Bond}, {Borrill}, {Bouchet}, {Boulanger},
  {Bucher}, {Burigana}, {Butler}, {Calabrese}, {Cardoso}, {Carron},
  {Challinor}, {Chiang}, {Chluba}, {Colombo}, {Combet}, {Contreras}, {Crill},
  {Cuttaia}, {de Bernardis}, {de Zotti}, {Delabrouille}, {Delouis}, {Di
  Valentino}, {Diego}, {Dor{\'e}}, {Douspis}, {Ducout}, {Dupac}, {Dusini},
  {Efstathiou}, {Elsner}, {En{\ss}lin}, {Eriksen}, {Fantaye}, {Farhang},
  {Fergusson}, {Fernandez-Cobos}, {Finelli}, {Forastieri}, {Frailis},
  {Fraisse}, {Franceschi}, {Frolov}, {Galeotta}, {Galli}, {Ganga},
  {G{\'e}nova-Santos}, {Gerbino}, {Ghosh}, {Gonz{\'a}lez-Nuevo}, {G{\'o}rski},
  {Gratton}, {Gruppuso}, {Gudmundsson}, {Hamann}, {Handley}, {Hansen},
  {Herranz}, {Hildebrandt}, {Hivon}, {Huang}, {Jaffe}, {Jones}, {Karakci},
  {Keih{\"a}nen}, {Keskitalo}, {Kiiveri}, {Kim}, {Kisner}, {Knox},
  {Krachmalnicoff}, {Kunz}, {Kurki-Suonio}, {Lagache}, {Lamarre}, {Lasenby},
  {Lattanzi}, {Lawrence}, {Le Jeune}, {Lemos}, {Lesgourgues}, {Levrier},
  {Lewis}, {Liguori}, {Lilje}, {Lilley}, {Lindholm}, {L{\'o}pez-Caniego},
  {Lubin}, {Ma}, {Mac{\'\i}as-P{\'e}rez}, {Maggio}, {Maino}, {Mandolesi},
  {Mangilli}, {Marcos-Caballero}, {Maris}, {Martin}, {Martinelli},
  {Mart{\'\i}nez-Gonz{\'a}lez}, {Matarrese}, {Mauri}, {McEwen}, {Meinhold},
  {Melchiorri}, {Mennella}, {Migliaccio}, {Millea}, {Mitra},
  {Miville-Desch{\^e}nes}, {Molinari}, {Montier}, {Morgante}, {Moss}, {Natoli},
  {N{\o}rgaard-Nielsen}, {Pagano}, {Paoletti}, {Partridge}, {Patanchon},
  {Peiris}, {Perrotta}, {Pettorino}, {Piacentini}, {Polastri}, {Polenta},
  {Puget}, {Rachen}, {Reinecke}, {Remazeilles}, {Renzi}, {Rocha}, {Rosset},
  {Roudier}, {Rubi{\~n}o-Mart{\'\i}n}, {Ruiz-Granados}, {Salvati}, {Sandri},
  {Savelainen}, {Scott}, {Shellard}, {Sirignano}, {Sirri}, {Spencer},
  {Sunyaev}, {Suur-Uski}, {Tauber}, {Tavagnacco}, {Tenti}, {Toffolatti},
  {Tomasi}, {Trombetti}, {Valenziano}, {Valiviita}, {Van Tent}, {Vibert},
  {Vielva}, {Villa}, {Vittorio}, {Wandelt}, {Wehus}, {White}, {White},
  {Zacchei}, \& {Zonca}}]{Planck2020}
{Planck Collaboration}, {Aghanim}, N., {Akrami}, Y., {et~al.} 2020, \aap, 641,
  A6

\bibitem[{{Potter} {et~al.}(2017){Potter}, {Stadel}, \&
  {Teyssier}}]{Potter:2016ttn}
{Potter}, D., {Stadel}, J., \& {Teyssier}, R. 2017, Computational Astrophysics
  and Cosmology, 4, 2

\bibitem[{{Raghunathan} {et~al.}(2020){Raghunathan}, {Nadathur}, {Sherwin}, \&
  {Whitehorn}}]{Raghunathan:2020}
{Raghunathan}, S., {Nadathur}, S., {Sherwin}, B.~D., \& {Whitehorn}, N. 2020,
  \apj, 890, 168

\bibitem[{{Ryden}(1995)}]{Ryden:1995}
{Ryden}, B.~S. 1995, \apj, 452, 25

\bibitem[{{S{\'a}nchez} {et~al.}(2017){S{\'a}nchez}, {Clampitt}, {Kovacs},
  {Jain}, {Garc{\'{\i}}a-Bellido}, {Nadathur}, {Gruen}, {Hamaus}, {Huterer},
  {Vielzeuf}, {Amara}, {Bonnett}, {DeRose}, {Hartley}, {Jarvis}, {Lahav},
  {Miquel}, {Rozo}, {Rykoff}, {Sheldon}, {Wechsler}, {Zuntz}, {Abbott},
  {Abdalla}, {Annis}, {Benoit-L{\'e}vy}, {Bernstein}, {Bernstein}, {Bertin},
  {Brooks}, {Buckley-Geer}, {Carnero Rosell}, {Carrasco Kind}, {Carretero},
  {Crocce}, {Cunha}, {D'Andrea}, {da Costa}, {Desai}, {Diehl}, {Dietrich},
  {Doel}, {Evrard}, {Fausti Neto}, {Flaugher}, {Fosalba}, {Frieman},
  {Gaztanaga}, {Gruendl}, {Gutierrez}, {Honscheid}, {James}, {Krause}, {Kuehn},
  {Lima}, {Maia}, {Marshall}, {Melchior}, {Plazas}, {Reil}, {Romer}, {Sanchez},
  {Schubnell}, {Sevilla-Noarbe}, {Smith}, {Soares-Santos}, {Sobreira},
  {Suchyta}, {Tarle}, {Thomas}, {Walker}, {Weller}, \& {DES
  Collaboration}}]{Sanchez:2016}
{S{\'a}nchez}, C., {Clampitt}, J., {Kovacs}, A., {et~al.} 2017, \mnras, 465,
  746

\bibitem[{Satpathy {et~al.}(2019)Satpathy, Croft, Ho, \& Li}]{Satpathy:2019}
Satpathy, S., Croft, R. A.~C., Ho, S., \& Li, B. 2019, \mnras, 484, 2148

\bibitem[{Satpathy {et~al.}(2017)}]{BOSS:2016ntk}
Satpathy, S. {et~al.} 2017, \mnras, 469, 1369

\bibitem[{{Schuster} {et~al.}(2023){Schuster}, {Hamaus}, {Dolag}, \&
  {Weller}}]{Schuster:2023}
{Schuster}, N., {Hamaus}, N., {Dolag}, K., \& {Weller}, J. 2023, \jcap, 05, 031

\bibitem[{{Scoccimarro} {et~al.}(2001){Scoccimarro}, {Feldman}, {Fry}, \&
  {Frieman}}]{Scoccimarro:2001}
{Scoccimarro}, R., {Feldman}, H.~A., {Fry}, J.~N., \& {Frieman}, J.~A. 2001,
  \apj, 546, 652

\bibitem[{{Seo} \& {Eisenstein}(2003)}]{Seo:2003}
{Seo}, H.-J. \& {Eisenstein}, D.~J. 2003, \apj, 598, 720

\bibitem[{Sinha \& Garrison(2019)}]{Corrfunc2}
Sinha, M. \& Garrison, L. 2019, in Software Challenges to Exascale Computing,
  ed. A.~Majumdar \& R.~Arora (Singapore: Springer Singapore), 3--20

\bibitem[{{Sinha} \& {Garrison}(2020)}]{Corrfunc1}
{Sinha}, M. \& {Garrison}, L.~H. 2020, \mnras, 491, 3022

\bibitem[{Slepian {et~al.}(2017)Slepian, Eisenstein, Beutler, Chuang, Cuesta,
  Ge, Gil-Mar{\'{\i}}n, Ho, Kitaura, McBride, Nichol, Percival,
  Rodr{\'{\i}}guez-Torres, Ross, Scoccimarro, Seo, Tinker, Tojeiro, \&
  Vargas-Maga{\~{n}}a}]{Slepian:2017}
Slepian, Z., Eisenstein, D.~J., Beutler, F., {et~al.} 2017, \mnras, 468, 1070

\bibitem[{{Tallada} {et~al.}(2020){Tallada}, {Carretero}, {Casals},
  {Acosta-Silva}, {Serrano}, {Caubet}, {Castander}, {C{\'e}sar}, {Crocce},
  {Delfino}, {Eriksen}, {Fosalba}, {Gazta{\~n}aga}, {Merino}, {Neissner}, \&
  {Tonello}}]{CosmoHub_Tallada}
{Tallada}, P., {Carretero}, J., {Casals}, J., {et~al.} 2020, Astronomy and
  Computing, 32, 100391

\bibitem[{{Tinker}(2007)}]{Tinker:2007}
{Tinker}, J.~L. 2007, \mnras, 374, 477

\bibitem[{{Torrado} \& {Lewis}(2021)}]{Torrado:2020dgo}
{Torrado}, J. \& {Lewis}, A. 2021, \jcap, 2021, 057

\bibitem[{Uhlemann {et~al.}(2017)Uhlemann, Pajer, Pichon, Nishimichi, Codis, \&
  Bernardeau}]{Uhlemann:2017}
Uhlemann, C., Pajer, E., Pichon, C., {et~al.} 2017, \mnras, 474, 2853

\bibitem[{Valogiannis \& Dvorkin(2022)}]{Valogiannis:2022b}
Valogiannis, G. \& Dvorkin, C. 2022, \prd, 106, 103509

\bibitem[{{Valogiannis} \& {Dvorkin}(2022)}]{Valogiannis:2022a}
{Valogiannis}, G. \& {Dvorkin}, C. 2022, \prd, 105, 103534

\bibitem[{{Verde} {et~al.}(2002){Verde}, {Heavens}, {Percival}, {Matarrese},
  {Baugh}, {Bland-Hawthorn}, {Bridges}, {Cannon}, {Cole}, {Colless}, {Collins},
  {Couch}, {Dalton}, {De Propris}, {Driver}, {Efstathiou}, {Ellis}, {Frenk},
  {Glazebrook}, {Jackson}, {Lahav}, {Lewis}, {Lumsden}, {Maddox}, {Madgwick},
  {Norberg}, {Peacock}, {Peterson}, {Sutherland}, \& {Taylor}}]{Verde:2002}
{Verde}, L., {Heavens}, A.~F., {Percival}, W.~J., {et~al.} 2002, \mnras, 335,
  432

\bibitem[{{Wang} {et~al.}(2004){Wang}, {Yang}, {Mo}, {van den Bosch}, \&
  {Chu}}]{Wang:2004}
{Wang}, Y., {Yang}, X., {Mo}, H.~J., {van den Bosch}, F.~C., \& {Chu}, Y. 2004,
  \mnras, 353, 287

\bibitem[{{Wang} {et~al.}(2022){Wang}, {Zhao}, {Koyama}, {Percival},
  {Takahashi}, {Hikage}, {Gil-Mar{\'\i}n}, {Hahn}, {Zhao}, {Zhang}, {Mu}, {Yu},
  {Zhu}, \& {Ge}}]{Wang:2022}
{Wang}, Y., {Zhao}, G.-B., {Koyama}, K., {et~al.} 2022, arXiv:2202.05248

\bibitem[{{White}(2016)}]{White:2016}
{White}, M. 2016, \jcap, 2016, 057

\bibitem[{{Wolk} {et~al.}(2015){Wolk}, {Carron}, \& {Szapudi}}]{Wolk:2015}
{Wolk}, M., {Carron}, J., \& {Szapudi}, I. 2015, \mnras, 454, 560

\bibitem[{{Woodfinden} {et~al.}(2022){Woodfinden}, {Nadathur}, {Percival},
  {Radinovic}, {Massara}, \& {Winther}}]{Woodfinden:2022}
{Woodfinden}, A., {Nadathur}, S., {Percival}, W.~J., {et~al.} 2022, \mnras,
  516, 4307

\bibitem[{Yang \& Saslaw(2011)}]{Yang:2011}
Yang, A. \& Saslaw, W.~C. 2011, \apj, 729, 123

\bibitem[{{Zhao} {et~al.}(2022){Zhao}, {Variu}, {He}, {Forero-S{\'a}nchez},
  {Tamone}, {Chuang}, {Kitaura}, {Tao}, {Yu}, {Kneib}, {Percival}, {Shan},
  {Zhao}, {Burtin}, {Dawson}, {Rossi}, {Schneider}, \& {de la
  Macorra}}]{Zhao:2021}
{Zhao}, C., {Variu}, A., {He}, M., {et~al.} 2022, \mnras, 511, 5492

\end{thebibliography}


\begin{appendix}

\section{Accounting for correlation between data and theory}
\label{sec:Appendix}

\begin{figure}
	\centering
	\includegraphics[width=\hsize]{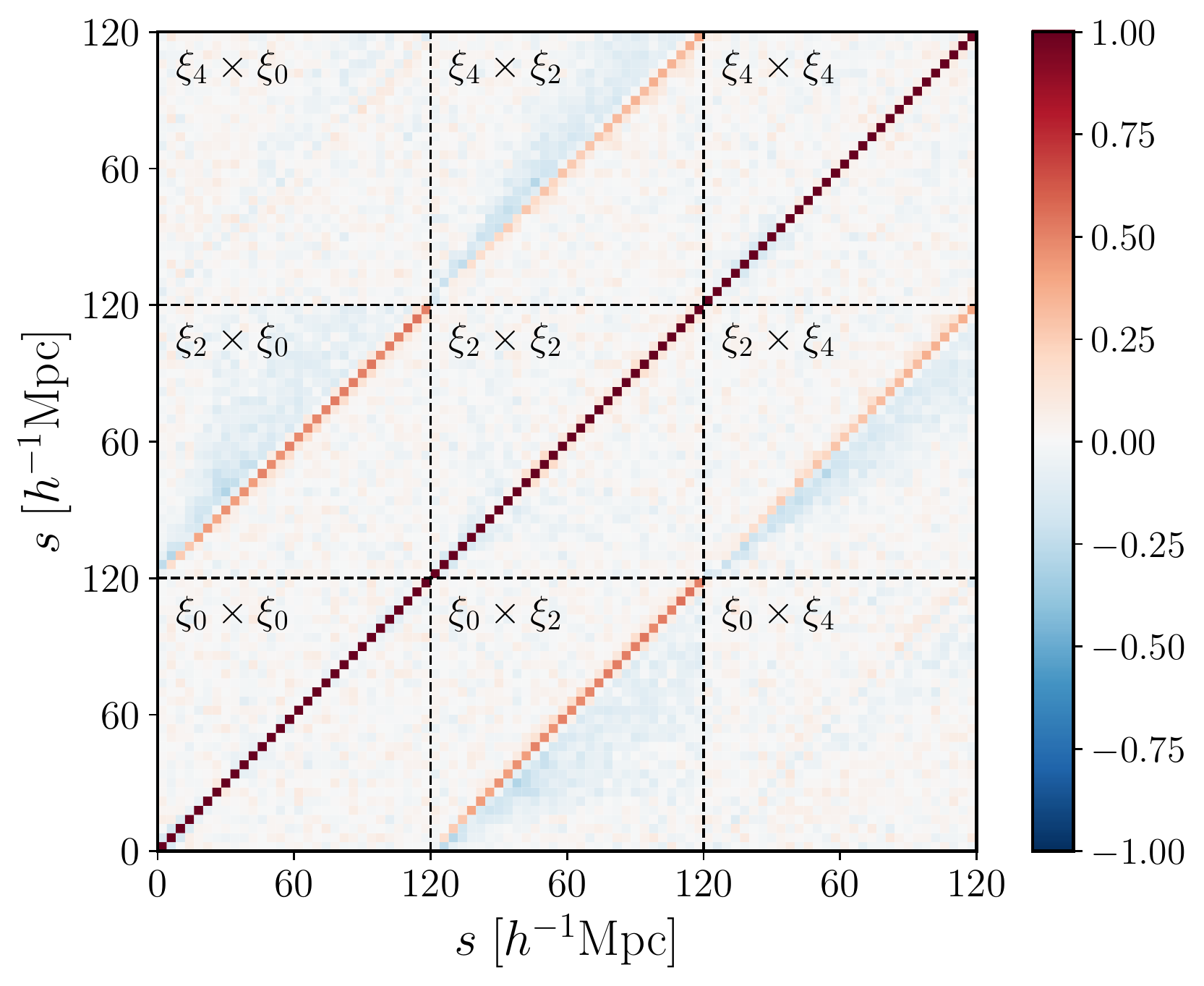}
	\caption{Normalised full covariance matrix $\tens{C}^{\rm tot}$ from the Flagship $\zeff = 1.0$ redshift bin, estimated using jackknife resampling. Dashed lines differentiate between monopole, quadrupole and hexadecapole blocks of the data vector.}
	\label{fig:fullcov}
\end{figure}

\begin{figure}
	\centering
	\includegraphics[width=\hsize]{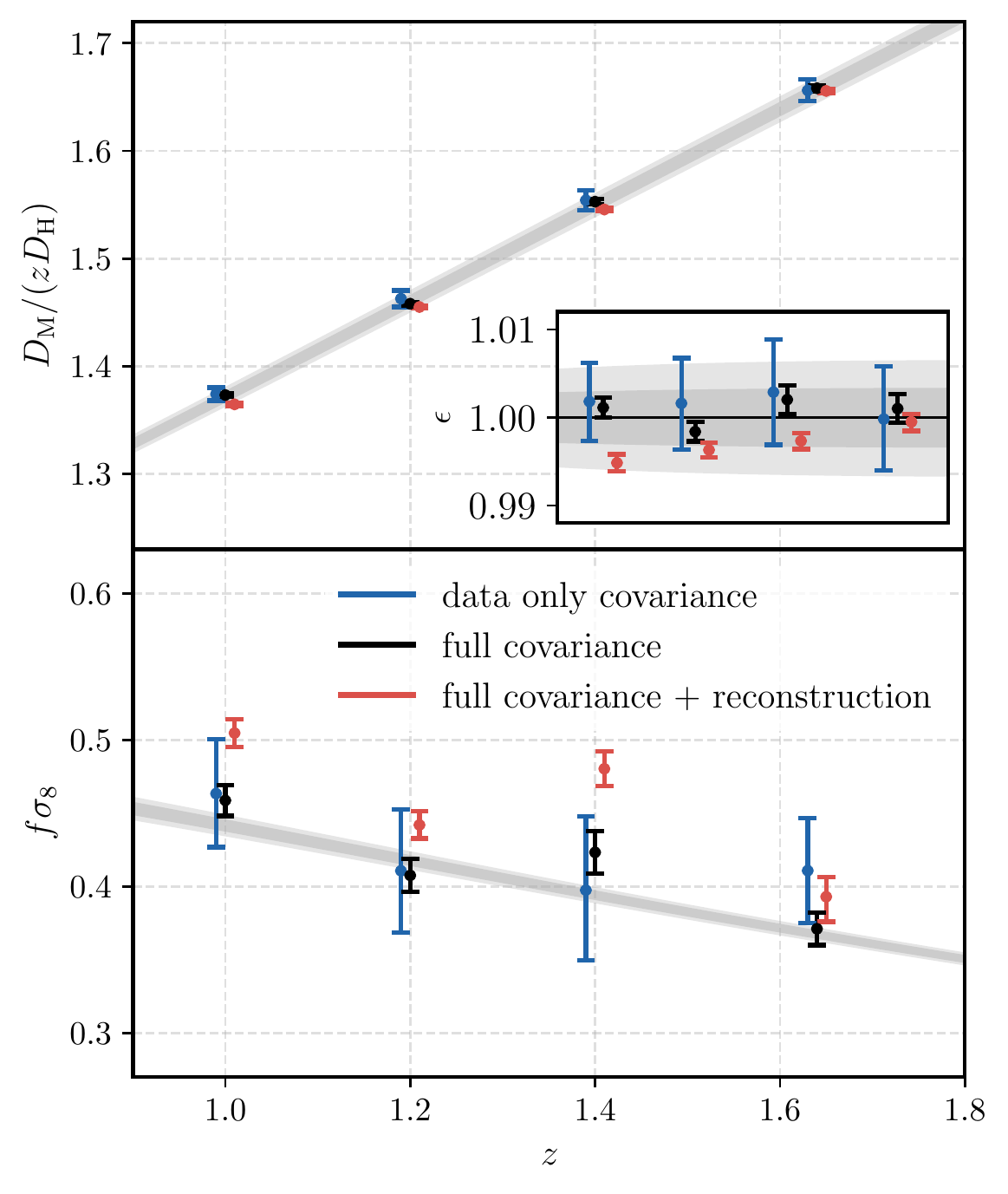}
	\caption{Redshift dependence of the measured cosmological parameters obtained from Flagship using different covariance prescriptions. The blue points show results for fits to the data vector for the perfect reconstruction case using $\tens{C}^{\rm data}$ and are the same as in \cref{fig:redshift_evolution}. Black points are for fits to the same data vector but using the full covariance $\tens{C}^{\rm tot}$. Light red points are for the fit to the realistic reconstruction case when using $\tens{C}^{\rm tot}$. The use of the full covariance $\tens{C}^{\rm tot}$ greatly reduces the statistical uncertainty but can introduce systematic offsets when combined with the use of reconstruction.}
	\label{fig:redshift_evolution_covmat}
\end{figure}

\begin{figure*}
	\centering
	\includegraphics[width=\hsize]{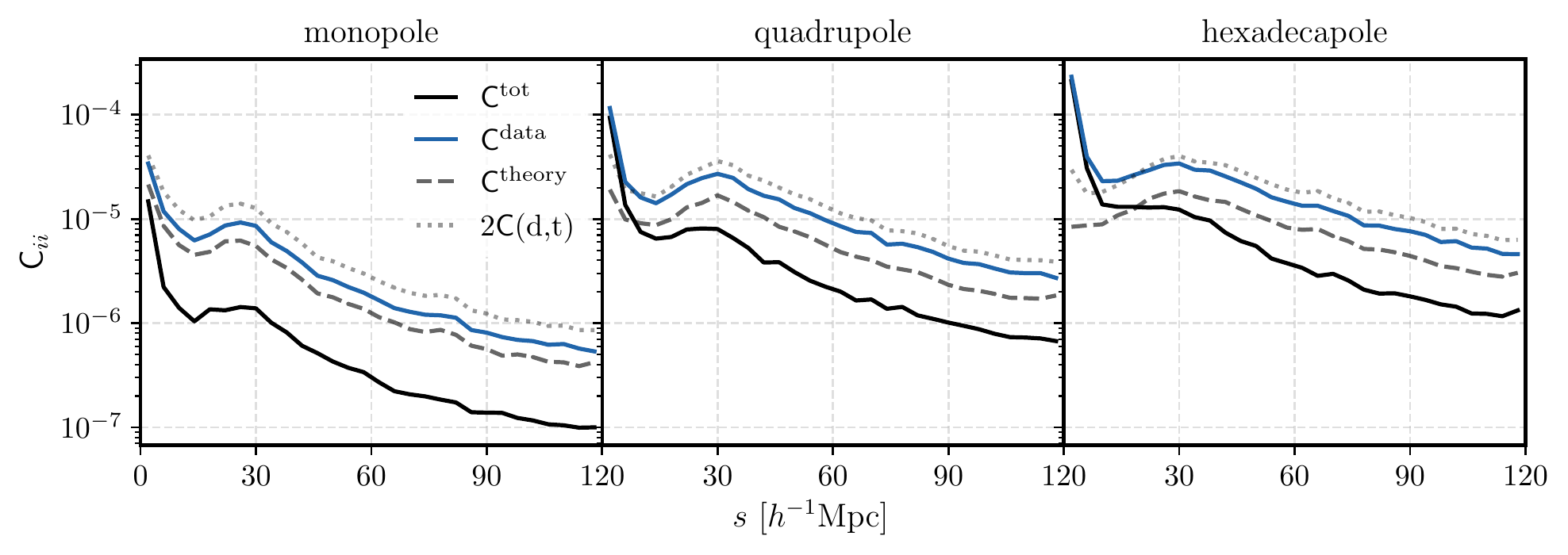}
	\caption{Diagonal elements of the full covariance matrix $\tens{C}^{\rm tot}$ (black lines), as well as the three different contributions from \cref{eq:DMTcov}, showing the reduced variance that results from properly accounting for correlation between data and theory. The three different panels represent the monopole, quadrupole and hexadecapole part, respectively. }
	\label{fig:covarianceterms}
\end{figure*}

\Cref{eq:covariance} defines the jackknife estimate of the covariance $\tens{C}^\mathrm{data}$ of the redshift-space data vector $\vec{\xi}^{\rm s}_\mathrm{data}$, which we have used for computing the likelihood in the main analysis. However, this is not the only source of uncertainty: the computation of the theory model $\vec{\xi}^{\rm s}_\mathrm{theory}$ depends on the multipoles of the real-space CCF $\xi^{\rm rr}_\ell$, which are themselves determined from measurements performed on the galaxy catalogue using the estimators described in \cref{subs:templates}. These estimators have an associated uncertainty, which is negligible when $\xi^{\rm rr}_\ell$ is estimated from the mean of a large number ($N\approx1000$) of mock realisations as done, for example, by \cite{Nadathur:2019c} and \cite{Woodfinden:2022}, but is significant in our case as the estimate is derived from a single Flagship realisation. Equally, since $\xi^{\rm rr}_\ell$ is estimated from the same Flagship void and galaxy catalogues which are used to measure the data vector, the uncertainties in $\vec{\xi}^{\rm s}_\mathrm{theory}$ and $\vec{\xi}^{\rm s}_\mathrm{data}$ will in principle be correlated with each other. 

In order to self-consistently account for these model uncertainties and correlations with the data vector, we can define a combined effective data vector $\vec{\xi}^{\rm tot}\equiv\vec{\xi}^{\rm s}_{\rm theory}-\vec{\xi}^{\rm s}_{\rm data}$ and estimate the covariance $\tens{C}\superscr{tot}$ using the same jackknife procedure described in \cref{subs:inference}. Then \cref{eq:likelihood} for the log-likelihood remains unchanged, provided that $\tens{C}\inv$ is understood to refer to the inverse covariance matrix of $\vec{\xi}^{\rm tot}$ and not $\vec{\xi}^{\rm s}_{\rm data}$ alone. We can write this schematically as: 
\begin{multline}
    \label{eq:DMTcov}
    \tens{C} \paren{{\rm data} - {\rm theory}} = 
          \tens{C}\paren{\rm data} 
        + \tens{C}\paren{\rm theory} \\ 
        -2\tens{C}\paren{\rm data, theory} \,,
\end{multline}
where the last term represents the cross-covariance between the data vector and the theory model. The familiar scenario corresponds to a situation where the second term on the right is negligible and the last term completely vanishes. This is the case by construction in previous studies in the literature \citep[e.g.][]{Nadathur:2020b, Woodfinden:2022}, since $\xi^{\rm rr}_\ell$ was taken as the mean from many mocks and thus data were theory are not correlated. For the method used in this work, given the limitations of only a single Flagship mock sample, this is not true. For future \Euclid data analyses, whether this term is important or not will depend on the choice of how to determine $\xi^{\rm rr}_\ell$.

For Flagship, we found a high degree of correlation between the model and the data vectors, which changes the correlation structure of the full covariance $\tens{C}^{\rm tot}$ with respect to $\tens{C}^{\rm data}$. This is shown in \cref{fig:fullcov}, and can be compared to \cref{fig:datacov}. Two features are immediately noticeable: off-diagonal correlations between the monopole and quadrupole components in particular are enhanced, and within each diagonal block of the covariance matrix, correlations between different radial bins are relatively suppressed. This behaviour arises because uncertainties in the measurement of the real and redshift-space monopoles $\xi^{\rm rr}_0$ and $\xi^{\rm rs}_0$ in each bin are naturally strongly correlated, as the real and redshift-space galaxy positions are closely correlated. In each radial bin, the monopole component $\xi^{\rm rr}_0$ strongly influences the corresponding modelled multipoles of $\vec\xi^{\rm s}_\mathrm{theory}$, thus creating the correlation between the elements of $\vec\xi^{\rm s}_\mathrm{theory}$ and the monopole component of $\vec\xi^{\rm s}_\mathrm{data}$.
 
Even more importantly, the strong correlation between real- and redshift-space multipole moments causes the amplitude of the diagonal elements of $\tens{C}^{\rm tot}$ to be significantly smaller than those of $\tens{C}^{\rm data}$. This is shown in \cref{fig:covarianceterms}. At first glance, it may seem unusual that accounting for additional uncertainty in the model effectively reduces the total uncertainty $\tens{C}^{\rm tot}$, but this is simply a reflection of the correlation that means that variations in $\vec\xi^{\rm s}_\mathrm{theory}-\vec\xi^{\rm s}_\mathrm{data}$ are reduced since uncertainties move both components in the same direction. It is intuitive that when $\vec\xi^{\rm s}_\mathrm{theory}$ and $\vec\xi^{\rm s}_\mathrm{data}$ are highly correlated with each other, models that deviate from observation should be more severely penalised in the log-likelihood, and this is achieved by the smaller values of $\tens{C}^{\rm tot}$. An equivalent way to view this is that when $\vec\xi^{\rm s}_\mathrm{theory}$ and $\vec\xi^{\rm s}_\mathrm{data}$ are both obtained from the same realisation of the initial conditions, the cosmic variance in them cancels out.

Changing the covariance matrix used in the likelihood evaluation in this way naturally propagates through to the recovered statistical uncertainty on the cosmological parameters. To estimate this, we performed all the model fits again using the appropriate full covariance $\tens{C}^{\rm tot}$ instead of $\tens{C}^{\rm data}$. The results obtained for $D_{\rm M}/D_{\rm H}$ and $f\sigma_8$ are shown in \cref{fig:redshift_evolution_covmat}, where the black points are for the perfect reconstruction scenario fit using $\tens{C}^{\rm tot}$, and the blue points are for fits to the same data using $\tens{C}^\mathrm{data}$. It is clear from this that correctly accounting for the correlation between model and data leads to a very significant increase in the statistical precision of the fit, roughly by a factor of 3 in both $D_{\rm M}/D_{\rm H}$ and $f\sigma_8$. The main results of our paper, which do not account for this effect, are therefore conservative overestimates of the statistical error that can be achieved with \Euclid if the real-space cross-correlation function is measured from the data instead of being taken from a simulation. 

However, the reduction in statistical uncertainty makes systematic errors more important. While in the perfect reconstruction scenario we are able to recover unbiased estimates of $D_{\rm M}/D_{\rm H}$ and $f\sigma_8$ at the higher precision, this is not so for the realistic reconstruction scenario (red points in \cref{fig:redshift_evolution_covmat}), where systematic offsets from the fiducial values are apparent. The source of these offsets is likely to be the residuals in the modelling of the monopole moment at small separations, an example of which can be seen in \cref{fig:realistic_recon_multipoles}. These are caused by the imperfections of the practical reconstruction technique. \Cref{fig:covarianceterms} shows that the relative reduction in covariance terms from using $\tens{C}^\mathrm{tot}$ instead of $\tens{C}^{\rm data}$ is largest for the monopole, so these residuals have a larger effect on the recovered fit. For the realistic reconstruction scenario, the full covariance matrix $\tens{C}^{\rm tot}$ therefore does not give a good estimate of the total error budget, and systematic uncertainties would have to be folded in. 

As we only have one Flagship simulation available, we cannot quantitatively assess the true size of these systematic uncertainties. However, it appears that the apparent large improvement in statistical precision would not be practically realised for \Euclid in the reconstruction scenarios, even if $\xi^{\rm rr}_\ell$ were estimated from the data and the full covariance $\tens{C}^{\rm tot}$ were used. Therefore the pessimistic forecasts we have obtained for the existing methodology using $\tens{C}^{\rm data}$ alone are more realistic.

\end{appendix}

\end{document}